\documentclass{article}

\usepackage{arxiv}

\usepackage{hyperref}       
\usepackage{url}            
\usepackage{booktabs}       
\usepackage{amsfonts}       
\usepackage{nicefrac}       
\usepackage{microtype}      
\usepackage{subcaption}
\usepackage{lipsum}
\usepackage{graphicx}
\usepackage{amsmath}
\usepackage{amsthm}
\usepackage{amssymb}

\newtheorem{definition}{Definition}
\newtheorem{remark}{Remark}
\newtheorem{theorem}{Theorem}
\newtheorem{lemma}{Lemma}

\newtheorem{proposition}{Proposition}

\graphicspath{ {./images/} }

\usepackage{tikz}
\usetikzlibrary{positioning,arrows.meta,fit}
\usepackage{pgfplots}
\pgfplotsset{compat=1.17}

\title{An Information-Theoretic Analysis of Information Hiding in Semantic Communication}

\author{
	Yanhao Huo \\
	College of Information Science and Technology \\
	Jinan University \\
	Guangzhou 510632, China \\
	\texttt{gzhu\_hyh@163.com} \\
	\And
	Shijun Xiang\thanks{Corresponding author.} \\
	College of Information Science and Technology \\
	Jinan University \\
	Guangzhou 510632, China \\
	\texttt{shijun\_xiang@qq.com} \\
}

\begin{document}
\maketitle
\begin{abstract}
Semantic communication shifts the transmission goal from the accurate recovery of syntactic symbols to the reliable conveyance of semantic content, yet a systematic information-theoretic characterization of information hiding at the semantic layer is still missing. Building on the semantic information theory of Niu-Zhang and the classical information hiding theory of Moulin-O'Sullivan, this paper establishes an information-theoretic framework for information hiding under semantic communication. The classical hiding model is extended to a two-layer structure (semantic host, syntactic host, syntactic composite signal, syntactic received, semantic received) bridged by synonymous mappings \(f_s,f_x\) and the inverse mapping \(f_y^{-1}\), subject to four distortion constraints: syntactic transparency, semantic fidelity, syntactic perturbation, and semantic destruction. The capacity is cast as a Stackelberg game with the hider moving first and the payoff \(J(Q,A)=I(\tilde U;\tilde Y\mid K)-I(\tilde U;\tilde S\mid K)\), yielding the single-letter maximin \(C^s=\max_Q\min_A J(Q,A)\), whose achievability and converse follow from random coding and joint typicality decoding. Relative to the classical capacity, comparison payoffs built from the up/down semantic mutual information give the endpoint sandwich \(C^s_{\mathrm{end},(0,1)}\le C^s\le C^s_{\mathrm{end},(1,0)}\) with the classical value \(C\) in the same interval, while closing semantic tolerance yields \(C^s_{\mathrm{no\text{-}tol}}\le C_{\mathrm{cl}}\) under a fair comparison. These results clarify that semantic hiding is not unconditionally superior to classical hiding. Binary, ternary, and Gaussian examples confirm the degeneration under one-to-one mappings and the capacity loss under coarsening.
\end{abstract}


\section{Introduction}
\label{sec:introduction}

Since Shannon laid the foundations of information theory in 1948~\cite{Shannon1948}, the design of communication systems has long centered on syntactic-level reliability, with channel capacity and rate-distortion functions providing precise theoretical limits for data compression and transmission~\cite{CoverThomas2006}. Within this framework, information hiding is formalized as a communication problem in which a message is embedded into a host signal and reliably transmitted over a noisy channel. Moulin and O'Sullivan characterized the hiding capacity as a zero-sum game between the hider and the attacker, derived a maximin expression with a mutual-information difference as the payoff function, and showed that this capacity admits closed-form solutions for binary and Gaussian sources~\cite{Moulin2003}. Subsequently, game-theoretic analyses, rate-distortion perspectives, and security constraints of information hiding have been studied systematically: Ettinger discussed game equilibria in steganalysis~\cite{Ettinger1998}, Somekh-Baruch and Merhav analyzed the capacity of watermarking with side information~\cite{SomekhBaruch2003}, Cohen and Lapidoth studied the Gaussian watermarking game~\cite{CohenLapidoth2002}, and Cachin provided an information-theoretic characterization of the security of steganographic systems from the relative-entropy perspective~\cite{Cachin1998}. Meanwhile, digital watermarking and fingerprinting, as two important applications of information hiding, have accumulated rich theoretical and algorithmic results on robustness, capacity, and collusion attacks~\cite{Cox1997,BonehShaw1998,Hartung1999,Petitcolas1999,Barni2001,Merhav2000}. The information-theoretic limits of information hiding are also closely related to classical problems such as coding with side information and dirty-paper coding~\cite{WynerZiv1976,GelfandPinsker1980,Costa1983,CsiszarKorner2011}.

However, in the classical information hiding theory described above, the ``host'' is always a syntactic object, such as pixels, transform coefficients, or sample points, whose distortion measures and capacity analyses do not involve the semantic content of the data.
In recent years, the rise of semantic communication has opened up a new dimension for communication theory.
As early as the year following the publication of Shannon's paper, Weaver pointed out that communication comprises three levels, namely syntactic, semantic, and pragmatic~\cite{Weaver1953}; yet a rigorous mathematical theory of the semantic level has long been lacking.
In 2024, Niu and Zhang established a mathematical framework for semantic information theory, with synonymous mappings as the core concept, and introduced semantic entropy, upper and lower semantic mutual information, semantic channel capacity, and semantic rate-distortion functions, proving that semantic measures enjoy directional advantages over their classical counterparts, such as \(H_s\le H\), \(C_s\ge C\), and \(R_s(D)\le R(D)\)~\cite{Niu2024}.
This framework shifts the focus of communication from ``whether symbols are accurately transmitted'' to ``whether meanings are accurately conveyed''.
Toward the system design of semantic communication, deep-learning-driven joint source-channel coding~\cite{Bourtsoulatze2019}, task-oriented communication~\cite{Gunduz2023}, semantic-aware networks~\cite{Shi2021}, and semantic communication architectures for 6G~\cite{Zhang2022,Strina2025} have been successively proposed; the standardization progress and multiple access techniques of semantic communication have also received attention~\cite{Niu2022,NativeAI2026}.

From a security perspective, the combination of semantic communication and information hiding has attracted attention in recent years.
The DRSW framework points out that traditional transform-domain watermarking fails during the semantic extraction process, and proposes a two-stage robust semantic watermark to protect both the semantic content and the reconstructed image simultaneously~\cite{Huo2025DRSW}.
WatCOM protects the intellectual property of semantic communication from the perspective of imperceptible watermarking~\cite{WatCOM2025}.
Signal-processing-driven watermarking schemes emphasize that watermark embedding should be jointly trained with the semantic codec to adapt to channel variations~\cite{Han2026Ownership}.
The steganography direction also has specialized designs tailored to semantic communication scenarios.
Huo et al.\ pointed out that existing robust steganographic methods cannot be correctly extracted in semantic communication because the embedding occurs at the pixel level or at the frequency-domain coefficient level, and proposed a GAN-based image semantic steganography framework that embeds the message directly into semantic features to ensure both successful extraction and that the semantic communication functionality remains unaffected~\cite{Huo2025ImageSS}.
Preliminary explorations of controllable semantic steganography have also appeared in the literature~\cite{Ni2026Stega}.
However, the aforementioned works primarily focus on algorithm design and experimental validation, and the capacity theory of semantic hiding, namely the maximum rate at which information can be reliably embedded under given semantic and syntactic distortion constraints, has not yet been systematically established.

To date, the two fields of semantic communication and information hiding have not yet established a systematic theoretical connection.
On the one hand, classical hiding capacity analysis cannot answer how the amount of hidden information changes when both the host and the receiver are assessed in terms of semantic equivalence classes rather than exact symbols.
On the other hand, although sandwich bounds on semantic mutual information in semantic information theory establish the ordering between semantic and syntactic quantities~\cite{Niu2024}, whether they apply to the payoff function of the hiding game and whether they can support the achievability and converse proofs have not yet been examined.
This gap leads to two consequences: the design of semantic hiding systems lacks capacity-based guidance, while the application boundaries of semantic information theory have not been fully explored.
The development of information theory itself indicates that when a new system of measures is introduced, its compatibility with existing capacity frameworks must be rigorously examined: the systematic treatment of information measures and coding theorems by Csisz\'ar and K\"orner~\cite{CsiszarKorner2011}, the classical analysis of coding with side information by Wyner-Ziv and Gel'fand-Pinsker~\cite{WynerZiv1976,GelfandPinsker1980}, and the dirty-paper coding of Costa~\cite{Costa1983} all provide methodological references for such an examination.

This paper aims to establish an information-theoretic framework for information hiding under semantic communication.
Specifically, we extend the classical hiding model to a two-layer structure consisting of the semantic host, the semantic composite signal, the syntactic composite signal, the syntactic receiver, and the semantic receiver, bridge the semantic and syntactic layers through synonymous mappings \(f_s,f_x,f_y\), and introduce four distortion constraints, namely syntactic transparency, semantic fidelity, syntactic perturbation, and semantic destruction.
On this basis, we characterize the hiding capacity as a Stackelberg game with the hider as the leader and the attacker as the follower, take the difference of two Shannon conditional mutual informations on semantic variables as the payoff function, derive a single-letter maximin expression for the capacity, and prove its achievability and converse along the lines of classical random coding and joint typicality decoding.
For comparison with the classical capacity, we further introduce a parameterized family of payoff functions based on the up/down semantic mutual information of Niu-Zhang, obtain the endpoint sandwich \(C^s_{\mathrm{end},(0,1)}\le C\le C^s_{\mathrm{end},(1,0)}\), and show, via a fair comparison, that when semantic tolerance is closed, carrier coarsening keeps the semantic hiding capacity from exceeding the classical capacity~\cite{Moulin2003}, thereby clarifying the applicable boundary of the claim that ``semantic capacity is greater than classical capacity'' in the hiding scenario.
Numerical examples for binary, \(q\)-ary, and Gaussian sources verify the degeneration property under one-to-one mappings and the capacity loss under coarsening mappings; whether semantic hiding outperforms classical hiding is jointly characterized by the endpoint sandwich and the upper bound obtained by closing semantic tolerance.

Our contributions are summarized as follows:
\begin{itemize}
	\item[(i)] \emph{Model:} On the basis of the classical hiding model, we introduce a two-layer characterization of the semantic and syntactic layers, present four distortion constraints bridged by synonymous mappings, and show that the model reduces to the classical form when the synonymous mappings degenerate to one-to-one mappings; this provides the modeling foundation for the subsequent capacity characterization.
	\item[(ii)] \emph{Game:} We characterize the semantic hiding capacity as a Stackelberg game with the hider as the leader and the attacker as the follower, point out that the attacker's feasible region depends on the hider's strategy and that the joint strategy space is non-rectangular, and derive the corresponding extremal structure $J_1\le J_2\le J_3$.
	\item[(iii)] \emph{Capacity:} We derive the single-letter maximin expression \(C^{s}=\max_{Q}\min_{A}J(Q,A)\) for the semantic hiding capacity, where \(J\) is the difference of two conditional mutual informations on semantic variables, and prove its achievability and converse.
	\item[(iv)] \emph{Comparison:} We introduce a family of comparison payoff functions based on up/down semantic mutual information, derive the endpoint sandwich for $C^s$ and its classical counterpart $C$, and prove that, under the fair comparison with closed semantic tolerance, the semantic capacity does not exceed the classical capacity $C_{\mathrm{cl}}$; these results clarify the applicable boundary of the claim that ``semantic capacity is greater than classical capacity'' and show that semantic hiding is not unconditionally superior to classical hiding.
\end{itemize}

This framework can be viewed both as a natural extension of classical information hiding theory to the semantic layer and as an analytical tool for the tradeoff between hiding capacity, transparency, and robustness in semantic communication systems.


\section{Mathematical Description of the Semantic Information Hiding Problem}
\label{sec:semantic_hiding_problem}

\subsection{Notation}
\label{subsec:notation}

This paper adopts the following notation. Uppercase letters such as \(X,\tilde{X}\) denote random variables, their realizations are denoted by lowercase letters such as \(x,\tilde{x}\), and their alphabets (value spaces) are denoted by calligraphic letters such as \(\mathcal{X},\tilde{\mathcal{X}}\). The symbol \(|\mathcal{X}|\) denotes the cardinality of the set \(\mathcal{X}\). Moreover, \(\mathcal{X}\times\mathcal{Y}\) denotes the Cartesian product, \(\mathcal{X}^n\) denotes the \(n\)-fold Cartesian power, and \(x^n=(x_1,\dots,x_n)\in\mathcal{X}^n\) denotes a sequence of length \(n\).

\begin{figure}[t]
	\centering
	\includegraphics[width=0.6\textwidth]{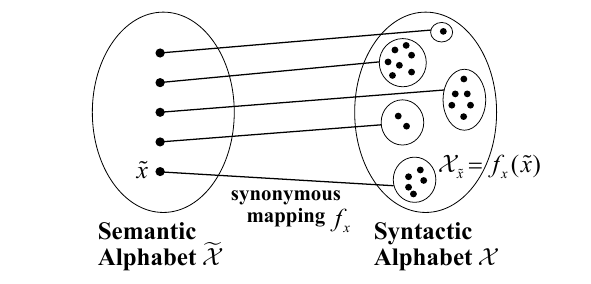}
	\caption{Illustration of the synonymous mapping \(f:\tilde{\mathcal{X}}\rightrightarrows\mathcal{X}\): each semantic symbol \(\tilde{x}\) corresponds to a synonymous class \(\mathcal{X}_{\tilde{x}}\subseteq\mathcal{X}\) in the syntactic alphabet, the synonymous classes are mutually disjoint, and the number of classes does not exceed the number of syntactic symbols, i.e., \(|\tilde{\mathcal{X}}|\le|\mathcal{X}|\).}
	\label{fig:synonymous_mapping}
\end{figure}

For a single-valued mapping \(f:\mathcal{X}\to\mathcal{Y}\), let \(f^{n}:\mathcal{X}^{n}\to\mathcal{Y}^{n}\) denote its elementwise extension, that is, \(f^{n}(x^{n})=\bigl(f(x_{1}),\dots,f(x_{n})\bigr)\) for \(x^{n}=(x_{1},\dots,x_{n})\in\mathcal{X}^{n}\).
Let \(\widetilde{\mathcal{X}}\) denote the semantic alphabet and \(\mathcal{X}\) the syntactic alphabet.
The synonymous mapping is defined as a set-valued mapping \(f_x:\widetilde{\mathcal{X}}\rightrightarrows\mathcal{X}\) with \(\tilde{x}\mapsto \mathcal{X}_{\tilde{x}}:=f_x(\tilde{x})\subseteq\mathcal{X}\), where \(\mathcal{X}_{\tilde{x}}\) denotes the synonymous class corresponding to the semantic symbol \(\tilde{x}\), as illustrated in Fig.~\ref{fig:synonymous_mapping}.
This mapping is one-to-many: a semantic symbol \(\tilde{x}\) may correspond to multiple syntactic symbols, that is, \(|\mathcal{X}_{\tilde{x}}|\) can be greater than \(1\); conversely, multiple distinct syntactic symbols \(x,x'\in\mathcal{X}_{\tilde{x}}\) correspond to the same semantic symbol \(\tilde{x}\).
Throughout this paper, \(f_x\) is treated as a set-valued correspondence, that is, \(X\in f_x(\tilde{X})\), rather than as a single-valued function; when \(|\mathcal{X}_{\tilde{x}}|=1\) holds for every \(\tilde{x}\), \(f_x\) degenerates to a one-to-one correspondence.
The family \(\bigl\{\mathcal{X}_{\tilde{x}}:\tilde{x}\in\widetilde{\mathcal{X}}\bigr\}\) forms a partition of \(\mathcal{X}\), that is,
\[
\bigcup_{\tilde{x}\in\widetilde{\mathcal{X}}}\mathcal{X}_{\tilde{x}}=\mathcal{X},
\qquad
\mathcal{X}_{\tilde{x}}\cap\mathcal{X}_{\tilde{x}'}=\varnothing
\quad(\tilde{x}\neq\tilde{x}'),
\]
and \(\mathcal{X}_{\tilde{x}}\neq\varnothing\) holds for every \(\tilde{x}\in\widetilde{\mathcal{X}}\).
In this case, each syntactic symbol \(x\in\mathcal{X}\) belongs to a unique synonymous class.
In the finite-alphabet case, \(|\widetilde{\mathcal{X}}|\le|\mathcal{X}|\), with equality if and only if every synonymous class is a singleton, that is, \(f_x\) degenerates to a one-to-one correspondence.
Under the partition condition, the inverse synonymous mapping \(f_x^{-1}:\mathcal{X}\to\widetilde{\mathcal{X}}\) is defined by \(f_x^{-1}(x)=\tilde{x}\iff x\in f_x(\tilde{x})\), so that \(\tilde{X}=f_x^{-1}(X)\).
For the synonymous mapping \(f_x\), its sequence extension is defined as \(f_x^{n}:\widetilde{\mathcal{X}}^{n}\rightrightarrows\mathcal{X}^{n}\) with \(f_x^{n}(\tilde{x}^{n}):=\prod_{i=1}^{n}f_x(\tilde{x}_i)\subseteq\mathcal{X}^{n}\), that is, \(x^{n}\in f_x^{n}(\tilde{x}^{n})\) if and only if \(x_i\in f_x(\tilde{x}_i)\) for all \(i=1,\dots,n\).
Since the partition condition ensures that each \(x\in\mathcal{X}\) belongs to a unique synonymous class, the inverse synonymous mapping can be extended elementwise to \(f_x^{-1,n}:\mathcal{X}^{n}\to\widetilde{\mathcal{X}}^{n}\) with \(f_x^{-1,n}(x^{n}):=\bigl(f_x^{-1}(x_1),\dots,f_x^{-1}(x_n)\bigr)\), so that \(\tilde{X}^{n}=f_x^{-1,n}(X^{n})\).
In what follows, all synonymous mappings satisfy the above partition condition, so their inverse mappings are single-valued functions.

In the memoryless setting with the semantic variable \(\tilde{X}\) as the source, the syntactic variable \(X\) is randomly generated from \(\tilde{X}\) within the synonymous class \(f_x(\tilde{x})\) via a fixed within-class selection mechanism \(p(x\mid\tilde{x})\), whose support is contained in \(f_x(\tilde{x})\), i.e., \(p(x\mid\tilde{x})>0\Rightarrow x\in f_x(\tilde{x})\). This paper sets this mechanism as a fixed prior agreed upon by both the sender and the receiver, which does not participate in optimization, and it is memoryless at the sequence level: \(p(x^n\mid\tilde{x}^n)=\prod_{i=1}^{n}p(x_i\mid\tilde{x}_i)\).

\(I(\cdot;\cdot)\) denotes the mutual information and \(H(\cdot\mid\cdot)\) denotes the conditional entropy; \(\log(\cdot)\) denotes the base-2 logarithm and \(\ln(\cdot)\) denotes the natural logarithm; \((x)^+=\max(x,0)\) denotes the positive part; \((\cdot)^{\mathrm{T}}\) denotes the transpose; \(\mathbb{E}(X)\) and \(\operatorname{Var}(X)\) denote the expectation and variance of a random variable \(X\), respectively; and \(\mathbb{I}(\cdot)\) denotes the indicator function.

\subsection{System Model}

\begin{figure}[t]
	\centering
	\begin{subfigure}[b]{0.60\textwidth}
		\centering
		\includegraphics[width=\textwidth]{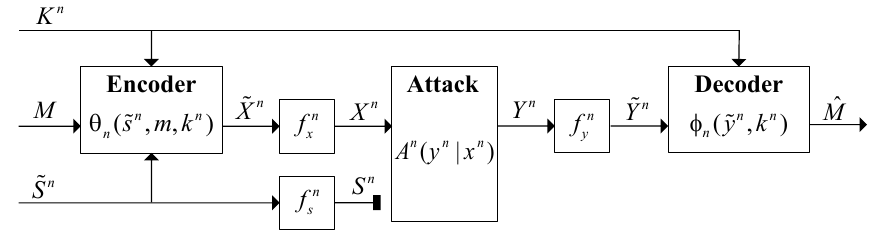}
		\caption{Hider's view}
		\label{fig:hider_view}
	\end{subfigure}
	\hfill
	\begin{subfigure}[b]{0.35\textwidth}
		\centering
		\includegraphics[width=\textwidth]{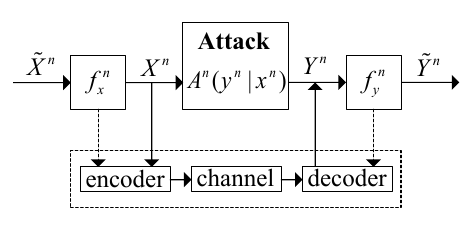}
		\caption{Attacker's view}
		\label{fig:attacker_view}
	\end{subfigure}
	\caption{Semantic information hiding model. (a) Hider's view: the embedding path is \(\tilde{S}^n\to\tilde{X}^n\to X^n\to Y^n\to\tilde{Y}^n\), and the virtual reference is \((\tilde{S}^n,S^n)\) with \(S^n\in f_s^n(\tilde{S}^n)\). The attack acts only on the actually transmitted composite signal \(X^n\), producing the attacked \(Y^n\) through the attack channel \(A^n(y^n|x^n)\). At the receiver, the evaluation is referenced directly against the original hosts \(S^n\) and \(\tilde{S}^n\). The side information $K^n$ is shared by the encoder and the decoder and is unknown to the attacker. (b) Attacker's view: the attacker only observes the syntactic composite signal $X^n$ and produces the attacked syntactic $Y^n$ through the attack channel $A^n(y^n|x^n)$; the semantic host $\tilde{S}^n$, the syntactic host $S^n$, and the side information $K^n$ are all invisible to the attacker.}
	\label{fig:semantic_hiding_models}
\end{figure}

Based on the semantic communication model proposed by Niu-Zhang~\cite{Niu2024} and the classical information hiding model of Moulin-O'Sullivan~\cite{Moulin2003}, this paper proposes a general semantic information hiding model, as shown in Fig.~\ref{fig:semantic_hiding_models}. The system consists of three components: the encoder, the channel attack, and the decoder. At the transmitter, the hider embeds a message \(M\) into the semantic host sequence \(\tilde{S}^n\); the encoder can observe the syntactic host \(S^n\) and obtain its semantic coarsening \(\tilde{S}^n=f_s^{-1,n}(S^n)\). Since embedding is performed at the semantic layer, the encoder \(\theta_n\) jointly uses the semantic host \(\tilde{S}^n\), the message \(M\), and the side information \(K^n\) to generate the semantic composite signal \(\tilde{X}^n\), which is then mapped through the extended synonymous mapping \(f_x^n\) to the syntactic composite signal \(X^n\) actually transmitted over the channel. Over the channel, the attacker applies an attack channel \(A^n(y^n|x^n)\) to transform \(X^n\) into the attacked syntactic received sequence \(Y^n\) (note that the attacker can only observe \(X^n\), while both \(\tilde{S}^n\) and \(K^n\) are invisible to it). At the receiver, the received \(Y^n\) is first recovered into the semantic received sequence \(\tilde{Y}^n\) through the extended inverse synonymous mapping \(f_y^{-1,n}\); then, the decoder \(\phi_n\) combines \(\tilde{Y}^n\) with the side information \(K^n\) to output an estimate \(\hat{M}\) of the message.

Specifically, suppose that the syntactic source generates a sequence \({S}^n=({S}_1,\dots,{S}_n)\) of syntactic random variables of length \(n\), taking values in \(\mathcal{S}^n\), where \(\mathcal{S}\) is a finite syntactic alphabet representing the multimedia host dataset in classical information hiding (e.g., discrete cosine transform (DCT) coefficients or pixels).
Under the extended inverse synonymous mapping \(f_s^{-1,n}:{\mathcal{S}^n\to\widetilde{\mathcal{S}}^n}\), the corresponding semantic host sequence \(\tilde{S}^n\) is given by \(\tilde{S}^n=f_s^{-1,n}({S}^n)=(f_s^{-1}({S}_1),\dots,f_s^{-1}({S}_n))\).
The syntactic source sequence is assumed to be independent and identically distributed (i.i.d.) according to a known probability mass function \(p(s^n)=\prod_{i=1}^n p(s_i)\).
The semantic host \(\tilde{S}^n\) is a deterministic coarsening of the syntactic source \(S^n\), and its probability measure is induced by the corresponding syntactic synonymous set \(f_s^n(\tilde{s}^n)=\prod_{i=1}^n f_s(\tilde{s}_i)\):
\[
p(\tilde{s}^n)=\sum_{s^n\in f_s^n(\tilde{s}^n)}p(s^n).
\]
Given the semantic host \(\tilde{s}^n\), the conditional distribution of the syntactic host is
\[
p(s^n\mid\tilde{s}^n)
=\frac{p(s^n)\,\mathbb{I}\bigl(s^n\in f_s^n(\tilde{s}^n)\bigr)}
{\sum_{s'^n\in f_s^n(\tilde{s}^n)}p(s'^n)}
=\frac{p(s^n)\,\mathbb{I}\bigl(\tilde{s}^n=f_s^{-1,n}(s^n)\bigr)}
{p(\tilde{s}^n)},
\]
whose support is contained in \(f_s^n(\tilde{s}^n)\).
In the symbol-by-symbol memoryless case,
\[
p(s^n\mid\tilde{s}^n)=\prod_{i=1}^n p(s_i\mid\tilde{s}_i),
\qquad
p(s_i\mid\tilde{s}_i)
=
\frac{p(s_i)\,\mathbb{I}\bigl(s_i\in f_s(\tilde{s}_i)\bigr)}
{\sum_{s'\in f_s(\tilde{s}_i)}p(s')}
=\frac{p(s_i)\,\mathbb{I}\bigl(\tilde{s}_i = f_s^{-1}(s_i)\bigr)}
{p(\tilde{s}_i)}.
\]

Let the random variable \(M\) be the message uniformly distributed over the message set \(\mathcal{M}\).
Let \(K^n=(K_1,\dots,K_n)\) be the side information sequence, taking values in the finite set \(\mathcal{K}^n\) according to a known distribution \(p(k^n)\).
\(K^n\) may be correlated with \(\tilde{S}^n\), characterized by the joint distribution \(p(\tilde{s}^n,k^n)\), and is shared by the encoder and the decoder while being unknown to the attacker.

In the basic model, the hider employs a deterministic encoder \(\theta_n:\tilde{\mathcal{S}}^n\times\mathcal{M}\times\mathcal{K}^n\to\tilde{\mathcal{X}}^n\), which maps the semantic host sequence \(\tilde{S}^n\), the message \(M\), and the side information \(K^n\) into a semantic composite signal
 \[
 \tilde{X}^n=\theta_n(\tilde{S}^n,M,K^n),
 \]
taking values in \(\tilde{\mathcal{X}}^n\).
Under the extended synonymous mapping \(f_x^n:\tilde{\mathcal{X}}^n\rightrightarrows\mathcal{X}^n\), the encoder does not directly select the syntactic composite signal \(X^n\); instead, \(X^n\) is randomly generated within the synonymous class \(f_x^n(\tilde{X}^n)\) via a fixed within-class selection mechanism \(p(x^n\mid\tilde{x}^n)=\prod_{i=1}^{n}p(x_i\mid\tilde{x}_i)\), whose support is contained in \(f_x^n(\tilde{X}^n)=\prod_{i=1}^{n}f_x(\tilde{X}_i)\), that is, symbol by symbol, \(p(x_i\mid\tilde{x}_i)>0\Rightarrow x_i\in f_x(\tilde{X}_i)\) for \(i=1,\dots,n\).

Subsequently, \(X^n\) passes through an attacker-controlled conditional transition probability distribution \(A^n(y^n|x^n)\), producing the attacked syntactic received sequence \(Y^n\) taking values in \(\mathcal{Y}^n\).
The received \(Y^n\) is then transformed through the extended inverse synonymous mapping \(f_y^{-1,n}:\mathcal{Y}^n\to\tilde{\mathcal{Y}}^n\) to obtain the semantic received sequence 
\( \tilde{Y}^n = f_y^{-1,n}(Y^n) = (f_y^{-1}(Y_1), \dots, f_y^{-1}(Y_n)) \).

Finally, the decoder \(\phi_n:\tilde{\mathcal{Y}}^n\times\mathcal{K}^n\to\mathcal{M}\) uses the semantic received sequence \(\tilde{Y}^n\) and the side information \(K^n\) to produce an estimate \(\hat{M}\) of the message.

\paragraph{Hider's View}
At the generative level, the syntactic source sequence \(S^n\) is the underlying random variable, and the semantic host sequence \(\tilde{S}^n=f_s^{-1,n}(S^n)\) is a deterministic coarsening of the syntactic source obtained through the inverse synonymous mapping, satisfying \(p(\tilde{s}^n)=\sum_{s^n\in f_s^n(\tilde{s}^n)}p(s^n)\).
At the operational level of the hider, however, since the embedding operation is performed at the semantic layer, this paper defines the semantic host \(\tilde{S}^n\) as the source sequence of the encoder, while the syntactic host \(S^n\) serves only as the underlying physical realization and the reference for distortion evaluation.
The hider generates the semantic composite signal \(\tilde{X}^n\) from \(\tilde{S}^n\), the message \(M\), and the side information \(K^n\), and then selects the actually transmitted syntactic composite signal \(X^n\) within its synonymous class according to the conditional distribution \(p(x^n\mid\tilde{x}^n)\).
To define the distortion and design the encoder in the subsequent game, the hider, in the design phase, takes the semantic host sequence \(\tilde{S}^n\) without any embedded message and its corresponding syntactic host \(S^n\) as the virtual reference baseline, as shown in Fig.~\ref{fig:hider_view}.
This baseline exists only in the hider's theoretical analysis; it involves no actual physical transmission and is not subjected to the channel attack, and the attacker does not know its concrete form.

\paragraph{Attacker's View}
To align with the classical communication framework, as shown in Fig.~\ref{fig:attacker_view}, the underlying physical coding, transmission, and decoding mechanisms guided by the synonymous mappings \(f_x^n\) and \(f_y^n\) (indicated by the dashed box in the figure) are implicitly included in the attack module but do not participate in optimization, serving only to convert the semantic composite signal into a physical signal and recover it at the receiver.
Therefore, at the model level, the attacker can only observe the syntactic composite signal \(X^n\) transmitted at the physical layer, and produces the attacked syntactic signal \(Y^n\) through the attack channel \(A^n(y^n|x^n)\). The attacker does not know the side information \(K^n\), the original semantic host \(\tilde{S}^n\), or the syntactic host \(S^n\), nor does it know that \(X^n\) is the composite version of \(S^n\).
Although the attacker knows the public structure of the synonymous mappings \(f_x^n\) and \(f_y^n\), owing to the lack of the side information \(K^n\) and the host knowledge, it cannot accurately infer the specific semantics implied by \(X^n\), nor can it prevent the legitimate receiver from recovering the correct semantics \(\tilde{Y}^n\) via \(f_y^{-1,n}\). This information asymmetry is the fundamental source of the security of information hiding.

Compared with the classical information hiding model~\cite{Moulin2003}, the semantic model superimposes on the syntactic variables \((S,X,Y,K)\) three semantic layers \(\tilde{\mathcal{S}},\tilde{\mathcal{X}},\tilde{\mathcal{Y}}\), three synonymous mappings \(f_s,f_x,f_y\) and their sequence extensions \(f_s^n,f_x^n,f_y^n\), as well as the virtual reference baseline \((\tilde{S}^n,S^n)\) in the hider's design phase, where \(S^n\in f_s^n(\tilde{S}^n)\). When the three synonymous mappings are all one-to-one mappings, the above elements degenerate to the identity, and the model reduces to the classical form.

\subsection{Distortion Constraints}
\label{subsec:distortion_constraints}

In this model, to ensure the end-to-end security of semantic information hiding (transparency, fidelity, robustness, and utility), the system must simultaneously satisfy four distortion constraints.
These four distortions can be divided into two categories: the embedding distortion of the hider and the channel distortion of the attacker.
All distortion constraints are given in the form of sequence-level expectations, and the single-symbol distortion functions are extended by \(1/n\sum_i\) under the memoryless assumption.
Before deriving the distortion constraints, we first present the following conventions:
(i) The selection of the syntactic composite signal within the synonymous class is memoryless, that is,
\[
p(x^n\mid\tilde{x}^n)=\prod_{i=1}^{n}p(x_i\mid\tilde{x}_i);
\]
(ii) The side information is correlated with the syntactic host only through the semantic host, that is, \(S^n\to\tilde{S}^n\to K^n\) forms a Markov chain (given \(\tilde{S}^n\), \(K^n\) and \(S^n\) are conditionally independent), so that for \(\tilde{s}^n=f_s^{-1,n}(s^n)\),
\[
p(s^n,k^n)=p(s^n)\,\frac{p(\tilde{s}^n,k^n)}{p(\tilde{s}^n)};
\]
(iii) The message \(M\) is independent of \((S^n,K^n)\), and \(p(m)=1/|\mathcal{M}|\).

\paragraph{Embedding Distortion of the Hider}
When embedding the message \(M\), the hider must simultaneously control the modifications at both the syntactic and the semantic layers.

\begin{definition}[Syntactic Distortion Function of the Hider]
	\label{def:syntactic_distortion_hider}
	The syntactic distortion function of the hider is a nonnegative function \(d_0:\mathcal{S}\times\mathcal{X}\to\mathbb{R}_+\), whose symbol-by-symbol extension on sequences of length \(n\) is defined as
	\[
	d_{0}^{n}(s^n,x^n)=\frac{1}{n}\sum_{i=1}^{n}d_0(s_i,x_i),
	\]
	and the corresponding sequence-level syntactic distortion constraint is
	\begin{equation}
		\label{eq:D0_expanded}
		\begin{aligned}
			\mathbb{E}\bigl[d_0^n(S^n,X^n)\bigr]
			&=\sum_{s^n,k^n}p(s^n,k^n)\sum_{m}\frac{1}{|\mathcal{M}|}\\
			&\qquad\times
			\frac{1}{n}\sum_{i=1}^{n}
			\sum_{x_i\in f_x(\tilde{x}_i)}
			p(x_i\mid\tilde{x}_i)\,
			d_0(s_i,x_i)
			\le D_0,
		\end{aligned}
	\end{equation}
	where \(\tilde{x}_i=\theta_n\bigl(f_s^{-1,n}(s^n),m,k^n\bigr)_i\).
	In \eqref{eq:D0_expanded}, the two arguments of \(d_0\), namely \(s_i\in\mathcal{S}\) and \(x_i\in\mathcal{X}\), are both syntactic symbols; the semantic composite signal \(\tilde{x}_i\) affects the syntactic distortion only indirectly through the synonymous class \(f_x(\tilde{x}_i)\) and the conditional distribution \(p(x_i\mid\tilde{x}_i)\).
	Equivalently, \eqref{eq:D0_expanded} can be rewritten as
	\begin{equation}
		\label{eq:D0_expanded_tilde}
		\begin{aligned}
			\mathbb{E}\bigl[d_0^n(S^n,X^n)\bigr]
			&=\sum_{\tilde{s}^n,k^n}p(\tilde{s}^n,k^n)\sum_{m}\frac{1}{|\mathcal{M}|}\\
			&\qquad\times
			\frac{1}{n}\sum_{i=1}^{n}
			\sum_{s_i\in f_s(\tilde{s}_i)}
			p(s_i\mid\tilde{s}_i)
			\sum_{x_i\in f_x(\tilde{x}_i)}
			p(x_i\mid\tilde{x}_i)\,
			d_0(s_i,x_i)
			\le D_0,
		\end{aligned}
	\end{equation}
	where
	\[
	p(\tilde{s}^n,k^n)
	=\sum_{s^n\in f_s^n(\tilde{s}^n)}p(s^n,k^n),
	\qquad
	p(s_i\mid\tilde{s}_i)
	=\frac{p(s_i)\,\mathbb{I}\bigl(s_i\in f_s(\tilde{s}_i)\bigr)}
	{\sum_{s'\in f_s(\tilde{s}_i)}p(s')},
	\]
	and \(\tilde{x}_i=\theta_n(\tilde{s}^n,m,k^n)_i\).
	
\end{definition}

\begin{remark}[Measure Equivalence]
	\label{rem:measure_equivalence}
	By convention (ii), \(S^n\to\tilde{S}^n\to K^n\) forms a Markov chain, that is, given \(\tilde{S}^n\), \(K^n\) and \(S^n\) are conditionally independent, so that for \(\tilde{s}^n=f_s^{-1,n}(s^n)\),
	\[
	p(s^n,k^n)=p(s^n)\,\frac{p(\tilde{s}^n,k^n)}{p(\tilde{s}^n)}.
	\]
	Moreover, by the determinism of \(\tilde{S}^n=f_s^{-1,n}(S^n)\), given the semantic host \(\tilde{s}^n\), the within-class conditional distribution of the syntactic host is
	\[
	p(s^n\mid\tilde{s}^n)
	=\frac{p(s^n)\,\mathbb{I}\bigl(s^n\in f_s^n(\tilde{s}^n)\bigr)}
	{p(\tilde{s}^n)},
	\]
	that is, the conditional probability is zero when \(s^n\notin f_s^n(\tilde{s}^n)\); moreover, under the Markov assumption, \(p(s^n\mid\tilde{s}^n,k^n)=p(s^n\mid\tilde{s}^n)\).
	Thus, for any function \(g(s^n,k^n)\), we have
	\[
	\mathbb{E}_{p(\tilde{s}^n,k^n)}
	\Big[\sum_{s^n}p(s^n\mid\tilde{s}^n)\,g(s^n,k^n)\Big]
	=\sum_{s^n,k^n}p(s^n,k^n)\,g(s^n,k^n)
	=\mathbb{E}_{p(s^n,k^n)}\big[g(S^n,K^n)\big].
	\]
	Therefore, the two-step evaluation in \eqref{eq:D0_expanded_tilde}, which first takes the expectation over the semantic distribution and then over the within-class conditional distribution, is numerically equivalent to the one-step evaluation in \eqref{eq:D0_expanded}, which takes the expectation directly over the syntactic distribution \(p(s^n,k^n)\).
\end{remark}

\begin{definition}[Semantic Distortion Function of the Hider]
	\label{def:semantic_distortion_hider}
	The semantic distortion function of the hider is a nonnegative function \(d_1:\widetilde{\mathcal{S}}\times\widetilde{\mathcal{X}}\to\mathbb{R}_+\), whose symbol-by-symbol extension on sequences of length \(n\) is defined as
	\[
	d_{1}^{n}(\tilde{s}^n,\tilde{x}^n)
	=\frac{1}{n}\sum_{i=1}^{n}d_1(\tilde{s}_i,\tilde{x}_i),
	\]
	and the corresponding sequence-level semantic distortion constraint is
	\begin{equation}
		\begin{aligned}
			\mathbb{E}[d_1^{n}(\tilde{S}^n,\tilde{X}^n)]
			&=\sum_{\tilde{s}^n,k^n}p(\tilde{s}^n,k^n)\sum_{m}\frac{1}{|\mathcal{M}|}
			\,d_1^{n}\bigl(\tilde{s}^n,\theta_n(\tilde{s}^n,m,k^n)\bigr)\\
			&=\sum_{\tilde{s}^n,k^n}p(\tilde{s}^n,k^n)\sum_{m}\frac{1}{|\mathcal{M}|}
			\left(\frac{1}{n}\sum_{i=1}^{n}
			d_1\bigl(\tilde{s}_i,\theta_n(\tilde{s}^n,m,k^n)_i\bigr)\right)
			\le D_1.
		\end{aligned}
		\label{eq:D1_expanded}
	\end{equation}
	Here, \(\mathbb{E}[d_1^{n}(\tilde{S}^n,\tilde{X}^n)]\) is the sequence-level semantic distortion expectation, which is computed by summing over all possible semantic hosts, side information, and messages.
	Since the encoder \(\theta_n\) is a deterministic mapping, the semantic composite signal sequence \(\tilde{x}^n\) is uniquely determined by \((\tilde{s}^n,m,k^n)\); therefore, the summation over \(\tilde{x}^n\) reduces to a single-point substitution, without the need to introduce an indicator function.
	The sequence distortion is further expanded into the average distortion over the \(n\) symbols, and its expectation is ultimately required not to exceed the maximum allowed semantic distortion \(D_1\).
\end{definition}

\begin{remark}[Instantiation of Semantic Distortion in Deep Learning]
	\label{rem:dl_semantic_distortion}
	In practical deep learning semantic communication systems, the syntactic distortion \(d_0(S,X)\) is typically used to constrain the symbol-by-symbol difference at the physical layer: the mean squared error (MSE) can be adopted for continuous-valued hosts (e.g., pixels and DCT coefficients), while the Hamming distance can be adopted for discrete symbols (e.g., bits and quantization indices).
	The semantic distortion \(d_1(\tilde{S},\tilde{X})\), in contrast, is an abstract distortion at the semantic symbol layer; in practical systems, it can be approximated by the deep feature distance extracted by a pretrained model from the corresponding syntactic signal, so as to more accurately reflect the semantic deviation in human perception or downstream tasks.
	For example, the deep feature distances of VGG/ResNet are commonly used for perceptual similarity measures, and the CLIP feature distance is commonly used for cross-modal semantic consistency measures.
	These specific deep metric methods are only potential instantiations of \(d_1\), and the theoretical framework of this paper imposes no restriction on the specific form of \(d_1\).
\end{remark}

\paragraph{Channel Distortion of the Attacker}

\begin{definition}[Syntactic Distortion Function of the Attacker]
	\label{def:syntactic_distortion_attacker}
	The syntactic distortion function of the attacker is a nonnegative function \(d_2:\mathcal{X}\times\mathcal{Y}\to\mathbb{R}_+\), whose symbol-by-symbol extension on sequences of length \(n\) is defined as
	\[
	d_{2}^{n}(x^n,y^n)=\frac{1}{n}\sum_{i=1}^{n}d_2(x_i,y_i),
	\]
	and the corresponding sequence-level attack distortion constraint is
	\begin{equation}
		\begin{aligned}
			\mathbb{E}[d_2^n(X^n,Y^n)]
			&=\sum_{x^n}p(x^n)\sum_{y^n}A^n(y^n\mid x^n)\,d_2^n(x^n,y^n)\\
			&=\sum_{x^n}p(x^n)\sum_{y^n}A^n(y^n\mid x^n)
			\left(\frac{1}{n}\sum_{i=1}^{n}d_2(x_i,y_i)\right)
			\le D_2.
		\end{aligned}
		\label{eq:D2_expanded}
	\end{equation}
	Here, \(p(x^n)\) is the marginal distribution of the syntactic composite signal sequence, jointly induced by the joint distribution \(p(\tilde{s}^n,k^n)\) of the semantic host and the side information, the equiprobable message distribution \(p(m)=1/|\mathcal{M}|\), the deterministic encoder \(\theta_n\), and the syntactic composite signal selection mechanism \(p(x^n\mid\tilde{x}^n)\) within the synonymous class:
	\[
	p(x^n)
	=
	\sum_{\tilde{s}^n,k^n}p(\tilde{s}^n,k^n)
	\sum_{m}\frac{1}{|\mathcal{M}|}
	\sum_{\tilde{x}^n}
	\mathbb{I}\bigl(\tilde{x}^n=\theta_n(\tilde{s}^n,m,k^n)\bigr)\,
	p(x^n\mid\tilde{x}^n),
	\]
	where the support of \(p(x^n\mid\tilde{x}^n)\) is contained in \(f_x^n(\tilde{x}^n)=\prod_{i=1}^{n}f_x(\tilde{x}_i)\).
	\(A^n(y^n\mid x^n)\) is the channel transition probability chosen by the attacker.
	Since the attacker can only observe the syntactic composite signal \(X^n\) and does not know the message \(M\) or the side information \(K^n\), this expectation is expanded only over the joint space of \(X^n\) and \(Y^n\), and the sequence-level distortion is expanded into the symbol-by-symbol average distortion, with its total expectation bounded by the maximum allowed attack distortion \(D_2\).
\end{definition}

\begin{definition}[Semantic Distortion Function of the Attacker]
	\label{def:semantic_distortion_attacker}
	The semantic distortion function of the attacker is a nonnegative function \(d_3:\widetilde{\mathcal{X}}\times\widetilde{\mathcal{Y}}\to\mathbb{R}_+\), whose symbol-by-symbol extension on sequences of length \(n\) is defined as
	\[
	d_{3}^{n}(\tilde{x}^n,\tilde{y}^n)
	=\frac{1}{n}\sum_{i=1}^{n}d_3(\tilde{x}_i,\tilde{y}_i),
	\]
	and the corresponding sequence-level semantic attack distortion constraint is
	\begin{equation}
		\begin{aligned}
			\mathbb{E}[d_3^{n}(\tilde{X}^n,\tilde{Y}^n)]
			&=\sum_{x^n}p(x^n)\sum_{y^n}A^n(y^n\mid x^n)\,
			d_3^{n}\bigl(f_x^{-1,n}(x^n),f_y^{-1,n}(y^n)\bigr)\\
			&=\sum_{x^n}p(x^n)\sum_{y^n}A^n(y^n\mid x^n)
			\left(\frac{1}{n}\sum_{i=1}^{n}
			d_3\bigl(f_x^{-1}(x_i),f_y^{-1}(y_i)\bigr)\right)
			\le D_3.
		\end{aligned}
		\label{eq:D3_expanded}
	\end{equation}
	Here, \(p(x^n)\) is the marginal distribution of the syntactic composite signal sequence, jointly induced by the joint distribution \(p(\tilde{s}^n,k^n)\) of the semantic host and the side information, the equiprobable message distribution \(p(m)=1/|\mathcal{M}|\), the deterministic encoder \(\theta_n\), and the syntactic composite signal selection mechanism \(p(x^n\mid\tilde{x}^n)\) within the synonymous class; the support of this selection mechanism is contained in \(f_x^n(\tilde{x}^n)=\prod_{i=1}^{n}f_x(\tilde{x}_i)\).
	\(A^n(y^n\mid x^n)\) is the physical channel transition probability chosen by the attacker.
	
	It should be particularly emphasized that the attacker can only observe and act on the syntactic composite signal \(X^n\), and cannot directly manipulate the semantic sequence.
	Therefore, the semantic-level attack effect is evaluated through the known fixed synonymous mappings \(f_x\) and \(f_y\): \(f_x^{-1,n}(x^n)\) denotes the semantic equivalence class sequence corresponding to the syntactic composite signal at the transmitter, while \(f_y^{-1,n}(y^n)\) denotes the semantic sequence recovered from the attacked syntactic signal through the inverse mapping.
	This constraint imposes an upper bound on the average damage that a semantic-aware attacker can inflict on the macroscopic semantics through physical perturbations: the attacker may perturb across different synonymous classes, but the expectation of the semantic distortion must not exceed \(D_3\).
	When \(D_3\) is small, the attack can only change the semantics to a small degree; when \(d_3(\tilde{x},\tilde{y})=\mathbb{I}(\tilde{x}\neq\tilde{y})\) and \(D_3=0\), cross-class perturbations are completely prohibited, and the attack can only flip within the same synonymous class.
\end{definition}

\paragraph{Summary of the Physical Meaning of the Four Constraints}
The four distortion constraints respectively restrict the behavior of the system from the transmitter side and the attacker side:
\begin{itemize}
	\item[(i)] \(D_0\) constrains the syntactic modification of the hider: the modification of the host pixels or coefficients when embedding the message must not be too large; otherwise, anomalies can be detected by human eyes or statistical detection, thereby breaking the transparency of the syntactic layer;
	\item[(ii)] \(D_1\) constrains the semantic modification of the hider: the semantic composite signal \(\tilde{X}\) must not deviate too far from the semantic host \(\tilde{S}\); otherwise, even if the syntactic-layer modification is small, the embedded semantics already deviates from the original meaning, thereby breaking the semantic fidelity;
	\item[(iii)] \(D_2\) constrains the syntactic destruction of the attacker: the attacker's modification of the composite signal must not be too large; otherwise, it is equivalent to turning the signal into noise, and the attack degenerates into meaningless destruction;
	\item[(iv)] \(D_3\) constrains the semantic destruction of the attacker: it limits the extent to which the attack causes the semantic received signal \(\tilde{Y}\) to deviate from the semantic composite signal \(\tilde{X}\); when \(d_3(\tilde{x},\tilde{y})=\mathbb{I}(\tilde{x}\neq\tilde{y})\), \(D_3\) directly limits the probability of crossing synonymous classes, and the smaller \(D_3\) is, the more limited the attack's destruction of the semantics.
\end{itemize}
When the synonymous mappings are one-to-one mappings, the semantic sequences correspond one-to-one to the syntactic sequences, the semantic constraints \(D_1,D_3\) can be equivalently characterized by the corresponding syntactic constraints \(D_0,D_2\), and the model reduces to classical information hiding.

\paragraph{Design Criteria}
Following the semantic communication framework of~\cite{Niu2024} and incorporating the requirements of information hiding for transparency and robustness, the semantic information hiding system obeys the following three criteria.

\begin{itemize}
	\item[(i)] \emph{Invisibility of semantic information.}
	Semantic information itself cannot be directly observed and can only be manifested indirectly through syntactic signals; therefore, the hider must simultaneously satisfy the syntactic transparency \(D_0\) and the semantic fidelity \(D_1\) (see Definition~\ref{def:syntactic_distortion_hider} and Definition~\ref{def:semantic_distortion_hider}), ensuring that the embedding operation does not change the basic meaning of the original data.
	
	\item[(ii)] \emph{Synonymy of semantic information.}
	The same semantic can be expressed by multiple syntactic signals within the same synonymous class; the synonymous mappings \(f_s,f_x,f_y\) characterizing this partition are fixed priors constructed from the shared background knowledge of the transmitter and the receiver, act elementwise on sequences, and do not participate in optimization.
	
	\item[(iii)] \emph{Dual security constraints.}
	Security is subject to constraints at both the syntactic and the semantic layers: the attacker is jointly limited by the syntactic perturbation upper bound \(D_2\) and the semantic destruction upper bound \(D_3\) (see Definition~\ref{def:syntactic_distortion_attacker} and Definition~\ref{def:semantic_distortion_attacker}); the hider's task is to design a robust coding strategy that can reliably extract the message under the above dual constraints.
\end{itemize}

\subsection{Rate and Capacity}
The rate of a semantic information hiding code is \(R=\frac{1}{n}\log|\mathcal{M}|\), which characterizes the average number of message bits carried per semantic symbol.
The average error probability is
\begin{equation}
	\begin{aligned}
		P_{e,n}
		&=\frac{1}{|\mathcal{M}|}\sum_{m\in\mathcal{M}}
		\sum_{\tilde{s}^n,k^n}p(\tilde{s}^n,k^n)
		\sum_{\tilde{x}^n}
		\mathbb{I}\bigl(\tilde{x}^n=\theta_n(\tilde{s}^n,m,k^n)\bigr)\\
		&\qquad\times
		\sum_{x^n\in f_x^n(\tilde{x}^n)}
		p(x^n\mid\tilde{x}^n)
		\sum_{y^n}A^n(y^n\mid x^n)\,
		\mathbb{I}\Bigl(\phi_n\bigl(f_y^{-1,n}(y^n),k^n\bigr)\neq m\Bigr).
	\end{aligned}
	\label{eq:error_probability_expanded}
\end{equation}
Here, \(p(\tilde{s}^n,k^n)\) is the joint distribution of the semantic host and the side information, and \(p(x^n\mid\tilde{x}^n)\) is the syntactic composite signal selection mechanism within the synonymous class, whose support is contained in \(f_x^n(\tilde{x}^n)=\prod_{i=1}^n f_x(\tilde{x}_i)\).
For a given message \(m\) and underlying realization \((\tilde{s}^n,k^n)\), the semantic composite signal is uniquely determined by the deterministic encoder as \(\tilde{x}^n=\theta_n(\tilde{s}^n,m,k^n)\); the actually transmitted syntactic composite signal \(X^n\) is selected within the synonymous class according to \(p(x^n\mid\tilde{x}^n)\).
After the receiver obtains \(y^n\) through the attack channel, it first obtains the semantic received sequence via the inverse synonymous mapping \(f_y^{-1,n}\), and then outputs a decision via the decoder \(\phi_n\).
The indicator function counts the decoding errors, and averaging over the uniformly distributed message finally yields \(P_{e,n}\).

\begin{definition}[Achievable Semantic Rate]
	\label{def:achievable_rate}
	Given a family of synonymous mappings \((f_s,f_x,f_y)\) and distortion thresholds \(D_0,D_1,D_2,D_3\), if there exists a sequence of semantic information hiding codes \((\mathcal{M},\theta_n,\phi_n)\) (\(n\ge1\)) satisfying the transmitter-side distortion constraints \(\mathbb{E}\bigl[d_0^n(S^n,X^n)\bigr]\le D_0\) and \(\mathbb{E}\bigl[d_1^n(\tilde{S}^n,\tilde{X}^n)\bigr]\le D_1\), and such that for all feasible attack channels \(A\in\mathcal A(D_2,D_3)\),
	\[
	\sup_{A\in\mathcal A(D_2,D_3)}
	P_{e,n}\bigl(A^{\otimes n}\bigr)\to 0
	\qquad(n\to\infty),
	\]
	then the rate \(R=\frac{1}{n}\log|\mathcal{M}|\) is said to be achievable under the attack distortion \((D_2,D_3)\).
	Here, \(A^{\otimes n}\) is the memoryless channel generated by the single-symbol channel \(A\), that is, \(A^{\otimes n}(y^n\mid x^n)=\prod_{i=1}^{n}A(y_i\mid x_i)\).
	\(\mathcal A(D_2,D_3)\) is the set of single-symbol attack channels satisfying \(\mathbb{E}\bigl[d_2(X,Y)\bigr]\le D_2\) and \(\mathbb{E}\bigl[d_3(\tilde{X},\tilde{Y})\bigr]\le D_3\), where \(X\sim p(x)\), \(Y\sim A(y\mid x)\), \(\tilde{X}=f_x^{-1}(X)\), \(\tilde{Y}=f_y^{-1}(Y)\), and \(p(x)\) is the code-induced single-symbol marginal of the composite signal (if the composite signal sequence is stationary, \(p(x)\) is its single-symbol marginal; in general, it is taken as the average of the per-position marginals \(\bar p(x)=\frac1n\sum_{i=1}^{n}p(x_i)\)).
	Under memoryless attacks and symbol-by-symbol distortion, the sequence-level attack constraint \(\frac1n\sum_i\mathbb{E}[d_2(X_i,Y_i)]\le D_2\) (and similarly for \(d_3\)) reduces to the above single-symbol constraint.
\end{definition}

\begin{definition}[Semantic Hiding Capacity]
	\label{def:semantic_hiding_capacity}
	Given a fixed family of synonymous mappings \((f_s,f_x,f_y)\) and distortion thresholds \(D_0,D_1,D_2,D_3\), the semantic hiding capacity is defined, under the above distortion constraints, as the supremum of all achievable rates \(R\) against the class of memoryless attack channels \(\{A^{\otimes n}\}_{n\ge1}\) generated by the single-symbol feasible channel set \(\mathcal A(D_2,D_3)\) (where \(A\in\mathcal A(D_2,D_3)\)):
	\begin{equation}
		C^s=\sup\{R:R\text{ is achievable}\}.
	\end{equation}
\end{definition}


\section{Semantic Information Hiding Game Framework}
\label{sec:semantic_hiding_game_framework}

The game in this paper is essentially a zero-sum game between the ``hider'' and the ``attacker''. The strategy of the hider is jointly constituted by the encoder \(\theta_n:\tilde{\mathcal{S}}^n\times\mathcal{M}\times\mathcal{K}^n\to\tilde{\mathcal{X}}^n\) and the decoder \(\phi_n:\tilde{\mathcal{Y}}^n\times\mathcal{K}^n\to\mathcal{M}\). The family of synonymous mappings \((f_s, f_x, f_y)\), as a physical prior known to both the transmitter and the receiver, does not participate in strategy optimization; the syntactic composite signal realization mechanism \(p(x^n\mid\tilde{x}^n)\) within the synonymous class is also a fixed mechanism agreed upon by the transmitter and the receiver (its memoryless form is given in convention (i)), and likewise does not participate in strategy optimization. Let the joint strategy space of the hider satisfying the transmitter-side distortion constraints \(D_0, D_1\) be denoted by:
\begin{equation}
	\mathcal{H}_n
	= \left\{ (\theta_n, \phi_n) \in \Theta_n \times \Phi_n :
	\mathbb{E}\bigl[d_0^n(S^n,X^n)\bigr] \le D_0, \
	\mathbb{E}\bigl[d_1^n(\tilde{S}^n,\tilde{X}^n)\bigr] \le D_1 \right\}.
\end{equation}

The strategy of the attacker is the attack channel \(A^n(y^n|x^n)\). Its feasible region is constrained not only by the physical attack distortion \(D_2\) but also by the semantic-aware distortion \(D_3\). Since the semantic quality at the receiver depends on the specific coding strategy of the hider, given the encoder \(\theta_n\), the conditional strategy space of the attacker is defined as:
\begin{equation}
	\mathcal{A}^n(\theta_n)
	= \left\{ A^n \in \mathcal{A}^n :
	\mathbb{E}\bigl[d_2^n(X^n,Y^n)\bigr] \le D_2, \
	\mathbb{E}\bigl[d_3^n(\tilde{X}^n,\tilde{Y}^n)\bigr] \le D_3 \right\}.
\end{equation}

Define an abstract payoff function \(J:\mathcal{H}_n\times\mathcal{A}^n\to\mathbb{R}\), which characterizes the efficiency or reliability of the hider's successful transmission of information. The hider maximizes the payoff function by choosing a strategy \((\theta_n, \phi_n) \in \mathcal{H}_n\), while the attacker minimizes the payoff function by choosing \(A^n \in \mathcal{A}^n(\theta_n)\). Since the attacker's feasible region \(\mathcal{A}^n(\theta_n)\) depends on the hider's strategy choice, the joint feasible region of this game exhibits a typical non-rectangular structure.

\begin{definition}[Nash Equilibrium]
	\label{def:nash_equilibrium}
	Given a fixed family of synonymous mappings \((f_s, f_x, f_y)\), a strategy profile \((\theta_n^*, \phi_n^*, A^{n*})\) is a Nash equilibrium of the game if and only if, for all feasible joint strategies \((\theta_n, \phi_n) \in \mathcal{H}_n\) of the hider and all feasible attack strategies \(A^n \in \mathcal{A}^n(\theta_n^*)\) of the attacker, the following holds:
	\begin{equation}
		\begin{aligned}
			J(\theta_n, \phi_n, A^{n*}) \leq J(\theta_n^*, \phi_n^*, A^{n*})
			\leq J(\theta_n^*, \phi_n^*, A^n).
		\end{aligned}
		\label{eq:nash_equilibrium}
	\end{equation}
\end{definition}

Since the attacker's feasible region \(\mathcal{A}^n(\theta_n)\) depends on the hider's strategy, the above Nash equilibrium is a generalized Nash equilibrium: when the hider deviates to \((\theta_n,\phi_n)\ne(\theta_n^*,\phi_n^*)\), the attacker's original strategy \(A^{n*}\) may no longer satisfy \(D_2,D_3\) (i.e., \(A^{n*}\notin\mathcal{A}^n(\theta_n)\)), and its existence must be argued separately.

\begin{remark}[Nonexistence of Saddle Point]
	\label{rem:saddle_nonexistence}
	Since the attacker's feasible region \(\mathcal{A}^n(\theta_n)\) depends on the hider's strategy \((\theta_n,\phi_n)\), the feasible region \(\mathcal{H}_n\times\mathcal{A}^n(\theta_n)\) of the game is generally not rectangular.
	A standard saddle point requires both strategy sets to be fixed and independent of the opponent's strategy, and to satisfy
	\[
	\max_{(\theta_n,\phi_n)\in\mathcal H_n}
	\min_{A^n\in\mathcal A^n}
	J(\theta_n,\phi_n,A^n)
	=
	\min_{A^n\in\mathcal A^n}
	\max_{(\theta_n,\phi_n)\in\mathcal H_n}
	J(\theta_n,\phi_n,A^n),
	\]
	where \(\mathcal A^n\) is independent of \((\theta_n,\phi_n)\).
	In the present game, if the common feasible set \(\mathcal A^n_{\mathrm{fix}}=\bigcap_{(\theta_n,\phi_n)\in\mathcal H_n}\mathcal A^n(\theta_n)\) is taken, then in general \(\mathcal A^n_{\mathrm{fix}}\subsetneq\mathcal A^n(\theta_n)\), the maximin value and the minimax value are attained on different feasible regions, and the two are generally unequal, so the saddle point is not guaranteed to exist.
	Therefore, this paper does not adopt the saddle point as the equilibrium concept, but instead uses the generalized Nash equilibrium of Definition~\ref{def:nash_equilibrium} as the analytical framework; its existence must be argued separately under specific conditions.
\end{remark}

In the game of this paper, the family of synonymous mappings \((f_s,f_x,f_y)\) is fixed as a known prior. Depending on whether the feasible region of the attack channel depends on the hider's strategy, two cases can be discussed:
\begin{itemize}
	\item[(i)] \emph{Ideal case (rectangular domain):} if the attack channel is independent of the hider's coding strategy (i.e., \(\mathcal{A}^n\) does not vary with \(\theta_n\)), then the feasible region is the rectangle \(\Lambda_1=\mathcal{H}_n\times\mathcal{A}^n\). In this case, the Nash equilibrium and the saddle point are equivalent, and the existence of the saddle point can be guaranteed by classical results such as Sion's theorem.
	\item[(ii)] \emph{Actual case (non-rectangular domain):} the feasible region is \(\Lambda_2=\{(\theta_n,\phi_n,A^n):A^n\in\mathcal{A}^n(\theta_n)\}\). Since the attacker's feasible strategy set varies with \(\theta_n\), the feasible region exhibits a typical non-rectangular structure. In this case, the classical saddle point is not guaranteed to exist, and one needs to introduce mixed strategies or rely on specific convexity conditions.
\end{itemize}
In summary, under the setting of a non-rectangular joint strategy space, the Nash equilibrium and the saddle point are not equivalent, and the saddle-point existence theorems cannot be directly applied; moreover, the hider and the attacker have different orders of moves and different information advantages, so the extremal values of the game also change accordingly.
Therefore, it is necessary to examine the game separately according to the order of moves: the first mover acts first, and the second mover can observe the first mover's action and make the optimal response, so the first mover is at an information disadvantage. The typical cases are as follows:
\begin{itemize}
	\item[(a)] \emph{Hider as the first mover:} the encoder and the decoder fully share information and jointly determine the joint strategy; then the attacker observes the composite signal and adopts the worst-case attack. This case corresponds to the maximin structure
	\[
	J_1=\max_{(\theta_n,\phi_n)\in\mathcal H_n}
	\min_{A^n\in\mathcal A^n(\theta_n)}J(\theta_n,\phi_n,A^n).
	\]
	As the first mover, the hider cannot foresee the attacker's specific action, and this value characterizes the conservative capacity lower bound that it can guarantee.
	
	\item[(b)] \emph{Fully sequential:} the encoder \(\theta_n\) is first determined and made public; after observing it, the attacker applies a single round of interference \(A^n\); the decoder \(\phi_n\) is then optimized with knowledge of the attacked state. This case corresponds to
	\[
	J_2=\max_{\theta_n:\ \mathbb{E}[d_0^n(S^n,X^n)]\le D_0,\ 
		\mathbb{E}[d_1^n(\tilde{S}^n,\tilde{X}^n)]\le D_1}
	\min_{A^n\in\mathcal A^n(\theta_n)}
	\max_{\phi_n\in\Phi_n}J(\theta_n,\phi_n,A^n).
	\]
	The internal joint coordination of the hider is broken, but the decoder has the ability of ex post remediation, constituting an intermediate extremal value between the upper and lower bounds.
	
	\item[(c)] \emph{Attacker as the first mover:} the attacker first makes its attack strategy public; after observing it, the hider, as the second mover, jointly optimizes the encoder and the decoder. If the attack strategy is required to satisfy \(D_2,D_3\) for all feasible strategies of the hider, then the outer attack feasible region is taken as \(\mathcal A^n_{\mathrm{fix}}\) (defined in Remark~\ref{rem:saddle_nonexistence}), and in this case
	\[
	J_3=\min_{A^n\in\mathcal A^n_{\mathrm{fix}}}
	\max_{(\theta_n,\phi_n)\in\mathcal H_n}J(\theta_n,\phi_n,A^n).
	\]
	The hider has the full information advantage of a second-mover observation, and this expression constitutes an optimistic upper bound of the system.
\end{itemize}
By the min-max inequality and \(\mathcal A^n_{\mathrm{fix}}\subseteq\mathcal A^n(\theta_n)\) (for all feasible \(\theta_n\)), we obtain \(J_1\le J_2\le J_3\), and this inequality chain holds under both rectangular and non-rectangular feasible regions.

Among the above three structures, cases (b) and (c) respectively characterize the special theoretical scenarios where the system design is split and where the attacker is known in advance. However, in practical communication and information hiding systems, the transmitter and the receiver must first jointly complete the design of the encoder and the decoder, and only then does the attacker observe the composite signal and launch the attack. Therefore, case (a) is the game architecture most consistent with physical reality.
Based on this architecture, we adopt its conservative lower bound \(J_1\) as the semantic hiding capacity bound of the system, so as to ensure the security reliability of the system under the worst-case attack:
\begin{equation}
	J^*
	=
	\max_{(\theta_n,\phi_n)\in\mathcal H_n}
	\min_{A^n\in\mathcal A^n(\theta_n)}
	J(\theta_n,\phi_n,A^n).
	\label{eq:stackelberg_value}
\end{equation}
Here, \(J^*\) corresponds to the maximin game value in which the hider first determines the joint strategy and the attacker then chooses the worst-case feasible attack after observing the composite signal, and it constitutes the core foundation of the subsequent capacity analysis.


\section{Semantic Hiding Capacity}
\label{sec:covert_attack_capacity}
\subsection{Randomized Strategy and Payoff Function}

In the abstract game established in Section~\ref{sec:semantic_hiding_game_framework}, the pure strategy of the hider is a deterministic coding-decoding pair \((\theta_n,\phi_n)\). If the pure strategy is restricted, the saddle point may not exist under the non-rectangular feasible region, and deterministic coding is vulnerable to targeted attacks by an attacker who knows the structure of the synonymous mappings. Therefore, it is necessary to allow the hider to choose the codebook in a randomized manner.
Specifically, the hider can use a random seed (whose distribution may be correlated with \(K^n\) but does not depend on the message \(M\)) to map the message to a family of semantic composite signal sequences, rather than a single sequence.
This random seed is denoted as the semantic auxiliary sequence \(\tilde{U}^n=(\tilde{U}_1,\dots,\tilde{U}_n)\), taking values in the finite semantic auxiliary alphabet \(\tilde{\mathcal{U}}^n\).
The joint distribution of \(\tilde{U}^n\) and \((\tilde{S}^n,K^n)\) is induced by the hider's coding strategy, and \(\tilde{U}^n\) participates in the decision at the decoder together with \(K^n\).

After introducing \(\tilde{U}\), the hider's strategy is no longer a deterministic mapping but a conditional distribution \(Q(\tilde{x},\tilde{u}\mid\tilde{s},k)\): given \(\tilde{s},k\), an auxiliary symbol is drawn according to \(p(\tilde{u}\mid\tilde{s},k)\), and then a semantic composite signal symbol is generated according to \(p(\tilde{x}\mid\tilde{s},\tilde{u},k)\).
The semantic composite signal \(\tilde{X}\) is bridged to the syntactic composite signal \(X\in f_x(\tilde{X})\) through the synonymous mapping \(f_x\), whose within-class selection is given by \(p(x\mid\tilde{x})\), and then enters the channel.
The attacker's strategy is still denoted as the syntactic-layer conditional distribution \(A(y\mid x)\), since the attacker can only observe \(X\).
Here, both \(Q\) and \(A\) are single-symbol conditional distributions; under the memoryless assumption, the sequence-level strategies of length \(n\) take the symbol-by-symbol product form
\[
Q^n(\tilde{x}^n,\tilde{u}^n\mid\tilde{s}^n,k^n)
=\prod_{i=1}^{n}Q(\tilde{x}_i,\tilde{u}_i\mid\tilde{s}_i,k_i),
\qquad
A^n(y^n\mid x^n)=\prod_{i=1}^{n}A(y_i\mid x_i),
\]
The subsequent maximin is taken over the single-symbol (memoryless) strategy space; the capacity result corresponds to the memoryless attack class, and the sequence-level strategies are obtained by memoryless extension.

After the hider chooses \(Q(\tilde{x},\tilde{u}\mid\tilde{s},k)\) and the attacker chooses \(A(y|x)\), their strategies are not independently constrained: the attacker's distortion constraints depend on the syntactic composite signal marginal jointly induced by \(Q\), the fixed within-class selection mechanism \(p(x\mid\tilde{x})\), and \(f_x\):
\[
p_Q(x)=\sum_{\tilde{u},\tilde{s},k}\sum_{\tilde{x}} Q(\tilde{x},\tilde{u}\mid\tilde{s},k)\,p(\tilde{s},k)\,p(x\mid\tilde{x}).
\]
Therefore, the joint strategy space of the game is generally not rectangular. The definitions of the covert channel, the attack channel, and the payoff function are given in turn below.

\begin{definition}[Covert Channel]
	\label{def:covert_channel}
	Given a family of synonymous mappings \((f_s,f_x,f_y)\) and distortion thresholds \(D_0,D_1\), a conditional distribution \(Q(\tilde{x},\tilde{u}\mid\tilde{s},k)\) is called a memoryless covert channel if it satisfies the semantic fidelity constraint
	\[
	\sum_{\tilde{x},\tilde{u},\tilde{s},k} d_1(\tilde{s},\tilde{x})\,Q(\tilde{x},\tilde{u}\mid\tilde{s},k)\,p(\tilde{s},k)\le D_1,
	\]
	and the syntactic transparency constraint
	\[
	\sum_{\tilde{x},\tilde{u},\tilde{s},k}\sum_{s\in f_s(\tilde{s})} p(s\mid\tilde{s})\sum_{x\in f_x(\tilde{x})} p(x\mid\tilde{x})\,d_0(s,x)\,Q(\tilde{x},\tilde{u}\mid\tilde{s},k)\,p(\tilde{s},k)\le D_0.
	\]
	Let the set of all memoryless \(Q\) satisfying the above two inequalities be denoted by \(\mathcal{Q}(D_0,D_1)=\bigl\{Q:\mathbb{E}[d_0^n(S^n,X^n)]\le D_0,\ \mathbb{E}[d_1^n(\tilde{S}^n,\tilde{X}^n)]\le D_1\bigr\}\). From the game-theoretic perspective, \(Q\) is the randomized strategy of the hider on the coding side, and \(\mathcal{Q}(D_0,D_1)\) is the strategy space of the hider on the coding side; the decoder \(\phi_n\) is determined by the joint typicality decoding rule specified later and does not participate in the optimization here. The two inequalities act only through the marginal of \(Q\) over \(\tilde{u}\), consistent with the sequence-level distortion formulation in Section~\ref{subsec:distortion_constraints}.
\end{definition}

\begin{definition}[Attack Channel]
	\label{def:attack_channel}
	Given \(Q\in\mathcal{Q}(D_0,D_1)\), a conditional distribution \(A(y|x)\) is called a memoryless attack channel feasible under this \(Q\) if it satisfies the syntactic perturbation constraint
	\[
	\sum_{x,y} d_2(x,y)\,A(y|x)\,p_Q(x)\le D_2,
	\]
	and the semantic destruction constraint
	\[
	\sum_{x,y} d_3\big(f_x^{-1}(x),f_y^{-1}(y)\big)\,A(y|x)\,p_Q(x)\le D_3.
	\]
	Let the set of all memoryless \(A\) satisfying the above two inequalities be denoted by \(\mathcal{A}(Q,D_2,D_3)=\bigl\{A:\mathbb{E}[d_2^n(X^n,Y^n)]\le D_2,\ \mathbb{E}[d_3^n(\tilde{X}^n,\tilde{Y}^n)]\le D_3\bigr\}\), which is the strategy space of the attacker given \(Q\). The semantic destruction constraint is the key distinction of this paper from the classical model: it limits the semantic deviation caused by the attacker at the semantic layer, preventing it from pushing the entire signal into the wrong synonymous class by exploiting the public synonymous mappings; when \(f_x,f_y\) are both one-to-one mappings, this constraint is isomorphic to the syntactic perturbation constraint, and \(\mathcal{A}(Q,D_2,D_3)\) reduces to the classical form.
\end{definition}

Thus, the joint strategy space of the game is
\begin{equation}
	\Lambda(D_0,D_1,D_2,D_3)
	=\bigl\{(Q,A):Q\in\mathcal{Q}(D_0,D_1),\ A\in\mathcal{A}(Q,D_2,D_3)\bigr\}.
	\label{eq:joint_strategy_space}
\end{equation}
Its non-rectangular structure is exactly the case discussed in Section~\ref{sec:semantic_hiding_game_framework}; when proving the existence of the saddle point later, this structure needs to be handled separately, and Sion's theorem under the rectangular domain cannot be directly applied.

\begin{remark}[Non-emptiness of the Strategy Spaces]
	\label{rem:nonempty}
	In what follows, it is always assumed that the two strategy spaces are non-empty: \(\mathcal{Q}(D_0,D_1)\neq\varnothing\) (for example, when \(\tilde{\mathcal X}=\tilde{\mathcal S}\) and \(d_0(s,s)=d_1(\tilde{s},\tilde{s})=0\), the ``zero embedding'' strategy is feasible), and for each \(Q\in\mathcal{Q}(D_0,D_1)\), \(\mathcal{A}(Q,D_2,D_3)\neq\varnothing\) (for example, when \(f_x=f_y\) and \(d_2(x,x)=d_3(\tilde{x},\tilde{x})=0\), the identity attack is feasible).
\end{remark}

\begin{definition}[Payoff Function]
	\label{def:payoff}
	Given a family of synonymous mappings \((f_s,f_x,f_y)\) and a strategy pair \((Q,A)\in\Lambda(D_0,D_1,D_2,D_3)\), where \(f_s,f_x,f_y\) are fixed as a prior known to both the transmitter and the receiver and do not participate in optimization, the payoff function is defined as
	\begin{equation}
		J(Q,A)
		= I(\tilde{U};\tilde{Y}\mid K)-I(\tilde{U};\tilde{S}\mid K),
		\label{eq:payoff}
	\end{equation}
	where \(I(\tilde{U};\tilde{Y}\mid K)\) and \(I(\tilde{U};\tilde{S}\mid K)\) are conditional mutual informations on semantic variables, namely,
	\[
	I(\tilde{U};\tilde{Y}\mid K)=H(\tilde{U}\mid K)-H(\tilde{U}\mid\tilde{Y},K),
	\qquad
	I(\tilde{U};\tilde{S}\mid K)=H(\tilde{U}\mid K)-H(\tilde{U}\mid\tilde{S},K),
	\]
	and \(H(\cdot\mid\cdot)\) is the conditional entropy on semantic variables.
	The joint distribution of \((\tilde{U},\tilde{S},K,\tilde{X},X,Y,\tilde{Y})\) is \(p(\tilde{s},k)\,Q(\tilde{x},\tilde{u}\mid\tilde{s},k)\,p(x\mid\tilde{x})\,A(y|x)\,\mathbb{I}\big(\tilde{y}=f_y^{-1}(y)\big)\), that is, \((\tilde{U},\tilde{S},K)\to\tilde{X}\to X\to Y\to\tilde{Y}\) forms a Markov chain.
\end{definition}

\begin{definition}[Equivalent Semantic Channel]
	\label{def:equivalent_semantic_channel}
	Given an attack channel \(A(y\mid x)\) and a fixed within-class selection mechanism \(p(x\mid\tilde{x})\), the equivalent semantic channel from the semantic composite signal \(\tilde{x}\) to the semantic received signal \(\tilde{y}\) is defined as
	\begin{equation}
		W_A(\tilde{y}\mid\tilde{x})
		=\sum_{x\in f_x(\tilde{x})} p(x\mid\tilde{x})
		\sum_{y:\,f_y^{-1}(y)=\tilde{y}} A(y\mid x),
		\label{eq:equivalent_semantic_channel}
	\end{equation}
	that is, first the semantic composite signal \(\tilde{x}\) is randomized into a syntactic composite signal \(x\) within the synonymous class \(f_x(\tilde{x})\) according to \(p(x\mid\tilde{x})\), then \(y\) is obtained through the attack channel \(A(y\mid x)\), and finally \(\tilde{y}\) is obtained by merging through the inverse synonymous mapping \(f_y^{-1}\).
	Consequently, the Markov chain \((\tilde{U},\tilde{S},K)\to\tilde{X}\to X\to Y\to\tilde{Y}\) can be compressed into the semantic-layer Markov chain \((\tilde{U},\tilde{S},K)\to\tilde{X}\to\tilde{Y}\), whose transition kernel is exactly \(W_A(\tilde{y}\mid\tilde{x})\); the memoryless extension at the sequence level is denoted by
	\[
	W_A^{\otimes n}(\tilde{y}^n\mid\tilde{x}^n)=\prod_{i=1}^{n}W_A(\tilde{y}_i\mid\tilde{x}_i).
	\]
	The joint typicality and single-letterization in the subsequent achievability and converse are all built on \(W_A^{\otimes n}\).
\end{definition}

\begin{proposition}[Cardinality Bound of the Auxiliary Variable]
	\label{prop:aux_cardinality}
	Given the semantic composite signal alphabet \(\widetilde{\mathcal{X}}\), let
	\[
	\widetilde{\Omega}=\{(\tilde{s},k):p(\tilde{s},k)>0\},
	\qquad
	\Omega=\{(s,k):p(s,k)>0\}
	\]
	denote the joint supports of the semantic host and the side information, and of the syntactic host and the side information, respectively.
	Let the alphabet of the semantic auxiliary variable \(\tilde{U}\) be \(\widetilde{\mathcal{U}}\).
	Then, for any conditional distribution \(Q(\tilde{x},\tilde{u}\mid\tilde{s},k)\) satisfying the distortion constraints and any fixed attack \(A\), there exists an auxiliary variable \(\tilde{U}'\) whose alphabet satisfies
	\[
	|\widetilde{\mathcal{U}}'|
	\le|\widetilde{\mathcal{X}}|\,|\widetilde{\Omega}|+1,
	\]
	and \(J(Q',A)=J(Q,A)\), where \(Q'\) is the corresponding distribution obtained by replacing \(\tilde{U}\) in \(Q\) with \(\tilde{U}'\).
	Furthermore, since \(|\widetilde{\mathcal{X}}|\le|\mathcal{X}|\) and \(\widetilde{\Omega}\) is the image of \(\Omega\) under \(\tilde{s}=f_s^{-1}(s)\) (so \(|\widetilde{\Omega}|\le|\Omega|\)), we have
	\[
	|\widetilde{\mathcal{U}}'|
	\le|\widetilde{\mathcal{X}}|\,|\widetilde{\Omega}|+1
	\le|\mathcal{X}|\,|\Omega|+1.
	\]
\end{proposition}

By Proposition~\ref{prop:aux_cardinality}, with the help of the concavity of the payoff function with respect to the auxiliary distribution and Carathéodory's theorem, the size of the auxiliary alphabet can be restricted to \(|\widetilde{\mathcal{U}}|\le|\widetilde{\mathcal{X}}|\,|\widetilde{\Omega}|+1\le|\mathcal{X}|\,|\Omega|+1\), without affecting the attainment of optimality (the proof is given in Appendix~\ref{app:aux_cardinality}).
In the above expression, the left end is a tight bound at the semantic layer, the right end is a relaxed bound at the syntactic layer, and the middle step is given by the merging property of the synonymous mappings; the gap between the two ends is determined by the merging degree of \(f_x\) and \(f_s\): the stronger the merging, the smaller the upper bound of the auxiliary alphabet; this upper bound is a sufficient condition, and the actually required size of the auxiliary alphabet may be smaller.

\subsection{Expression of the Semantic Hiding Capacity}
\label{subsec:capacity_expression}

Definition~\ref{def:semantic_hiding_capacity} in Section~\ref{sec:semantic_hiding_problem} has already given the semantic hiding capacity \(C^s\) in the form of the supremum of achievable rates. On this basis, by instantiating the payoff function as \(J(Q,A)\) of Definition~\ref{def:payoff} and carrying over the optimality argument of memoryless strategies, this supremum can be written in a single-letter maximin form.

\begin{definition}[Expression of the Semantic Hiding Capacity]
	\label{def:capacity_expression}
	Given a family of synonymous mappings \(f_{sxy}=(f_s,f_x,f_y)\) and distortion thresholds \(D_0,D_1,D_2,D_3\), the semantic hiding capacity is defined as the maximin value of the payoff function over the joint strategy space:
	\begin{equation}
		C^s(D_0,D_1,D_2,D_3)
		=\max_{Q\in\mathcal{Q}(D_0,D_1)}\ \min_{A\in\mathcal{A}(Q,D_2,D_3)}
		J(Q,A),
		\label{eq:capacity}
	\end{equation}
	where \(\mathcal{Q}(D_0,D_1)\) and \(\mathcal{A}(Q,D_2,D_3)\) are given by Definition~\ref{def:covert_channel} and Definition~\ref{def:attack_channel}, respectively, and the joint strategy space \(\Lambda(D_0,D_1,D_2,D_3)\) is given by \eqref{eq:joint_strategy_space}. If the synonymous mappings are allowed to participate in the design, then an outer maximization over \(f_{sxy}\) is further applied (in which case the two strategy spaces also vary with \(f_{sxy}\)):
	\begin{equation}
		C^s_{\mathrm{joint}}(D_0,D_1,D_2,D_3)
		=\max_{f_{sxy}}\ \max_{Q\in\mathcal{Q}(D_0,D_1;f_{sxy})}\ \min_{A\in\mathcal{A}(Q,D_2,D_3;f_{sxy})}
		J(Q,A;f_{sxy}).
		\label{eq:capacity_joint}
	\end{equation}
\end{definition}

\eqref{eq:capacity} and \eqref{eq:capacity_joint} are the analytic forms of the supremum in Definition~\ref{def:semantic_hiding_capacity} under the memoryless strategy class.
Its maximin structure is consistent with \(J^*\) in \eqref{eq:stackelberg_value} of Section~\ref{sec:semantic_hiding_game_framework}: \(Q\) corresponds to the hider's coding-side strategy, \(A\) corresponds to the attacker's strategy, \(\mathcal{Q}(D_0,D_1)\) and \(\mathcal{A}(Q,D_2,D_3)\) are their respective strategy spaces satisfying the distortion constraints, and the joint strategy space \(\Lambda(D_0,D_1,D_2,D_3)\) is generally non-rectangular.
\eqref{eq:capacity} corresponds to the basic model where \(f_{sxy}\) is fixed as a known prior, and \eqref{eq:capacity_joint} corresponds to the extended case where the synonymous mappings can participate in the design; the subsequent analysis is based primarily on \eqref{eq:capacity}.

\subsection{Existence of the Expression of the Semantic Hiding Capacity}
\label{subsec:capacity_existence}

\eqref{eq:capacity} writes the semantic hiding capacity as \(\max_Q\min_A J(Q,A)\). For this expression to be a genuine maximum (rather than a supremum that is not attained), two things must be proved: the inner minimum \(\min_A J\) is attainable for each \(Q\), and the outer maximum \(\max_Q\min_A J\) is attainable; this section proves them in turn.

\paragraph{Convexity of the Payoff Function}

\begin{proposition}[Convexity of \(J\)]
	\label{prop:payoff_convexity}
	For fixed \(p(\tilde{s},k)\), the following hold:
	\begin{enumerate}
		\item[(i)] For fixed \(Q\), \(J(Q,A)\) is a convex function of \(A(y|x)\) (the proof is given in Appendix~\ref{app:payoff_convexity});
		\item[(ii)] For fixed \(p(\tilde{x}\mid\tilde{s},\tilde{u},k)\) and \(A\), \(J\) is a concave function of \(p(\tilde{u}\mid\tilde{s},k)\) (the proof is given in Appendix~\ref{app:payoff_convexity_ii});
		\item[(iii)] For fixed \(p(\tilde{u}\mid\tilde{s},k)\) and \(A\), \(J\) is a convex function of \(p(\tilde{x}\mid\tilde{s},\tilde{u},k)\); on the simplex without distortion constraints, its maximum is attained at some zero-one distribution (the proof is given in Appendix~\ref{app:payoff_convexity_iii}).
	\end{enumerate}
\end{proposition}

\paragraph{Attainability of the Inner Minimum}

By Proposition~\ref{prop:payoff_convexity}(i), \(J\) is convex in \(A\); by Definition~\ref{def:attack_channel} and the finite alphabet assumption, \(\mathcal{A}(Q,D_2,D_3)\) is a convex compact set (the proof is given in Appendix~\ref{app:inner_min_attained}).
Under the finite alphabet, \(J\) is continuous in \(A\), hence lower semicontinuous; the infimum of a lower semicontinuous convex function on a convex compact set is attainable, so for each \(Q\in\mathcal{Q}(D_0,D_1)\), there exists \(A^*(Q)\in\mathcal{A}(Q,D_2,D_3)\) such that
\[
\min_{A\in\mathcal{A}(Q,D_2,D_3)}J(Q,A)=J(Q,A^*(Q)).
\]

\paragraph{Attainability of the Outer Maximum}

Under the finite alphabet assumption and the boundedness of the distortion functions, \(\mathcal{Q}(D_0,D_1)\) is a compact convex set.
The value function \(W(Q)=\min_A J(Q,A)\) is continuous in \(Q\): the attack feasible-set correspondence \(Q\mapsto\mathcal{A}(Q,D_2,D_3)\) is continuous. The upper semicontinuity is given by the closed graph (\(A\ge0\), \(\sum_y A(y|x)=1\), and the two distortion constraints are all continuous functions of \((Q,A)\)); the lower semicontinuity is given by an interior-point argument (in the typical case of Remark~\ref{rem:nonempty} with \(D_2,D_3>0\) and an anchor channel \(A_0\) strictly feasible for all \(Q\) (e.g., the identity channel when \(f_x=f_y\)), for any \(A\in\mathcal{A}(Q,D_2,D_3)\) and \(\lambda\in(0,1)\), the distortion of \(A_\lambda=(1-\lambda)A+\lambda A_0\) does not exceed \((1-\lambda)D_2,(1-\lambda)D_3\), and it remains feasible for sufficiently close \(Q'\) with \(A_\lambda\to A\); the degenerate case with zero thresholds is covered by taking limits of a positive threshold sequence).
Therefore, by Berge's maximum theorem, \(W(Q)\) is continuous (the proof is given in Appendix~\ref{app:outer_max_attained}); a continuous function attains its maximum on a compact set, so there exists \(Q^*\in\mathcal{Q}(D_0,D_1)\) such that
\[
\max_{Q\in\mathcal{Q}(D_0,D_1)}\min_{A\in\mathcal{A}(Q,D_2,D_3)}J(Q,A)
=\min_{A\in\mathcal{A}(Q^*,D_2,D_3)}J(Q^*,A).
\]
That is, the outer maximum is attainable.

\paragraph{Existence of the Capacity Expression}

By the attainability of the inner and outer extrema, the right-hand side \(\max_Q\min_A J\) of \eqref{eq:capacity} exists (rather than being a supremum that is not attained).
Here, the cardinality of the auxiliary variable is guaranteed by Proposition~\ref{prop:aux_cardinality} and Remark~\ref{rem:minmax_cardinality}, and does not grow with \(n\).

\subsection{Achievability and Converse}
\label{subsec:achievability_converse}

\begin{proposition}[Achievability]
	\label{prop:achievability}
	For any \(R<C^s\), there exists a sequence of semantic information hiding codes \((\mathcal{M},\theta_n,\phi_n)\) such that
	\[
	\sup_{A\in\mathcal{A}(\theta_n,D_2,D_3)}P_{e,n}(A^{\otimes n})\to 0,\qquad n\to\infty,
	\]
	where \(\mathcal{A}(\theta_n,D_2,D_3)\) is the set of memoryless attack channels satisfying the distortion constraints \(D_2,D_3\) under the composite signal distribution induced by the encoder \(\theta_n\).
\end{proposition}

\begin{proof}[Proof sketch]
	By the existence of the capacity expression, fix a covert channel \(Q^*\in\mathcal{Q}(D_0,D_1)\) attaining \(\max_Q\min_A J\) and its corresponding worst-case attack \(A^*\in\mathcal{A}(Q^*,D_2,D_3)\).
	The construction uses classical random coding and universal (compound-channel) joint typicality decoding: for each message, several auxiliary sequences are generated according to the conditional distribution \(p^*(\tilde u\mid k)\); the encoder selects an auxiliary sequence jointly typical with \((\tilde{s}^n,k^n)\) and generates the semantic composite signal accordingly; the decoder searches the codebook for an auxiliary sequence jointly typical with \((y^n,k^n)\). The decoding decision does not depend on the actual attack \(A\), but only on \(Q^*\) and the attack class \(\mathcal{A}(Q^*,D_2,D_3)\).
	For any \(A\in\mathcal{A}(\theta_n,D_2,D_3)\), the error probability is controlled by the joint typicality lemma (the semantic tuple is jointly typical according to the joint distribution induced by \(Q^{*\otimes n}\) and the equivalent semantic channel \(W_A^{\otimes n}\)), and it decays to zero exponentially when \(R<I_{Q^*,A}(\tilde{U};\tilde{Y}\mid K)-I_{Q^*}(\tilde{U};\tilde{S}\mid K)-\epsilon\).
	Since \(A^*\) minimizes \(I_{Q^*,A}(\tilde{U};\tilde{Y}\mid K)\), taking a finite \(\epsilon\)-net over the attack class can uniformly bound the error probability of any \(A\) by the bound written in terms of \(A^*\), and the composite signal distribution induced by \(\theta_n\) is given by \(Q^*\) in the typicality sense, so the error probability still decays to zero exponentially under the worst-case attack.
	The complete construction and error probability estimation are given in Appendix~\ref{app:achievability}.
\end{proof}

\begin{proposition}[Converse]
	\label{prop:converse}
	Let \((\mathcal{M},\theta_n,\phi_n)\) be a sequence of semantic information hiding codes satisfying the distortion constraints, with code rate \(R=\frac{1}{n}\log|\mathcal{M}|\), such that \(P_{e,n}(A^{\otimes n})\to 0\) for any memoryless attack \(A\in\mathcal{A}(\theta_n,D_2,D_3)\). Then
	\[
	R\le C^s.
	\]
\end{proposition}

\begin{proof}[Proof sketch]
	Let \((\mathcal M,\theta_n,\phi_n)\) be a sequence of feasible codes such that \(P_{e,n}(A^{\otimes n})\to0\) for any memoryless attack \(A^{\otimes n}\) (where \(A\) satisfies \(\mathbb E[d_2^n]\le D_2\) and \(\mathbb E[d_3^n]\le D_3\) simultaneously under the code-induced composite signal distribution).
	By time-averaging the single-letter distributions induced by the code at each time instant, we obtain a single-letter covert channel \(Q\in\mathcal Q(D_0,D_1)\).
	For any single-letter attack \(A\in\mathcal A(Q,D_2,D_3)\), its product attack \(A^{\otimes n}(y^n|x^n)=\prod_iA(y_i|x_i)\) satisfies \(\mathbb E[d_2^n]\le D_2\) and \(\mathbb E[d_3^n]\le D_3\) (the distortion is linear in the time average), so \(A^{\otimes n}\in\mathcal A^n(\theta_n)\), and hence \(P_{e,n}(A^{\otimes n})\to0\).
	By Fano's inequality and the Markov chain \((\tilde U^n,\tilde S^n,K^n)\to\tilde X^n\to X^n\to Y^n\to\tilde Y^n\) (compressed into the semantic-layer chain \((\tilde U^n,\tilde S^n,K^n)\to\tilde X^n\to\tilde Y^n\) via the equivalent semantic channel \(W_A^{\otimes n}\)), \(nR\) is upper-bounded by \(I(\tilde U^n;\tilde Y^n\mid K^n)-I(\tilde U^n;\tilde S^n\mid K^n)+1+nRP_{e,n}\).
	By single-letterizing the mutual informations term by term (Lemma~\ref{lem:single_letter}) and letting \(n\to\infty\), we obtain \(R\le J(Q_A,A)\).
	Note that \(Q_A\) given by the single-letterization generally varies with \(A\), since the auxiliary variable contains \(Y^{i-1}\), so one cannot directly take the infimum over \(A\) on the same \(Q\).
	Finally, using Kakutani's fixed-point theorem on the compact convex feasible attack set, this per-attack upper bound is closed into \(\min_A\) on the same \(Q\), yielding \(R\le\max_Q\min_AJ(Q,A)=C^s\).
	The complete derivation is given in Appendix~\ref{app:converse}.
\end{proof}

Combining Proposition~\ref{prop:achievability} and Proposition~\ref{prop:converse}, the achievability gives that any \(R<C^s\) is achievable, and the converse gives that any reliable code satisfies \(R\le C^s\); therefore, \(C^s\) is the supremum of achievable rates, i.e., the semantic hiding capacity.

\subsection{Semantic Hiding Capacity Theorem}
\label{subsec:capacity_theorem}

\begin{theorem}[Semantic Hiding Capacity]
	\label{thm:semantic_capacity}
	Under the setting of finite alphabets, a memoryless source, and memoryless attacks, given a family of synonymous mappings \(f_{sxy}=(f_s,f_x,f_y)\) and distortion thresholds \(D_0,D_1,D_2,D_3\), the semantic hiding capacity is
	\begin{equation}
		C^s(D_0,D_1,D_2,D_3)
		=\max_{Q\in\mathcal{Q}(D_0,D_1)}\ \min_{A\in\mathcal{A}(Q,D_2,D_3)}
		J(Q,A),
		\label{eq:capacity_theorem}
	\end{equation}
	where \(J(Q,A)=I(\tilde{U};\tilde{Y}\mid K)-I(\tilde{U};\tilde{S}\mid K)\), and \(\mathcal{Q}(D_0,D_1)\) and \(\mathcal{A}(Q,D_2,D_3)\) are the strategy spaces of the covert channel and the attack channel, respectively.
	Any \(R<C^s\) is achievable; any code satisfying the distortion constraints whose error probability under the worst-case attack tends to zero satisfies \(R\le C^s\).
	The auxiliary variable can be taken with \(|\tilde{\mathcal{U}}|\le|\tilde{\mathcal{X}}|\,|\tilde{\Omega}|+1\le|\mathcal{X}|\,|\Omega|+1\) (Proposition~\ref{prop:aux_cardinality} and Remark~\ref{rem:minmax_cardinality}).
\end{theorem}

\begin{proof}
	By Section~\ref{subsec:capacity_existence}, both the inner minimum and the outer maximum in \eqref{eq:capacity_theorem} are attainable, so the right-hand side of this equation is well-defined and is denoted by \(C^s\). By Proposition~\ref{prop:achievability}, for any \(R<C^s\), there exists a code sequence whose error probability under the worst-case attack tends to zero, i.e., \(C^s\) is achievable; by Proposition~\ref{prop:converse}, any code satisfying the distortion constraints whose error probability under the worst-case attack tends to zero satisfies \(R\le C^s\). Combining the two, \(C^s\) is exactly the supremum of achievable rates.
\end{proof}

In Theorem~\ref{thm:semantic_capacity}, \eqref{eq:capacity_theorem} gives the maximin form of the capacity, and the following two sentences give the operational meaning of this form. When \(f_{sxy}\) is a one-to-one mapping, \eqref{eq:capacity_theorem} reduces to the classical hiding capacity of Moulin-O'Sullivan~\cite{Moulin2003} (subject to the conditions of Remark~\ref{rem:d0d1}).

\begin{remark}[Cardinality Bound under Maximin]
	\label{rem:minmax_cardinality}
	The bound of Proposition~\ref{prop:aux_cardinality} preserves \(J(Q,A)\) under a fixed attack, whereas \eqref{eq:capacity} requires preserving the inner minimum \(\min_A J(Q,A)\).
	The rigorous approach is as follows: take a finite \(\delta\)-net \(\{A_1,\dots,A_{N_\delta}\}\) on the compact set \(\mathcal A(Q,D_2,D_3)\), and use the support lemma to simultaneously preserve \(p(\tilde x\mid\tilde s,k)\) and \(J(Q,A_j)\) (\(j=1,\dots,N_\delta\)), obtaining the bound \(|\tilde{\mathcal U}|\le|\tilde\Omega|(|\tilde{\mathcal X}|-1)+N_\delta+2\).
	By the uniform continuity of \(J\) in \(A\), \(|\min_A J(Q',A)-\min_A J(Q,A)|\) tends to zero uniformly as \(\delta\to0\).
	Letting \(\delta\to0\) shows that the supremum over all finite alphabets equals the maximum over the bounded alphabet class, so the value of \(C^s\) in Theorem~\ref{thm:semantic_capacity} is independent of the specific form of the auxiliary alphabet bound.
	The main text retains the tight bound under a fixed attack for notational simplicity.
\end{remark}

\subsection{Comparison with the Classical Hiding Capacity}
\label{subsec:comparison_classic}

The payoff function \(J(Q,A)\) defined earlier is based on the semantic Shannon mutual information and suffices for proving the achievability and the converse, but it has no direct ordering relationship with the classical capacity \(C\).
To characterize the gains and losses of semantic hiding relative to classical hiding, this section introduces, on the basis of \(J\), a parameterized family constituted by the up/down semantic mutual information, uses its endpoint capacities to give a sandwich of \(C^s\) and \(C\), and discusses the degeneration under one-to-one mappings and the capacity degeneration after closing semantic tolerance.
This section does not redefine the capacity; all capacities refer to the main capacity \(C^s\) of Section~\ref{subsec:capacity_theorem} or its endpoint counterparts.

For comparison with the classical capacity, let \(U\) denote the syntactic-layer auxiliary variable corresponding to the semantic auxiliary variable \(\tilde U\).
To avoid additional conventions, this paper takes the auxiliary variable as not participating in the synonymous coarsening, that is, \(U=\tilde U\) (the auxiliary mapping is the identity): only the host, the composite signal, and the receiver variables \(S,X,Y\) are coarsened into \(\tilde S,\tilde X,\tilde Y\) through the synonymous mappings.
Hence \(I(U;Y\mid K)=I(\tilde U;Y\mid K)\) and \(I(U;S\mid K)=I(\tilde U;S\mid K)\); the ``syntactic correspondence'' in Remark~\ref{rem:up_down_semantic_mi} is \(S\leftrightarrow\tilde S\) and \(Y\leftrightarrow\tilde Y\), while the correspondence between the two layers of the auxiliary variable is \(\tilde U\leftrightarrow\tilde U\).

\paragraph{Family of Comparison Payoff Functions}
Let \(I_s\) and \(I^s\) denote the down and up semantic mutual information of Niu-Zhang (defined in Remark~\ref{rem:up_down_semantic_mi}), which satisfy \(I_s\le I\le I^s\), where \(I\) can be the semantic Shannon mutual information or the mutual information between the corresponding syntactic variables (both lie in the same interval).
For \(\lambda,\mu\in[0,1]\), define the comparison payoff function
\begin{equation}
	\begin{aligned}
		J^{\mathrm{cmp}}_{\lambda,\mu}(Q,A)
		={}&\big[(1-\lambda)I_s+\lambda I^s\big](\tilde{U};\tilde{Y}\mid K)\\
		&-\big[(1-\mu)I_s+\mu I^s\big](\tilde{U};\tilde{S}\mid K).
	\end{aligned}
	\label{eq:comparison_payoff}
\end{equation}
\(J^{\mathrm{cmp}}_{\lambda,\mu}\) is used only for comparison with the classical capacity and does not replace the main payoff function \(J\).
The four endpoints are respectively
\[
\begin{aligned}
	J^{\mathrm{cmp}}_{0,0}&=I_s(\tilde{U};\tilde{Y}\mid K)-I_s(\tilde{U};\tilde{S}\mid K),&
	J^{\mathrm{cmp}}_{0,1}&=I_s(\tilde{U};\tilde{Y}\mid K)-I^s(\tilde{U};\tilde{S}\mid K),\\
	J^{\mathrm{cmp}}_{1,0}&=I^s(\tilde{U};\tilde{Y}\mid K)-I_s(\tilde{U};\tilde{S}\mid K),&
	J^{\mathrm{cmp}}_{1,1}&=I^s(\tilde{U};\tilde{Y}\mid K)-I^s(\tilde{U};\tilde{S}\mid K).
\end{aligned}
\]
Since the payoff is the receiver term minus the host term, the payoff is larger when the receiver term is larger and the host term is smaller; therefore, \((1,0)\) gives the largest payoff and \((0,1)\) gives the smallest payoff, that is, \(J^{\mathrm{cmp}}_{0,1}\le J^{\mathrm{cmp}}_{\lambda,\mu}\le J^{\mathrm{cmp}}_{1,0}\) holds for any \(\lambda,\mu\in[0,1]\).

\begin{remark}[Definition of Up/Down Semantic Mutual Information]
	\label{rem:up_down_semantic_mi}
	For any two semantic variables \(\tilde{A},\tilde{B}\) and their syntactic counterparts \(A,B\), define
	\[
	\begin{aligned}
		I^s(\tilde{A};\tilde{B})&=H(A)+H(B)-H(\tilde{A},\tilde{B}),\\
		I_s(\tilde{A};\tilde{B})&=H(\tilde{A})+H(\tilde{B})-H(A,B),
	\end{aligned}
	\]
	which are called the up semantic mutual information and the down semantic mutual information, respectively.
	According to the convention of Niu-Zhang, the semantic variables are the images of the syntactic variables after synonymous class merging, so that
	\[
	I_s(\tilde{A};\tilde{B})\le I(\tilde{A};\tilde{B})\le I(A;B)\le I^s(\tilde{A};\tilde{B}),
	\]
	where \(I(\tilde{A};\tilde{B})\) is the semantic Shannon mutual information and \(I(A;B)\) is the syntactic mutual information; both lie within \([I_s,I^s]\).
	This definition follows the construction of Niu-Zhang: \(I^s\) is given by the difference between the syntactic marginal entropy and the semantic joint entropy, and \(I_s\) is given by the difference between the semantic marginal entropy and the syntactic joint entropy.
	When the synonymous mappings are one-to-one mappings, the up and down semantic mutual informations coincide with the classical mutual information.
	Replacing \(H(\cdot)\) with the conditional entropy \(H(\cdot\mid K)\) yields the conditional version, and the above inequality still holds.
	According to the convention of Niu-Zhang, the down semantic mutual information takes the nonnegative part (its original expression may be negative).
\end{remark}

\paragraph{Endpoint Capacities and Sandwich}
Let the maximin values corresponding to the endpoints be
\begin{equation}
	C^s_{\mathrm{end},i}=\max_{Q\in\mathcal{Q}(D_0,D_1)}\ \min_{A\in\mathcal{A}(Q,D_2,D_3)}J^{\mathrm{cmp}}_{i}(Q,A),
	\qquad i\in\{(0,1),(1,0)\}.
	\label{eq:endpoint_capacity}
\end{equation}
More generally, for any \((\lambda,\mu)\in[0,1]^2\), let \(C^s_{\mathrm{end},(\lambda,\mu)}=\max_{Q\in\mathcal{Q}(D_0,D_1)}\min_{A\in\mathcal{A}(Q,D_2,D_3)}J^{\mathrm{cmp}}_{\lambda,\mu}(Q,A)\).
Since \(J^{\mathrm{cmp}}_{0,1}\le J^{\mathrm{cmp}}_{1,0}\) holds pointwise, it is preserved after taking the maximin, yielding the endpoint sandwich
\begin{equation}
	C^s_{\mathrm{end},(0,1)}\ \le\ C^s_{\mathrm{end},(1,0)}.
	\label{eq:endpoint_sandwich}
\end{equation}
The physical meaning of the endpoints is as follows: \(C^s_{\mathrm{end},(1,0)}\) corresponds to taking the up semantic mutual information for the receiver term and the down semantic mutual information for the host term, giving an optimistic upper bound of the semantic-layer capacity; \(C^s_{\mathrm{end},(0,1)}\) corresponds to the opposite choice, giving a conservative lower bound.
The relationship between the main capacity \(C^s\) and the endpoint capacities follows directly from \(I_s\le I\le I^s\):
\begin{equation}
	C^s_{\mathrm{end},(0,1)}\ \le\ C^s\ \le\ C^s_{\mathrm{end},(1,0)},
	\label{eq:main_vs_endpoint}
\end{equation}
that is, the main capacity is sandwiched between the two endpoint capacities.

\paragraph{Degeneration and Sandwich with the Classical Capacity}
Let the classical payoff function be
\[
J_{\mathrm{cl}}(Q,A)=I(U;Y\mid K)-I(U;S\mid K),
\]
where \(U,Y,S\) are the syntactic-layer variables corresponding to \(\tilde{U},\tilde{Y},\tilde{S}\).
To make the pointwise comparison on the same feasible region, this section defines the ``classical capacity'' as the maximin of the classical payoff function over the semantic feasible region:
\begin{equation}
	C=\max_{Q\in\mathcal{Q}(D_0,D_1)}\ \min_{A\in\mathcal{A}(Q,D_2,D_3)}J_{\mathrm{cl}}(Q,A).
	\label{eq:classic_common_domain}
\end{equation}

\begin{remark}[Pointwise Difference between Semantic and Classical Payoffs]
	\label{rem:pointwise_diff}
	When the main capacity \(C^s\) and the classical payoff value \(C\) of \eqref{eq:classic_common_domain} share the same feasible region and are evaluated at the same \((Q,A)\), the pointwise difference between the semantic and classical payoffs can be decomposed as
	\begin{equation}
		\begin{aligned}
			J(Q,A)-J_{\mathrm{cl}}(Q,A)
			={}&\big[I(\tilde U;\tilde Y\mid K)-I(U;Y\mid K)\big]
			-\big[I(\tilde U;\tilde S\mid K)-I(U;S\mid K)\big]\\
			={}&-\underbrace{\big[I(U;Y\mid K)-I(\tilde U;\tilde Y\mid K)\big]}_{\text{receiver-term coarsening loss}\;\ge0}
			+\underbrace{\big[I(U;S\mid K)-I(\tilde U;\tilde S\mid K)\big]}_{\text{host-term contraction gain}\;\ge0},
		\end{aligned}
		\label{eq:pointwise_diff}
	\end{equation}
	The two nonnegativities come from the data processing inequality (\(\tilde Y=f_y^{-1}(Y)\), \(\tilde S=f_s^{-1}(S)\), \(\tilde U=U\)).
\end{remark}

\begin{remark}[Feasible Region of the Classical Capacity]
	\label{rem:classic_domain}
	If the ``classical capacity'' is defined over the classical feasible region \(\mathcal{Q}^{\mathrm{cl}}\) (which allows \(Q^{\mathrm{cl}}(x,u\mid s,k)\) to depend directly on the complete \(s\)), then, since \(\mathcal{Q}(D_0,D_1)\subseteq\mathcal{Q}^{\mathrm{cl}}\), its value can be strictly larger than \(C\) of \eqref{eq:classic_common_domain}.
	This section adopts the definition \eqref{eq:classic_common_domain} over the same feasible region so as to enable the pointwise comparison with the endpoints; when \(f_s,f_x\) and the auxiliary mappings are all one-to-one mappings, and \(d_1\equiv d_0\), \(D_1\ge D_0\) and \(d_3\equiv d_2\), \(D_3\ge D_2\) (or \(D_1,D_3\) are non-binding), the two feasible regions coincide, and \(C\) is exactly the classical capacity of Moulin-O'Sullivan~\cite{Moulin2003} (see Remark~\ref{rem:d0d1}).
\end{remark}

\begin{proposition}[Degeneration under One-to-One Mappings]
	\label{prop:degeneracy}
	If \(f_s,f_x,f_y\) and the auxiliary mappings are all one-to-one mappings, then \(C^s=C\), and for any \(\lambda,\mu\in[0,1]\), \(C^s_{\mathrm{end},(\lambda,\mu)}=C\).
\end{proposition}

\begin{proof}
	Under one-to-one mappings, the synonymous classes degenerate to singletons, and the semantic entropy equals the syntactic entropy, so \(I_s=I=I^s\), \(J^{\mathrm{cmp}}_{\lambda,\mu}=J_{\mathrm{cl}}\) holds for any \(\lambda,\mu\), and both the endpoint capacities and the main capacity reduce to \(C\).
\end{proof}

\begin{remark}[Distortion Merging under One-to-One Mappings]
	\label{rem:d0d1}
	When \(f_s,f_x\) are both one-to-one mappings, the semantic alphabet and the syntactic alphabet correspond one-to-one; even after identifying \(\widetilde{\mathcal{S}}\times\widetilde{\mathcal{X}}\) with \(\mathcal{S}\times\mathcal{X}\) under the one-to-one mappings, the distortion functions \(d_0\) and \(d_1\) may still differ.
	Therefore, the claim that ``the semantic constraints coincide with the syntactic constraints, and the model returns to classical hiding'' requires the additional conditions \(d_1\equiv d_0\) and \(D_1\ge D_0\) (as well as \(d_3\equiv d_2\) and \(D_3\ge D_2\)), or \(D_1,D_3\) being non-binding so that they do not constitute additional constraints.
	The equality \(C^s=C\) in Proposition~\ref{prop:degeneracy} does not depend on the above assumptions: the two share the same feasible region \(\mathcal Q(D_0,D_1)\), so \(C^s=C\) only states that the semantic capacity equals the capacity over the same feasible region, and does not involve the comparison with the classical capacity.
	However, the claim that ``\(C\) is exactly the classical capacity of Moulin-O'Sullivan'' requires \(\mathcal Q(D_0,D_1)=\mathcal Q^{\mathrm{cl}}\) and \(\mathcal A(Q,D_2,D_3)=\mathcal A(Q,D_2)\), where \(\mathcal A(\cdot)\) denotes the feasible strategy set of the attacker; that is, it additionally requires \(d_1\equiv d_0\) and \(D_1\ge D_0\) (or \(D_1\) non-binding), as well as \(d_3\equiv d_2\) and \(D_3\ge D_2\) (or \(D_3\) non-binding).
\end{remark}

For general mappings, from \(I_s\le I\le I^s\) and the definition of the endpoint capacities, \(C\) lies between the two endpoint capacities.

\begin{proposition}[Endpoint Sandwich with the Classical Capacity]
	\label{prop:vs_classic}
	For any family of synonymous mappings and distortion thresholds,
	\begin{equation}
		C^s_{\mathrm{end},(0,1)}\ \le\ C\ \le\ C^s_{\mathrm{end},(1,0)}.
		\label{eq:vs_classic}
	\end{equation}
\end{proposition}

\begin{proof}
	By Remark~\ref{rem:up_down_semantic_mi}, \(I_s\le I\le I^s\), so pointwise \(J^{\mathrm{cmp}}_{0,1}\le J_{\mathrm{cl}}\le J^{\mathrm{cmp}}_{1,0}\). The strategy spaces of all three are the same \(\mathcal{Q}(D_0,D_1)\) and \(\mathcal{A}(Q,D_2,D_3)\), and the inequality is preserved after taking the maximin, yielding \eqref{eq:vs_classic}.
\end{proof}

Proposition~\ref{prop:vs_classic} gives an endpoint bound, rather than an assertion that ``the main capacity \(C^s\) is larger or smaller than \(C\)''.
By \eqref{eq:main_vs_endpoint} and \eqref{eq:vs_classic}, both \(C^s\) and \(C\) lie within the same interval \([C^s_{\mathrm{end},(0,1)},C^s_{\mathrm{end},(1,0)}]\), and the endpoint interval is jointly determined by the synonymous merging of the receiver side and the host side: the \((1,0)\) endpoint replaces the receiver term with the larger \(I^s(\tilde{U};\tilde{Y}\mid K)\), so \(C^s_{\mathrm{end},(1,0)}\) is raised; the \((0,1)\) endpoint replaces the (subtracted) host term with the larger \(I^s(\tilde{U};\tilde{S}\mid K)\), so \(C^s_{\mathrm{end},(0,1)}\) is lowered.
The main capacity \(C^s\) and the classical value \(C\) both lie in this interval, and their relative order is jointly determined by the merging degrees of \(f_y\) and \(f_s\), but cannot be determined by the endpoints alone.

To obtain a comparison with a determined direction, in what follows we close the semantic tolerance on the main capacity \(C^s\) and examine the capacity relationship between semantic hiding and classical hiding in that case.

\begin{proposition}[Capacity Degeneration after Closing Semantic Tolerance]
	\label{prop:no_tolerance}
	Closing semantic tolerance means performing the comparison fairly: the extraction accuracy is the same (both sides adopt the syntactic-layer payoff function \(J_{\mathrm{cl}}(Q,A)=I(U;Y\mid K)-I(U;S\mid K)\)), the attack is the same (\(D_3\) is non-binding, i.e., for each \(Q\in\mathcal{Q}(D_0,D_1)\), \(\mathcal{A}(Q,D_2,D_3)=\mathcal{A}(Q,D_2)\), so the attack feasible region is consistent with the classical model), and only the embedding domain differs (the semantic side is \(\mathcal{Q}(D_0,D_1)\), and the classical side is \(\mathcal{Q}^{\mathrm{cl}}\)).
	Let the capacity after closing semantic tolerance be
	\[
	C^s_{\mathrm{no\text{-}tol}}
	=\max_{Q\in\mathcal{Q}(D_0,D_1)}\ \min_{A\in\mathcal{A}(Q,D_2,D_3)}J_{\mathrm{cl}}(Q,A),
	\]
	and let the classical capacity over the complete classical feasible region be
	\[
	C_{\mathrm{cl}}=\max_{Q\in\mathcal{Q}^{\mathrm{cl}}}\ \min_{A\in\mathcal{A}(Q,D_2)}J_{\mathrm{cl}}(Q,A),
	\]
	where \(\mathcal{Q}^{\mathrm{cl}}\) is the classical feasible region.
	Then, under the above comparison,
	\[
	C^s_{\mathrm{no\text{-}tol}}\ \le\ C_{\mathrm{cl}},
	\]
	The necessary and sufficient condition for equality is: there exists an optimal classical covert channel attaining \(C_{\mathrm{cl}}\) that lies in the image \(\Phi(\mathcal Q(D_0,D_1))\) of the semantic feasible region (the proof is given in Appendix~\ref{app:no_tolerance}).
	Two sufficient conditions are respectively: (i) \((f_s,f_x)\) and the auxiliary mappings are all one-to-one mappings, and \(d_1\equiv d_0\) with \(D_1\ge D_0\) (or \(D_1\) non-binding), in which case \(\Phi\) is surjective and \(\Phi(\mathcal Q(D_0,D_1))=\mathcal{Q}^{\mathrm{cl}}\); (ii) some optimal covert channel attaining \(C_{\mathrm{cl}}\) neither exploits the within-class details of \(S\) nor uses syntactic composite signal symbols outside the image of \(f_x\), and its pushforward-induced semantic distortion satisfies \(\mathbb E[d_1]\le D_1\) (automatically true when \(D_1\) is non-binding).
\end{proposition}

\begin{remark}[Premise of the Fair Comparison]
	\label{rem:no_tolerance_fair}
	The ``same attack'' is the premise for the fair comparison to hold: if \(D_3\) is binding, the semantic-side attacker is additionally restricted by \(d_3\le D_3\) beyond \(d_2\le D_2\), and the contraction of the attack domain is favorable to the hider (\(\min_A\) is raised), which may make \(C^s_{\mathrm{no\text{-}tol}}\) exceed \(C_{\mathrm{cl}}\).
	For example, with binary Hamming distortion and \(D_3=0\), the attack is forced to be the identity, and \(C^s_{\mathrm{no\text{-}tol}}=H(D_0)\), whereas with \(D_2=1/2\), the classical attack can make \(Y\) independent of \(X\), giving \(C_{\mathrm{cl}}=0\).
	When \(D_3\) is non-binding (e.g., \(d_3\le d_2\) and \(D_3\ge D_2\)), the attack feasible region is consistent with the classical one, and the inequality of this proposition comes entirely from the embedding domain contraction (carrier coarsening loss), which is exactly the comparison with ``only the embedding domain differing''.
\end{remark}

\begin{remark}[Meaning of ``Closing Semantic Tolerance'']
	\label{rem:no_tolerance_meaning}
	Here, ``closing semantic tolerance'' means that both sides return to the syntactic-layer evaluation: the host term uses \(S\) instead of \(\tilde S\), and the receiver term uses \(Y\) instead of \(\tilde Y\), while taking the fair comparison with the same attack (\(D_3\) non-binding) and keeping the embedding domain \(\mathcal{Q}(D_0,D_1)\) unchanged.
	It differs from ``closing only the receiver-side tolerance'' (taking \(f_y\) as the identity, which only changes \(\tilde Y\) back to \(Y\) while the host term remains \(\tilde S\), giving the payoff function \(I(\tilde U;Y\mid K)-I(\tilde U;\tilde S\mid K)\)).
	Therefore, the inequality of Proposition~\ref{prop:no_tolerance} comes from the feasible region contraction caused by \(f_s,f_x\) (carrier coarsening loss), and has nothing to do with the receiver-side tolerance of \(f_y\); if \(D_3\) is binding, the attack side is no longer the same, the comparison is no longer fair, and the conclusion may be reversed (see Remark~\ref{rem:no_tolerance_fair}).
\end{remark}

Proposition~\ref{prop:no_tolerance} gives a comparison with a determined direction at the endpoint of ``closing tolerance'': after closing tolerance, the semantic payoff function degenerates to the classical \(J_{\mathrm{cl}}\), and the gap comes from the feasible region contraction caused by the \(f_s\) coarsening. The semantic encoder can only generate \(X\) through \(\tilde{S}\), so the selectable \(Q\) is a subset of the classical feasible region; hence the same payoff function is maximized over a smaller set, and its value cannot be larger.
It is not contradictory to the endpoint sandwich \(C\le C^s_{\mathrm{end},(1,0)}\): the latter is the optimistic endpoint with the tolerance fully enabled, and the former is the conservative endpoint with the tolerance closed.
Therefore, one cannot derive the main capacity \(C^s\le C_{\mathrm{cl}}\) from Proposition~\ref{prop:no_tolerance}; it only states that, at the endpoint of ``closing tolerance'', semantic hiding is not better than classical hiding.

\section{Semantic Hiding Capacity under Typical Sources}
\label{sec:examples}

The preceding sections have characterized the semantic hiding capacity as \(C^s=\max_{Q}\min_A J(Q,A)\) and established a comparison framework with the classical capacity.
This section uses binary, \(q\)-ary Hamming, and Gaussian sources as examples to illustrate how \(C^s\) depends on the host distribution, the synonymous mappings, and the distortion.
The exposition strictly distinguishes three kinds of conclusions: exact ones (the classical capacity, the one-to-one mapping degeneration, and the \(L_s,L_x,L_y\to0\) limit), strict bounds (the endpoint sandwich and the upper bound of closing tolerance), and approximate demonstrations (the effective binarization surrogate under coarsening).
After coarsening, \(C^s\) generally has no closed form; the numerical values labeled ``approximate'' are used only to show trends and are not treated as the exact capacity.

\subsection{Binary Source}
\label{subsec:binary}

Let the syntactic host \(S\) follow the Bernoulli distribution \(p_S(0)=p_S(1)=1/2\), \(\mathcal X=\mathcal Y=\{0,1\}\), and \(d_0,d_2\) be the Hamming distance, with \(D_0,D_2\le1/2\).
Take \(f_s,f_x,f_y\) all as one-to-one mappings, and take \(D_1,D_3\) sufficiently large (or \(d_1\equiv d_0\), \(d_3\equiv d_2\) with \(D_1\ge D_0\), \(D_3\ge D_2\)) so that the semantic constraints are non-binding (see Remark~\ref{rem:d0d1}); then \(\tilde S=S\), \(\tilde X=X\), \(\tilde Y=Y\), \(I\) degenerates to the syntactic mutual information, and \(C^s\) degenerates to the classical hiding capacity.
In what follows, take the private watermarking \(K=\tilde S\) (the decoder fully knows the host), so the host term \(I(\tilde U;\tilde S\mid K)=0\).
The classical binary result is given by Moulin-O'Sullivan:
\begin{equation}
	C=H(D_0\star D_2)-H(D_2),
	\label{eq:binary_classic}
\end{equation}
where \(H(t)=-t\log t-(1-t)\log(1-t)\) and \(D_0\star D_2=D_0(1-D_2)+D_2(1-D_0)\); the optimal covert channel is \(X=S\oplus Z\) (\(Z\sim\mathrm{Bern}(D_0)\) independent of \(S\)), and the optimal attack is \(Y=X\oplus W\) (\(W\sim\mathrm{Bern}(D_2)\) independent of \(X\)).
By Proposition~\ref{prop:degeneracy}, under one-to-one mappings \(C^s=C\), so \eqref{eq:binary_classic} is the semantic hiding capacity of this example.

\begin{remark}[Side Information Assumption]
	\label{rem:binary_private}
	\eqref{eq:binary_classic} is the classical capacity of private watermarking (\(K=\tilde S\), the decoder fully knows the host), in which case the host term \(I(\tilde U;\tilde S\mid\tilde S)=0\), the payoff function reduces to \(I(X;Y\mid S)\), and the optimal solution is \(X=S\oplus Z\).
	If \(K\) is independent of \(\tilde S\) (blind watermarking), the host term \(I(\tilde U;\tilde S\mid K)\) is generally nonzero, and \(C^s\) is lower than in the private case; its closed-form solution must be solved separately for blind watermarking and is not elaborated in this section.
\end{remark}

The binary alphabet has only two symbols, and the only possible coarsening merges the two classes into one, in which case the capacity is zero. Therefore, the binary source is used only to verify the degeneration proposition and does not exhibit the coarsening loss.

\subsection{\(q\)-ary Hamming Source}
\label{subsec:qary_hamming}

Let the syntactic host \(S\) be uniformly distributed over \(\mathcal S=\{0,1,\dots,q-1\}\), \(\mathcal X=\mathcal Y=\{0,\dots,q-1\}\), and \(d_0,d_2\) be the Hamming distance, with \(D_0,D_2\le(q-1)/q\).
Let the \(q\)-ary error entropy be
\[
h_q(t)=-(1-t)\log(1-t)-t\log\frac{t}{q-1},\qquad 0\le t\le\frac{q-1}{q},
\]
and the \(q\)-ary Hamming convolution be
\[
D_0\star_q D_2=D_0+D_2-\frac{q}{q-1}D_0D_2 .
\]
When \(q=2\), \(\star_q\) reduces to the binary \(D_0\star D_2\).

\paragraph{One-to-One Mapping Benchmark}
Take \(f_s,f_x,f_y\) as one-to-one mappings, and take \(D_1,D_3\) sufficiently large (or \(d_1\equiv d_0\), \(d_3\equiv d_2\) with \(D_1\ge D_0\), \(D_3\ge D_2\)) so that the semantic constraints are non-binding; then \(C^s\) reduces to the \(q\)-ary classical hiding capacity
\begin{equation}
	C_q=h_q(D_0\star_q D_2)-h_q(D_2).
	\label{eq:C3}
\end{equation}
\eqref{eq:C3} is exactly the classical \(q\)-ary Hamming result (private watermarking, \(K=\tilde S\)).

\paragraph{Coarsening to Two Classes}
To exhibit the loss caused by coarsening, a general (not necessarily uniform) source distribution \(p_i=\Pr[S=i]\) is allowed below (the uniform case is \(p_i=1/q\), which is the premise of the \(C_q\) benchmark in the previous paragraph).
Take the synonymous partition \(\tilde s_0\leftrightarrow\{0\}\) and \(\tilde s_1\leftrightarrow\{1,\dots,q-1\}\), and take the isomorphic partition for the composite signal and the receiver.
Let \(p_i=\Pr[S=i]\), and let \(\bar p = \sum_{i\ge1}p_i\) be the total probability of the class \(\tilde s_1\); then \(p(\tilde s_0)=p_0\) and \(p(\tilde s_1)=\bar p\).
The semantic encoder cannot distinguish the \(q-1\) syntactic symbols within the class \(\tilde s_1\), and \(C^s\) is a constrained maximin, which generally has no closed form.
Two pieces of rigorous information can be given.
By Proposition~\ref{prop:no_tolerance} (under the fair comparison with \(D_3\) non-binding, see Remark~\ref{rem:no_tolerance_fair}), let
\[
C^s_{\mathrm{no\text{-}tol}}=\max_{Q\in\mathcal Q(D_0,D_1)}\min_A J_{\mathrm{cl}}(Q,A),
\]
then \(C^s_{\mathrm{no\text{-}tol}}\le C_q\).
Note the side information conventions on the two sides: the semantic side takes private watermarking \(K=\tilde S\) (the surrogate model also implicitly assumes this setting), while the classical benchmark \(C_q\) takes private watermarking \(K=S\).
The rigorous argument proceeds in two steps: first, by Proposition~\ref{prop:no_tolerance} (with the same side information \(K=\tilde S\)), \(C^s_{\mathrm{no\text{-}tol}}\le C_{\mathrm{cl}}(K{=}\tilde S)\); then, by the monotonicity of the decoder side information (a decoder knowing the complete host \(S\) can ignore the within-class information, any code feasible for \(K=\tilde S\) is also feasible for \(K=S\), and the attack feasible region does not change with \(K\), so \(C_{\mathrm{cl}}(K{=}\tilde S)\le C_{\mathrm{cl}}(K{=}S)=C_q\)), combining the two yields \(C^s_{\mathrm{no\text{-}tol}}\le C_q\).
By Proposition~\ref{prop:vs_classic}, the main capacity satisfies
\[
C^s_{\mathrm{end},(0,1)}\le C^s\le C^s_{\mathrm{end},(1,0)}.
\]
These two only give an interval, not the specific numerical value after coarsening.

\paragraph{Effective Binarization Surrogate (Approximate)}
To show the coarsening trend, project the two-class source onto a binary host \(\tilde S\in\{\tilde s_0,\tilde s_1\}\): the encoder changes the semantic class with probability \(\delta_0\) (cross-class modification), the attack changes the semantic class with probability \(\delta_2\), and the surrogate capacity takes the binary Hamming form
\[
C^s_{\mathrm{proxy}}=H(\delta_0\star\delta_2)-H(\delta_2),
\qquad
\delta_0\star\delta_2=\delta_0(1-\delta_2)+\delta_2(1-\delta_0).
\]
where the cross-class probability is given approximately by the syntactic budget allocation: the cross-class rewritable probability mass is \(\min(p_0,\bar p)\), and the within-class rewritable probability mass is \(\bar q=\min_{i\ge1}p_i\); then
\[
\delta_0=\min\big(D_0,D_1,p_0,\bar p\big)\cdot\frac{\min(p_0,\bar p)}{\min(p_0,\bar p)+\bar q},
\qquad
\delta_2=\min(D_2,D_3).
\]
This surrogate only keeps the cross-class modification and treats the within-class modification as budget occupation, so it is only a trend approximation: it compresses the \(q\)-ary problem into a binary one, and the budget allocation of \(\delta_0\) is heuristic.
The exact \(C^s\) requires numerically solving the maximin under the \(q\)-ary constraints.
When \(\bar q\to0\) (some \(p_i\to0\), \(i\ge1\)), \(\delta_0\to\min(D_0,D_1,p_0,\bar p)\), and if \(D_0\) is the binding constraint then \(\delta_0\to D_0\), so the additional loss caused by coarsening tends to zero; when \(p_1=\cdots=p_{q-1}=\bar p/(q-1)\), \(\bar q=\bar p/(q-1)\) is maximal, and the loss is maximal.

\subsection{Gaussian Source}
\label{subsec:gaussian}

Let the syntactic host \(S\sim\mathcal N(0,\sigma^2)\), the composite signal and receiver alphabets be the real line, and \(d_0,d_2\) be the squared error, with thresholds \(D_0,D_2\).
Take the synonymous mapping \(f_s\) to partition the real line into equal-length intervals \(\{\Omega_i\}\) of length \(L_s\), each interval corresponding to one semantic symbol, with the probability of the semantic host given by the density integration:
\[
p(\tilde s_i)=\int_{\Omega_i}p_S(s)\,ds,\qquad p_S(s)=\frac{1}{\sqrt{2\pi\sigma^2}}e^{-s^2/(2\sigma^2)}.
\]
Similarly, take \(f_x,f_y\) as equal-length interval partitions of lengths \(L_x,L_y\).

\paragraph{Gaussian Capacity (Private Solution, Same Value under Blind Watermarking)}
Take private watermarking \(K=\tilde S\) (the decoder knows the host).
Moulin-O'Sullivan~\cite{Moulin2003} gave, via the maximin game under squared-error distortion and the average distortion constraint, the Gaussian capacity of private watermarking (\(K=S\)) (their Theorem 5.1; Cohen and Lapidoth~\cite{CohenLapidoth2002} independently gave the same maximin solution and corrected an error in the original derivation of MO. MO once argued optimality via the Nash equilibrium in the \(a=1\) case, but this property is insufficient to derive maximin optimality, see the footnote of that section in MO); under a Gaussian carrier and squared-error distortion, the capacity value of blind watermarking is the same (Moulin-O'Sullivan Theorem 5.3; Chen first obtained this conclusion in the related problem with fixed additive Gaussian noise attacks).
This capacity is given by:
\begin{equation}
	C_G=\frac12\log\left(1+\frac{[(2a-1)\sigma^2-D_2+D_0][D_0-(a-1)^2\sigma^2]}{[D_0+(2a-1)\sigma^2]D_2}\right),
	\label{eq:gaussian_classic}
\end{equation}
where \(a\) is the maximizer of \(f(a)=\frac{[(2a-1)\sigma^2-D_2+D_0][D_0-(a-1)^2\sigma^2]}{[D_0+(2a-1)\sigma^2]D_2}\) over the interval \(\big(\max(1,(\sigma^2+D_2-D_0)/(2\sigma^2)),\,1+\sqrt{D_0}/\sigma\big)\) (the endpoints are taken as limits).
The optimal covert channel is \(X=aS+Z\), with \(Z\sim\mathcal N(0,D_0-(a-1)^2\sigma^2)\) independent of \(S\); the optimal attack is \(A^*(y|x)=\mathcal N(\beta^{-1}x,\beta^{-1}D_2)\), with \(\beta=\sigma_x^2/(\sigma_x^2-D_2)\) and \(\sigma_x^2=D_0+(2a-1)\sigma^2\).
When \(\sigma^2+D_0\le D_2\) (the host is relatively weak compared to the distortion threshold), the feasible interval of \(a\) is empty, and the capacity is zero.
In the small-distortion limit \(D_0,D_2\to0\) (with \(\sigma^2\) fixed), \(a\to1\) and \(\beta\to1\), and
\[
C_G\sim\frac12\log\left(1+\frac{D_0}{D_2}\right).
\]

\begin{remark}[Optimal Strategy for Blind Watermarking]
	\label{rem:gaussian_blind}
	For blind watermarking (\(K\perp S\)), the optimal covert channel attaining the same capacity is not \(X=aS+Z\) itself, but a Costa-type auxiliary structure: \(U=W+\alpha S\) (\(W\perp S\)) together with the host scaling \(X=aS+W\), making \(I(U;Y)-I(U;S)\) reach the same level as the private value (Moulin-O'Sullivan Theorem 5.3).
	Therefore, the numerical value of \eqref{eq:gaussian_classic} is the same for the two cases, but the strategies attaining it differ, so one should not attribute the original source of this formula to \(K\perp S\) on this basis.
\end{remark}

By Proposition~\ref{prop:degeneracy}, in the limit \(L_s,L_x,L_y\to0\), \(C^s=C_G\).

\paragraph{Non-One-to-One Mapping}
When \(L>0\), the semantic layer is a coarsening of the syntactic layer; the encoder can only distinguish synonymous intervals and cannot distinguish their internal details, so the feasible region contracts.
According to the continuous semantic information measure of Niu-Zhang (the synonymous interval length enters the semantic entropy as a logarithmic term, and the divergent parts cancel in the mutual information), \(I\) and \(C^s\) can be computed; by Proposition~\ref{prop:no_tolerance}, \(C^s_{\mathrm{no\text{-}tol}}\le C_G\) (with the same side information convention as in the \(q\)-ary case: first apply the proposition with the same side information \(K=\tilde S\), then transition to \(C_G\) with \(K=S\) via the monotonicity of the decoder side information), the main capacity is bounded by the endpoint sandwich, and generally has no determined sign.
The specific numerical value depends on \(L_s,L_x,L_y,\sigma^2,D_0,D_2\) and requires numerical solution.

\subsection{Numerical Comparison and Discussion}
\label{subsec:numerical}

\begin{figure}[h]
	\centering
	\includegraphics[width=0.7\textwidth]{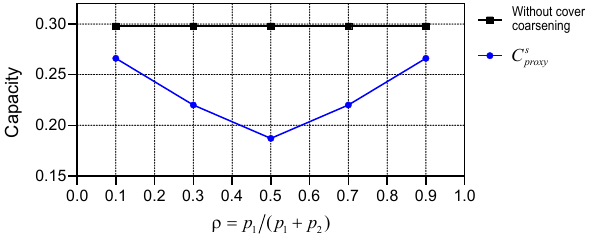}
	\caption{U-shaped trend of the coarsening loss under the effective binarization surrogate (approximate numerical values, not the exact \(C^s\)).}
	\label{fig:numerical}
\end{figure}

Table~\ref{tab:ternary_numerical} and Figure~\ref{fig:numerical} give the numerical values of the ternary two-class source under the effective binarization surrogate: \(p_0=0.4\), \(D_0=D_2=0.1\), \(D_3=0.05\), \(D_1\) sufficiently large, and \(\rho=p_1/(p_1+p_2)\) varying over \(\{0.1,0.3,0.5,0.7,0.9\}\).
In the surrogate model, without carrier coarsening (\(\delta_0=D_0\) and the attack \(\delta_2=0.05\) unchanged), \(H(0.1\star0.05)-H(0.05)\approx0.298\); after coarsening, \(\delta_0\) drops to \(0.057\sim0.087\), the capacity decreases accordingly, and it is lowest at \(\rho=0.5\) (the most uniform within the class).
If the attacker's semantic constraint is also removed (the fully classical surrogate: \(\delta_0=D_0=0.1\) and \(\delta_2=D_2=0.1\)), the surrogate value is \(H(0.1\star0.1)-H(0.1)\approx0.211\); the baseline column in the table fixes \(\delta_2=0.05\), measuring only the carrier coarsening loss and not directly comparing with the fully classical value.

\begin{table}[h]
	\centering
	\caption{Numerical values of the ternary two-class source under the effective binarization surrogate (\(p_0=0.4\), \(D_0=D_2=0.1\), \(\delta_2=0.05\))}
	\label{tab:ternary_numerical}
	\begin{tabular}{ccccc}
		\hline
		\(\rho\) & \(\bar q\) & \(\delta_0\) & \(C^s_{\mathrm{proxy}}\) & without carrier coarsening (same \(\delta_2\)) \\
		\hline
		0.1 & 0.06 & 0.087 & 0.266 & 0.298 \\
		0.3 & 0.18 & 0.069 & 0.220 & 0.298 \\
		0.5 & 0.30 & 0.057 & 0.187 & 0.298 \\
		0.7 & 0.18 & 0.069 & 0.220 & 0.298 \\
		0.9 & 0.06 & 0.087 & 0.266 & 0.298 \\
		\hline
	\end{tabular}
\end{table}

Table~\ref{tab:ternary_numerical} gives three observations.
First, at \(\rho=0.5\), the within-class distribution is the most uniform, the within-class modification has the strongest ability to absorb the budget, the cross-class \(\delta_0\) is the smallest, and the capacity is the lowest.
Second, as \(\rho\to0,1\), one within-class symbol tends to zero, \(\bar q\to0\), \(\delta_0\to D_0\), and the capacity recovers.
Third, under the surrogate comparison with the attack (\(\delta_2=0.05\)) unchanged, the carrier coarsening lowers the capacity by \(0.032\sim0.111\), showing that the synonymous merging indeed consumes the embeddable redundancy; if the attacker is not constrained by \(D_3\) (the fully classical surrogate \(\delta_2=D_2=0.1\), with benchmark \(0.211\)), the gap from \(C^s_{\mathrm{proxy}}\) is the net effect of the carrier coarsening loss and the attack tightening gain.
It must be emphasized that these numerical values belong to the surrogate model: the exact \(C^s\) has no closed form, and the absolute values cannot be directly compared with \(C_q\).

Reviewing the three conclusions proved in Section~\ref{subsec:comparison_classic}, the examples of this section respectively confirm:
\begin{itemize}
	\item[(i)] under one-to-one mappings (or the Gaussian case \(L_s,L_x,L_y\to0\)), \(C^s\) degenerates to the classical capacity, and the endpoints coincide with the main capacity (Proposition~\ref{prop:degeneracy});
	\item[(ii)] under general mappings, the endpoint sandwich \(C^s_{\mathrm{end},(0,1)}\le C^s\le C^s_{\mathrm{end},(1,0)}\) holds, and \(C\) lies in the same interval (Proposition~\ref{prop:vs_classic});
	\item[(iii)] when closing semantic tolerance (under the fair comparison with \(D_3\) non-binding), \(C^s_{\mathrm{no\text{-}tol}}\le C_q\) (Proposition~\ref{prop:no_tolerance}).
\end{itemize}

The source of the coarsening loss is twofold: at the transmitter, only \(X\in f_x(\tilde X)\) can be generated, so the selectable syntactic composite signal set contracts, reducing the embeddable redundancy (a loss); at the receiver, the synonymous merging of \(f_y\) tightens the attack through the \(D_3\) constraint (a gain), but at the same time coarsens the receiver term.
In this example, \(f_y\) takes the same isomorphic two-class partition as \(f_x,f_s\), and the surrogate model uses \(\delta_0\) to characterize the reduction of the embedding budget by the carrier coarsening, and \(\delta_2=\min(D_2,D_3)=0.05\) to characterize the attack strength under the semantic constraint.
The numerical values show that the net effect is a capacity decrease, i.e., in this example the carrier coarsening loss exceeds the attack tightening gain.

The reason why the coarsened \(C^s\) has no closed form is that coarsening turns the unconstrained \(\max_Q\min_A\) into a constrained convex-concave game where ``\(X\) is constrained by \(f_x\), and the host observation is constrained by \(f_s\)''; its optimal solution depends on the specific \(f\) and \(p_S\), and cannot be reduced to a single scalar optimization as in the classical Hamming case.
Therefore, except for special symmetric cases, numerical solution is needed.

\begin{remark}[Difference from the Niu-Zhang Comparison]
	\label{rem:vs_niuzhang}
	When Niu-Zhang proved \(C\le C_s\)~\cite{Niu2024}, the classical and semantic capacities were maximized over the same input distribution \(p(x)\), and the semantic layer is only a coarsening of the same \(X,Y\), so the pointwise \(I\le I^s\) directly gives \(C\le C_s\).
	The hiding problem is different: the semantic and classical models correspond to two different feasible regions (and there is an adversarial attack), so the pointwise inequality can only give a max-min relationship over the same feasible region; therefore, there is no unconditional ordering conclusion, but only the endpoint sandwich and the conditional lower bound of Section~\ref{subsec:comparison_classic}.
\end{remark}

\section{Special Cases and Capacity Properties}
\label{sec:special_cases}

This section discusses the simplified forms and basic properties of the main capacity \(C^s\) in common scenarios.
In what follows, the payoff function \(J(Q,A)=I(\tilde U;\tilde Y\mid K)-I(\tilde U;\tilde S\mid K)\) is taken as the example, where \(I\) is the Shannon conditional mutual information on semantic variables; the up/down semantic mutual information convention used in the comparison with the classical capacity is given in Section~\ref{subsec:comparison_classic}.

\subsection{Private Semantic Watermarking}
\label{subsec:private}

If the decoder fully knows the semantic host, i.e., \(K=\tilde S\) (private watermarking), then the host term \(I(\tilde U;\tilde S\mid\tilde S)=0\), and the payoff function simplifies to
\begin{equation}
	J(Q,A)=I(\tilde U;\tilde Y\mid\tilde S),
	\label{eq:private_payoff}
\end{equation}
with the corresponding capacity
\begin{equation}
	C^s_{\mathrm{private}}=\max_{Q\in\mathcal Q(D_0,D_1)}\min_{A\in\mathcal A(Q,D_2,D_3)}I(\tilde U;\tilde Y\mid\tilde S).
	\label{eq:private_capacity}
\end{equation}
This is exactly the setting adopted by the numerical examples (binary, \(q\)-ary, and Gaussian) in Section~\ref{sec:examples}: there, \(K=\tilde S\) is taken so that the host term vanishes, and the capacity is the private form of \eqref{eq:private_capacity}; the main capacity theorem~\ref{thm:semantic_capacity} holds for arbitrary \(K\), and the private case is its special case.

When \(f_s,f_x,f_y\) are all one-to-one mappings, \(\tilde S=S\), \(\tilde Y=Y\), \(\tilde U=U\), and \(I(\tilde U;\tilde Y\mid\tilde S)\) degenerates to the syntactic mutual information \(I(U;Y\mid S)\).
However, for \eqref{eq:private_capacity} to reduce to the private watermarking capacity of Moulin-O'Sullivan
\[
C=\max_{p(x\mid s)}\min_A I(X;Y\mid S)
\]
the feasible regions must be consistent: \(\mathcal Q(D_0,D_1)=\mathcal Q^{\mathrm{cl}}\) and \(\mathcal A(Q,D_2,D_3)=\mathcal A(Q,D_2)\), i.e., the semantic constraints \(D_1,D_3\) are both non-binding.
This is equivalent to (see Remark~\ref{rem:d0d1}) \(d_1\equiv d_0\) with \(D_1\ge D_0\) (or \(D_1\) non-binding), as well as \(d_3\equiv d_2\) with \(D_3\ge D_2\) (or \(D_3\) non-binding).
Otherwise, a binding \(D_1\) contracts \(\mathcal Q(D_0,D_1)\), and a binding \(D_3\) contracts \(\mathcal A(Q,D_2,D_3)\), and the two act in opposite directions on the maximin, so \eqref{eq:private_capacity} and \(C\) have no unconditional ordering relationship.

\subsection{Blind Semantic Watermarking}
\label{subsec:blind}

If the decoder does not know the semantic host, i.e., \(K\perp\tilde S\), it is called blind semantic watermarking.
The payoff function remains
\begin{equation}
	J(Q,A)=I(\tilde U;\tilde Y\mid K)-I(\tilde U;\tilde S\mid K).
	\label{eq:blind_payoff}
\end{equation}
The encoder still needs to use \(\tilde S\) to generate the composite signal, so \(I(\tilde U;\tilde S\mid K)\) is generally nonzero.
The decoder in the private case additionally knows \(\tilde S\), i.e., has more information, so
\[
C^s_{\mathrm{blind}}\le C^s_{\mathrm{private}}.
\]
The gap between the two is determined by the host term \(I(\tilde U;\tilde S\mid K)\): when the encoder barely uses the host information, the gap tends to zero, and it is maximal when the host information is fully exploited.

\subsection{No-Attack Case}
\label{subsec:no_attack}

If no attack exists, i.e., \(A\) is the identity channel \(Y=X\), and the receiver side and the transmitter side take the same partition \(f_y=f_x\) (equivalently \(f_y^{-1}=f_x^{-1}\), so that \(\tilde Y=\tilde X\)), then the payoff function simplifies to
\begin{equation}
	J(Q)=I(\tilde U;\tilde X\mid K)-I(\tilde U;\tilde S\mid K).
	\label{eq:no_attack_payoff}
\end{equation}
with the corresponding capacity
\begin{equation}
	C^s_{\mathrm{no\text{-}attack}}=\max_{Q\in\mathcal Q(D_0,D_1)}J(Q).
	\label{eq:no_attack_capacity}
\end{equation}
In this case, the problem reduces to a source-coding-type problem of ``embedding \(\tilde X\) under the host and distortion constraints'': if \(K=\tilde S\) is taken so that the host term is zero, then \(J=I(\tilde U;\tilde X\mid\tilde S)\), and the capacity is determined by the embedding distortion \(D_0,D_1\).
It must be emphasized that this is not the same object as the semantic channel capacity \(C_s=\max I^s(\tilde X;\tilde Y)\) of Niu-Zhang: \(C_s\) is defined in terms of \(I^s\) and is a pure channel capacity (with no auxiliary variable and no distortion constraints), whereas \(J\) uses the semantic Shannon mutual information \(I\) and explicitly contains the auxiliary variable \(\tilde U\), the host term \(I(\tilde U;\tilde S\mid K)\), and the distortion constraints \(D_0,D_1\).
Therefore, \(C^s_{\mathrm{no\text{-}attack}}\) is an ``embedding capacity with host and distortion constraints'', and cannot be directly identified with or compared against the semantic channel capacity.

\subsection{Lower Bound on the Capacity}
\label{subsec:lower_bound}

Restricting \(\tilde U=\tilde X\) yields a lower bound on the main capacity.
Substituting into \(J\) and expanding with conditional entropy,
\[
J=H(\tilde X\mid\tilde S,K)-H(\tilde X\mid\tilde Y,K),
\]
hence
\begin{equation}
	C^s\ \ge\ \max_{Q}\Big[H(\tilde X\mid\tilde S,K)-\max_A H(\tilde X\mid\tilde Y,K)\Big].
	\label{eq:lower_bound}
\end{equation}
Here, \(\max_Q\) is restricted to the covert channel class with \(\tilde U=\tilde X\), i.e., ``the capacity when taking \(\tilde U=\tilde X\)'', which is thus a lower bound on the main capacity; it provides a computable conservative estimate when the exact capacity is difficult to obtain.

\subsection{Monotonicity and Convexity of the Capacity}
\label{subsec:monotone_convex}

\begin{proposition}[Monotonicity]
	\label{prop:capacity_monotone_prop}
	\(C^s\) is nondecreasing in \(D_0,D_1\), and nonincreasing in \(D_2,D_3\).
\end{proposition}

\begin{proof}
	Increasing \(D_0,D_1\) enlarges \(\mathcal Q(D_0,D_1)\), so the outer \(\max\) is nondecreasing; increasing \(D_2,D_3\) enlarges each \(\mathcal A(Q,D_2,D_3)\), so the inner \(\min\) is nonincreasing, and hence \(C^s=\max_Q\min_A J\) is nonincreasing in \(D_2,D_3\).
\end{proof}

\begin{proposition}[Convexity]
	\label{prop:capacity_convex}
	If the attack class \(\mathcal A(Q,D_2,D_3)\) is defined only by the distortion constraints, without additional structural constraints, then \(C^s\) is a convex function of \(D_2\).
\end{proposition}

\begin{proof}
	Take arbitrary \(D_2,D_2'\) and \(\theta\in[0,1]\), and let \(D_2^\theta=\theta D_2+(1-\theta)D_2'\).
	For any \(Q\), let \(A^*\) and \(A^{*\prime}\) attain the minimum on \(\mathcal A(Q,D_2,D_3)\) and \(\mathcal A(Q,D_2',D_3)\), respectively (the inner minimum is attainable, see Appendix~\ref{app:inner_min_attained}), and let \(A_\theta=\theta A^*+(1-\theta)A^{*\prime}\).
	By the linearity of the distortion constraints in \(A\), \(\mathbb E_{A_\theta}[d_2]\le D_2^\theta\) and \(\mathbb E_{A_\theta}[d_3]\le D_3\), so \(A_\theta\in\mathcal A(Q,D_2^\theta,D_3)\); moreover, by Proposition~\ref{prop:payoff_convexity}(i), \(J(Q,\cdot)\) is convex in \(A\), so
	\[
	\min_{A\in\mathcal A(Q,D_2^\theta,D_3)}J(Q,A)\le J(Q,A_\theta)
	\le\theta J(Q,A^*)+(1-\theta)J(Q,A^{*\prime}).
	\]
	Taking \(\max\) over \(Q\) yields \(C^s(D_2^\theta)\le\theta C^s(D_2)+(1-\theta)C^s(D_2')\).
\end{proof}

\begin{remark}
	\label{rem:basic_properties}
	Basic properties of the capacity:
	\begin{enumerate}
		\item[(i)] \(C^s\le\log|\tilde{\mathcal Y}|\): since \(I(\tilde U;\tilde Y\mid K)\le H(\tilde Y\mid K)\le\log|\tilde{\mathcal Y}|\), and \(C^s\) is the maximin value of \(J\);
		\item[(ii)] If \(D_0=0\) (assuming \(d_0\) satisfies \(d_0(s,x)=0\iff x=s\)), then \(X=S\) almost surely, no information can be embedded, and \(C^s=0\);
		\item[(iii)] If there exists a fixed \(y^*\) such that the deterministic attack \(A(y\mid x)=\delta(y-y^*)\) (mapping all composite signals to the same \(y^*\)) satisfies \(\mathbb E[d_2]\le D_2\) and \(\mathbb E[d_3]\le D_3\) for every \(Q\in\mathcal Q(D_0,D_1)\), then \(\tilde Y\) is independent of \((\tilde U,\tilde S)\), and \(J\le0\); further taking \(Q\) with \(\tilde U\perp\tilde S\) gives \(C^s\ge0\), so \(C^s=0\).
	\end{enumerate}
\end{remark}

\section{Conclusion}
\label{sec:conclusion}

This paper is based on the semantic information theory of Niu-Zhang and the classical information hiding theory of Moulin-O'Sullivan, and establishes an information-theoretic analysis framework for information hiding under semantic communication.

At the model level, classical hiding is extended into a two-layer structure of ``semantic host, syntactic host, syntactic composite signal, syntactic received, semantic received'', the synonymous mappings \(f_s,f_x,f_y\) bridge the semantic layer and the syntactic layer, and four distortion constraints are introduced, namely syntactic transparency, semantic fidelity, syntactic perturbation, and semantic destruction; when the synonymous mappings degenerate to one-to-one mappings, the model returns to the classical form.

At the game level, the semantic hiding capacity is characterized as a zero-sum game between the hider and the attacker, it is argued that the Stackelberg structure with the hider as the first mover and the attacker as the second mover is most consistent with real systems, and it is pointed out that the attack feasible strategy depends on the hider's strategy and the joint strategy space is non-rectangular, so Sion's theorem under the rectangular domain cannot be directly applied.

At the capacity level, the semantic auxiliary variable \(\tilde U\) is introduced to realize randomized coding, the payoff function is taken as the difference of Shannon conditional mutual informations on semantic variables, \(J(Q,A)=I(\tilde U;\tilde Y\mid K)-I(\tilde U;\tilde S\mid K)\), the single-letter maximin expression \(C^s=\max_Q\min_A J(Q,A)\) of the semantic hiding capacity is given, and its achievability and converse are proved along the path of classical random coding and joint typicality decoding.

At the comparison level, a family of comparison payoff functions based on the up/down semantic mutual information is introduced, yielding the endpoint sandwich \(C^s_{\mathrm{end},(0,1)}\le C^s\le C^s_{\mathrm{end},(1,0)}\) and \(C^s_{\mathrm{end},(0,1)}\le C\le C^s_{\mathrm{end},(1,0)}\), where \(C\) is the classical payoff value over the same semantic feasible region; it is also proved that, when adopting the classical payoff function, \(C^s_{\mathrm{no\text{-}tol}}\le C_{\mathrm{cl}}\) (\(C_{\mathrm{cl}}\) being the classical capacity over the complete classical feasible region).
This clarifies the applicable boundary of the claim that ``the semantic capacity is larger than the classical capacity'' in the hiding scenario: semantic hiding is not unconditionally better; the endpoint sandwich can only bound the interval and cannot determine the order alone, and the upper bound after closing semantic tolerance can only give a comparison with a determined direction, so neither guarantees that semantic hiding is better than classical hiding.

At the numerical example level, the binary source verifies the degeneration property under one-to-one mappings (\(C^s=C\)); the \(q\)-ary Hamming source gives the exact \(q\)-ary classical benchmark; the ternary two-class source shows that carrier coarsening causes capacity loss, for which we give the endpoint sandwich, the upper bound with the classical payoff function, and the approximate numerical values under the effective binarization surrogate; the Gaussian source gives the classical capacity of blind watermarking and the degeneration limit \(L_s,L_x,L_y\to0\).
The special cases of private watermarking, blind watermarking, and no-attack, as well as basic properties such as the monotonicity and convexity of the capacity, are also discussed.

Future work can proceed in three directions: first, a rigorous analysis of the existence of the saddle point under continuous alphabets or non-binding constraints; second, the exact computation of the main capacity \(C^s\) under coarsening mappings, as well as the achievability of the endpoint capacities \(C^s_{\mathrm{end},(1,0)},C^s_{\mathrm{end},(0,1)}\) and whether their gap from the main capacity vanishes as the code length grows; third, the capacity behavior under block-correlated hosts and learned synonymous mappings, as well as the Bayesian extension when the knowledge base is random.

Overall, this paper provides an operational information-theoretic framework for information hiding under semantic communication, incorporates the gains and losses of semantic synonymous merging into the same game, and gives the achievable characterization of the capacity, the comparison boundary with the classical capacity, and a qualitative criterion for when semantic hiding may be better.

\bibliographystyle{unsrt}  
\bibliography{references}  

@ARTICLE{11112680,
	author={Ezzine, Rami and Wiese, Moritz and Deppe, Christian and Boche, Holger},
	journal={IEEE Transactions on Information Theory}, 
	title={Common Randomness Generation From Finite Compound Sources Aided by One-Way Communication}, 
	year={2025},
	volume={71},
	number={10},
	pages={7715-7734},
	doi={10.1109/TIT.2025.3595704}}

@article{Shannon1948,
	author = {Shannon, C. E.},
	title = {A Mathematical Theory of Communication},
	journal = {Bell System Technical Journal},
	volume = {27},
	number = {3},
	pages = {379-423},
	doi = {https://doi.org/10.1002/j.1538-7305.1948.tb01338.x},
	url = {https://onlinelibrary.wiley.com/doi/abs/10.1002/j.1538-7305.1948.tb01338.x},
	eprint = {https://onlinelibrary.wiley.com/doi/pdf/10.1002/j.1538-7305.1948.tb01338.x},
	year = {1948}
}

@inbook{CoverThomas2006,
	author    = {Cover, Thomas M. and Thomas, Joy A.},
	publisher = {John Wiley \& Sons, Ltd},
	isbn = {9780471748823},
	title = {Rate Distortion Theory},
	booktitle = {Elements of Information Theory},
	chapter = {10},
	pages = {301-346},
	doi = {https://doi.org/10.1002/047174882X.ch10},
	url = {https://onlinelibrary.wiley.com/doi/abs/10.1002/047174882X.ch10},
	eprint = {https://onlinelibrary.wiley.com/doi/pdf/10.1002/047174882X.ch10},
	year = {2005}
}

@ARTICLE{Moulin2003,
	author={Moulin, P. and O'Sullivan, J.A.},
	journal={IEEE Transactions on Information Theory}, 
	title={Information-theoretic analysis of information hiding}, 
	year={2003},
	volume={49},
	number={3},
	pages={563-593},
	doi={10.1109/TIT.2002.808134}}

@inproceedings{Ettinger1998,
	author    = {Ettinger, J. M.},
	title     = {Steganalysis and Game Equilibria},
	booktitle = {Information Hiding},
	series    = {Lecture Notes in Computer Science},
	volume    = {1525},
	pages     = {319--328},
	publisher = {Springer},
	year      = {1998}
}

@article{SomekhBaruch2003,
	author  = {Somekh-Baruch, Anelia and Merhav, Neri},
	title   = {On the Capacity Game of Public Watermarking Systems},
	journal = {IEEE Transactions on Information Theory},
	volume  = {50},
	number  = {3},
	pages   = {511--524},
	year    = {2004}
}

@article{CohenLapidoth2002,
	author  = {Cohen, Aaron S. and Lapidoth, Amos},
	title   = {The Gaussian Watermarking Game},
	journal = {IEEE Transactions on Information Theory},
	volume  = {48},
	number  = {6},
	pages   = {1639--1667},
	year    = {2002}
}

@inproceedings{Cachin1998,
	author    = {Cachin, Christian},
	title     = {An Information-Theoretic Model for Steganography},
	booktitle = {Information Hiding},
	series    = {Lecture Notes in Computer Science},
	volume    = {1525},
	pages     = {306--318},
	publisher = {Springer},
	year      = {1998}
}

@article{Cox1997,
	author  = {Cox, Ingemar J. and Kilian, Joe and Leighton, F. Thomson and Shamoon, Talal},
	title   = {Secure Spread Spectrum Watermarking for Multimedia},
	journal = {IEEE Transactions on Image Processing},
	volume  = {6},
	number  = {12},
	pages   = {1673--1687},
	year    = {1997}
}

@inproceedings{BonehShaw1998,
	author    = {Boneh, Dan and Shaw, James},
	title     = {Collusion-Secure Fingerprinting for Digital Data},
	booktitle = {Advances in Cryptology -- CRYPTO'95},
	series    = {Lecture Notes in Computer Science},
	volume    = {963},
	pages     = {452--465},
	publisher = {Springer},
	year      = {1995}
}

@article{Hartung1999,
	author  = {Hartung, Frank and Kutter, Martin},
	title   = {Multimedia Watermarking Techniques},
	journal = {Proceedings of the IEEE},
	volume  = {87},
	number  = {7},
	pages   = {1079--1107},
	year    = {1999}
}

@inproceedings{Petitcolas1999,
	author    = {Petitcolas, Fabien A. P. and Anderson, Ross J. and Kuhn, Markus G.},
	title     = {Information Hiding -- A Survey},
	booktitle = {Proceedings of the IEEE},
	volume    = {87},
	number    = {7},
	pages     = {1062--1078},
	year      = {1999}
}

@inproceedings{Barni2001,
	author    = {Barni, Mauro and Bartolini, Franco and De Rosa, Alessia and Piva, Alessandro},
	title     = {Capacity of the Watermark-Channel: How Many Bits Can Be Hidden Within a Digital Image?},
	booktitle = {Security and Watermarking of Multimedia Contents},
	series    = {Proc. SPIE},
	volume    = {3657},
	pages     = {437--448},
	year      = {1999}
}

@article{Merhav2000,
	author  = {Merhav, Neri},
	title   = {On Random Coding Error Exponents of Watermarking Codes},
	journal = {IEEE Transactions on Information Theory},
	volume  = {46},
	number  = {4},
	pages   = {1470--1476},
	year    = {2000}
}

@article{WynerZiv1976,
	author  = {Wyner, Aaron D. and Ziv, Jacob},
	title   = {The Rate-Distortion Function for Source Coding with Side Information at the Decoder},
	journal = {IEEE Transactions on Information Theory},
	volume  = {22},
	number  = {1},
	pages   = {1--10},
	year    = {1976}
}

@article{GelfandPinsker1980,
	author  = {Gel'fand, S. I. and Pinsker, M. S.},
	title   = {Coding for Channel with Random Parameters},
	journal = {Problems of Control and Information Theory},
	volume  = {9},
	number  = {1},
	pages   = {19--31},
	year    = {1980}
}

@article{Costa1983,
	author  = {Costa, Max H. M.},
	title   = {Writing on Dirty Paper},
	journal = {IEEE Transactions on Information Theory},
	volume  = {29},
	number  = {3},
	pages   = {439--441},
	year    = {1983}
}

@book{CsiszarKorner2011,
	author    = {Csisz{\'a}r, Imre and K{\"o}rner, J{\'a}nos},
	title     = {Information Theory: Coding Theorems for Discrete Memoryless Systems},
	edition   = {2nd},
	publisher = {Cambridge University Press},
	year      = {2011}
}

@article{Weaver1953,
	author  = {Weaver, Warren},
	title   = {Recent Contributions to the Mathematical Theory of Communication},
	journal = {ETC: A Review of General Semantics},
	volume  = {10},
	number  = {4},
	pages   = {261--281},
	year    = {1953}
}

@ARTICLE{Niu2024,
	author={Shao, Yulin and Cao, Qi and Gündüz, Deniz},
	journal={IEEE Transactions on Mobile Computing}, 
	title={A Theory of Semantic Communication}, 
	year={2024},
	volume={23},
	number={12},
	pages={12211-12228},
	doi={10.1109/TMC.2024.3406375}}

@article{Bourtsoulatze2019,
	author  = {Bourtsoulatze, Eirina and Kurka, David Burth and G{\"u}nd{\"u}z, Deniz},
	title   = {Deep Joint Source-Channel Coding for Wireless Image Transmission},
	journal = {IEEE Transactions on Cognitive Communications and Networking},
	volume  = {5},
	number  = {3},
	pages   = {567--579},
	year    = {2019}
}

@article{Gunduz2023,
	author  = {G{\"u}nd{\"u}z, Deniz and Qin, Zhijin and Aguerri, Inaki Estella and others},
	title   = {Beyond Transmitting Bits: Context, Semantics, and Task-Oriented Communications},
	journal = {IEEE Journal on Selected Areas in Communications},
	volume  = {41},
	number  = {1},
	pages   = {5--41},
	year    = {2023}
}

@article{Shi2021,
	author  = {Shi, Guangming and Xiao, Yang and Li, Yingyu and Xie, Xuan},
	title   = {From Semantic Communication to Semantic-Aware Networking: Model, Architecture, and Open Problems},
	journal = {IEEE Communications Magazine},
	volume  = {59},
	number  = {8},
	pages   = {44--50},
	year    = {2021}
}

@article{Zhang2022,
	author  = {Zhang, Ping and Xu, Wenjun and Gao, Hui and others},
	title   = {Toward Wisdom-Evolutionary and Primitive-Concise 6G: A New Paradigm of Semantic Communication Networks},
	journal = {Engineering},
	volume  = {8},
	number  = {1},
	pages   = {60--73},
	year    = {2022}
}

@article{Strina2025,
	author  = {Strina, A. and others},
	title   = {Semantic Communication for Image Transmission},
	journal = {IEICE ESS Fundamentals Review},
	volume  = {19},
	number  = {2},
	pages   = {70--77},
	year    = {2025}
}

@article{Niu2022,
	author  = {Niu, Kai and Dai, Jincheng and Yao, Shengshi and others},
	title   = {A Paradigm Shift Towards Semantic Communications},
	journal = {IEEE Communications Magazine},
	volume  = {60},
	number  = {11},
	pages   = {113--119},
	year    = {2022}
}

@ARTICLE{NativeAI2026,
	author={Zhang, Ping and Xu, Xiaodong and Sun, Mengying and Gao, Haixiao and Ma, Nan and Wang, Xiaoyun and Zhang, Ruichen and Wang, Jiacheng and Niyato, Dusit},
	journal={IEEE Communications Standards Magazine}, 
	title={Toward Native AI in 6G Standardization: The Roadmap of Semantic Communication}, 
	year={2026},
	volume={10},
	number={4},
	pages={245-255},
	doi={10.1109/MCOMSTD.2026.3669900}}

@ARTICLE{Huo2025DRSW,
	author={Huo, Yanhao and Xiang, Shijun},
	journal={IEEE Transactions on Circuits and Systems for Video Technology}, 
	title={DRSW: Dual-Stage Robust Semantic Watermarking for Image Semantic Communication}, 
	year={2026},
	volume={36},
	number={3},
	pages={3261-3274},
	doi={10.1109/TCSVT.2025.3613856}}

@INPROCEEDINGS{WatCOM2025,
	author={Yang, Xiao and He, Yuanhang and Li, Gaolei and Li, Jianhua},
	booktitle={ICC 2025 - IEEE International Conference on Communications}, 
	title={WatCOM: Unconscious Watermarking for Semantic Communication Intellectual Property Protection}, 
	year={2025},
	volume={},
	number={},
	pages={1181-1187},
	doi={10.1109/ICC52391.2025.11160994}}

@INPROCEEDINGS{Han2026Ownership,
	author={Han, Boyu and Yang, Xiao and Li, Gaolei and Yang, Zhaohui and Liu, Yuchen and Li, Jianhua},
	booktitle={ICC 2026 - IEEE International Conference on Communications}, 
	title={Ownership-Protected Semantic Communication via Signal Processing-Driven Robust Watermark}, 
	year={2026},
	volume={},
	number={},
	pages={1-6},
	doi={10.1109/ICC59461.2026.11587337}}

@article{Huo2025ImageSS,
	author  = {Huo, Yanhao and Xiang, Shijun and Luo, Xiangyang and Zhang, Xinpeng},
	title   = {Image Semantic Steganography: A Way to Hide Information in Semantic Communication},
	journal = {IEEE Transactions on Circuits and Systems for Video Technology},
	volume  = {35},
	number  = {2},
	pages   = {1951--1960},
	year    = {2025},
	doi     = {10.1109/TCSVT.2024.3476689}
}

@article{Ni2026Stega,
	author  = {Ni, Shouxiang and Liu, Guolian and others},
	title   = {Controllable Steganography for Robust and Efficient Semantic Communication Systems},
	journal = {IEEE Communications Magazine},
	volume  = {64},
	number  = {1},
	pages   = {160--166},
	year    = {2026}
}

\appendix

\section{Proof of Proposition~\ref{prop:payoff_convexity}}

\subsection{Proof of Proposition~\ref{prop:payoff_convexity}(i)}
\label{app:payoff_convexity}

\begin{proof}
	Fix \(p(\tilde{s},k)\) and the covert channel \(Q(\tilde{x},\tilde{u}\mid\tilde{s},k)\). The payoff function is
	\[
	J(Q,A)=I(\tilde{U};\tilde{Y}\mid K)-I(\tilde{U};\tilde{S}\mid K).
	\]
	The second term \(I(\tilde{U};\tilde{S}\mid K)\) depends only on \(p(\tilde{s},k)\) and \(Q\), and is independent of the attack channel \(A(y|x)\), so it suffices to prove that the first term \(I(\tilde{U};\tilde{Y}\mid K)\) is convex in \(A\).
	
	Write the conditional mutual information in the form of ``weighted over \(k\), weighted over \(\tilde{u}\), with the inner layer being the KL divergence of \(\tilde{Y}\)'':
	\begin{equation}
		I(\tilde{U};\tilde{Y}\mid K)
		=\sum_{k}p(k)\sum_{\tilde{u}}p(\tilde{u}\mid k)\,
		D\big(p(\tilde{y}\mid\tilde{u},k)\,\big\|\,p(\tilde{y}\mid k)\big),
		\label{eq:app_mi_kl}
	\end{equation}
	where
	\[
	D\big(p(\tilde{y}\mid\tilde{u},k)\,\big\|\,p(\tilde{y}\mid k)\big)
	=\sum_{\tilde{y}}p(\tilde{y}\mid\tilde{u},k)\log\frac{p(\tilde{y}\mid\tilde{u},k)}{p(\tilde{y}\mid k)}.
	\]
	\eqref{eq:app_mi_kl} is equivalent to the joint KL form, because
	\[
	\log\frac{p(\tilde{y}\mid\tilde{u},k)}{p(\tilde{y}\mid k)}
	=\log\frac{p(\tilde{u},\tilde{y}\mid k)}{p(\tilde{u}\mid k)p(\tilde{y}\mid k)}.
	\]
	
	Next, expand \(p(\tilde{y}\mid\tilde{u},k)\) and \(p(\tilde{y}\mid k)\) into \(A\). By the Markov chain \((\tilde{U},\tilde{S},K)\to\tilde{X}\to X\to Y\to\tilde{Y}\), we have
	\[
	p(\tilde{y}\mid\tilde{u},k)
	=\sum_{\tilde{x}}p(\tilde{x}\mid\tilde{u},k)\sum_{x\in f_x(\tilde{x})} p(x\mid\tilde{x})\ \sum_{y:\,f_y^{-1}(y)=\tilde{y}}A(y|x),
	\]
	and
	\[
	p(\tilde{y}\mid k)
	=\sum_{\tilde{u}}p(\tilde{u}\mid k)\,p(\tilde{y}\mid\tilde{u},k)
	=\sum_{\tilde{u}}p(\tilde{u}\mid k)\sum_{\tilde{x}}p(\tilde{x}\mid\tilde{u},k)
	\sum_{x\in f_x(\tilde{x})} p(x\mid\tilde{x})\ \sum_{y:\,f_y^{-1}(y)=\tilde{y}}A(y|x).
	\]
	After fixing \(Q\), both \(p(\tilde{u}\mid k)\) and \(p(\tilde{x}\mid\tilde{u},k)\) are independent of \(A\), so \(p(\tilde{y}\mid\tilde{u},k)\) and \(p(\tilde{y}\mid k)\) are both linear functions of \(A(y|x)\):
	\[
	p(\tilde{y}\mid\tilde{u},k)=\sum_{x,y}c_{\tilde{u},k}(\tilde{y};x,y)\,A(y|x),
	\qquad
	p(\tilde{y}\mid k)=\sum_{x,y}c_{k}(\tilde{y};x,y)\,A(y|x),
	\]
	where the coefficients \(c_{\tilde{u},k}\) and \(c_k\) are determined by \(p(\tilde{u}\mid k)\), \(p(\tilde{x}\mid\tilde{u},k)\), and \(f_x,f_y\), and do not depend on \(A\).
	
	Let \(A_1,A_2\) be any two feasible attack channels, \(\theta\in[0,1]\), and let \(A_\theta=\theta A_1+(1-\theta)A_2\). By the linearity of \(A\mapsto p(\tilde{y}\mid\tilde{u},k)\) and \(A\mapsto p(\tilde{y}\mid k)\),
	\[
	p_\theta(\tilde{y}\mid\tilde{u},k)=\theta\,p_1(\tilde{y}\mid\tilde{u},k)+(1-\theta)\,p_2(\tilde{y}\mid\tilde{u},k),
	\qquad
	p_\theta(\tilde{y}\mid k)=\theta\,p_1(\tilde{y}\mid k)+(1-\theta)\,p_2(\tilde{y}\mid k).
	\]
	The KL divergence \(D(p\|q)\) is jointly convex in \((p,q)\), so
	\[
	D\big(p_\theta(\tilde{y}\mid\tilde{u},k)\,\big\|\,p_\theta(\tilde{y}\mid k)\big)
	\le\theta\,D\big(p_1(\tilde{y}\mid\tilde{u},k)\,\big\|\,p_1(\tilde{y}\mid k)\big)
	+(1-\theta)\,D\big(p_2(\tilde{y}\mid\tilde{u},k)\,\big\|\,p_2(\tilde{y}\mid k)\big).
	\]
	Weighting over \(\tilde{u}\) by \(p(\tilde{u}\mid k)\) and over \(k\) by \(p(k)\), with weights independent of \(A\), the convexity is preserved, so
	\[
	I_\theta(\tilde{U};\tilde{Y}\mid K)
	\le\theta\,I_1(\tilde{U};\tilde{Y}\mid K)+(1-\theta)\,I_2(\tilde{U};\tilde{Y}\mid K),
	\]
	i.e., \(I(\tilde{U};\tilde{Y}\mid K)\) is convex in \(A\). This completes the proof.
\end{proof}

\subsection{Proof of Proposition~\ref{prop:payoff_convexity}(ii)}
\label{app:payoff_convexity_ii}

\begin{proof}
	Fix \(p(\tilde{s},k)\), \(p(\tilde{x}\mid\tilde{s},\tilde{u},k)\), and the attack channel \(A(y|x)\); at this point only \(p(\tilde{u}\mid\tilde{s},k)\) is variable.
	Let \(p_1(\tilde{u}\mid\tilde{s},k)\) and \(p_2(\tilde{u}\mid\tilde{s},k)\) be any two covert channels, \(\theta\in[0,1]\), and let the mixture be
	\[
	p_\theta(\tilde{u}\mid\tilde{s},k)=\theta p_1(\tilde{u}\mid\tilde{s},k)+(1-\theta)p_2(\tilde{u}\mid\tilde{s},k).
	\]
	Introduce a time-slot variable \(Q\sim\mathrm{Bern}(\theta)\) independent of \((\tilde{S},K)\), let \(\tilde{U}=\tilde{U}_Q\) (taking \(p_1\) when \(Q=1\), and \(p_2\) otherwise), and then generate \(\tilde{X},X,Y,\tilde{Y}\) according to the fixed \(p(\tilde{x}\mid\tilde{s},\tilde{u},k)\) and \(A\).
	
	By the chain rule, first eliminate \(Q\):
	\begin{align}
		I(\tilde{U};\tilde{Y}\mid K)
		&=\theta I(\tilde{U}_1;\tilde{Y}_1\mid K)+(1-\theta)I(\tilde{U}_2;\tilde{Y}_2\mid K)
		+I(Q;\tilde{Y}\mid K)-I(Q;\tilde{Y}\mid\tilde{U},K),\\
		I(\tilde{U};\tilde{S}\mid K)
		&=\theta I(\tilde{U}_1;\tilde{S}\mid K)+(1-\theta)I(\tilde{U}_2;\tilde{S}\mid K)
		-I(Q;\tilde{S}\mid\tilde{U},K),
	\end{align}
	where \(Q\perp(\tilde{S},K)\) is used, so \(I(Q;\tilde{S}\mid K)=0\).
	Subtracting the two equations gives
	\begin{equation}
		J(p_\theta)=\theta J(p_1)+(1-\theta)J(p_2)+\Delta,
		\qquad
		\Delta=I(Q;\tilde{Y}\mid K)-I(Q;\tilde{Y}\mid\tilde{U},K)+I(Q;\tilde{S}\mid\tilde{U},K).
		\label{eq:app_ii_mix}
	\end{equation}
	
	Next, prove \(\Delta\ge0\). By the Markov chain \(Q\to(\tilde{U},\tilde{S},K)\to\tilde{X}\to X\to Y\to\tilde{Y}\), \(I(Q;\tilde{Y}\mid\tilde{U},\tilde{S},K)=0\).
	Applying the chain rule twice to \(I(Q;\tilde{Y},\tilde{S}\mid\tilde{U},K)\):
	\[
	I(Q;\tilde{Y}\mid\tilde{U},K)+I(Q;\tilde{S}\mid\tilde{U},\tilde{Y},K)
	=I(Q;\tilde{S}\mid\tilde{U},K)+I(Q;\tilde{Y}\mid\tilde{U},\tilde{S},K)
	=I(Q;\tilde{S}\mid\tilde{U},K),
	\]
	hence
	\[
	I(Q;\tilde{Y}\mid\tilde{U},K)-I(Q;\tilde{S}\mid\tilde{U},K)
	=-I(Q;\tilde{S}\mid\tilde{U},\tilde{Y},K)\le0,
	\]
	therefore
	\[
	\Delta=I(Q;\tilde{Y}\mid K)+I(Q;\tilde{S}\mid\tilde{U},\tilde{Y},K)\ge0,
	\]
	both terms being nonnegative mutual informations.
	Substituting into \eqref{eq:app_ii_mix} gives
	\[
	J\big(\theta p_1+(1-\theta)p_2\big)\ge \theta J(p_1)+(1-\theta)J(p_2),
	\]
	i.e., \(J\) is concave in \(p(\tilde{u}\mid\tilde{s},k)\). This completes the proof.
\end{proof}

\subsection{Proof of Proposition~\ref{prop:payoff_convexity}(iii)}
\label{app:payoff_convexity_iii}

\begin{proof}
	Fix \(p(\tilde{s},k)\), \(p(\tilde{u}\mid\tilde{s},k)\), and the attack channel \(A(y|x)\). At this point, only \(p(\tilde{x}\mid\tilde{s},\tilde{u},k)\) of the covert channel \(Q\) remains variable.
	The payoff function is
	\[
	J(Q,A)=I(\tilde{U};\tilde{Y}\mid K)-I(\tilde{U};\tilde{S}\mid K).
	\]
	The second term \(I(\tilde{U};\tilde{S}\mid K)\) depends only on \(p(\tilde{s},k)\) and \(p(\tilde{u}\mid\tilde{s},k)\), is independent of \(p(\tilde{x}\mid\tilde{s},\tilde{u},k)\), and is a constant, so it suffices to prove that the first term is convex in \(p(\tilde{x}\mid\tilde{s},\tilde{u},k)\).
	
	\paragraph{The First Term Is Convex in \(p(\tilde{x}\mid\tilde{s},\tilde{u},k)\)}
	Since \(p(\tilde{u}\mid\tilde{s},k)\) and \(A\) are fixed, the marginal \(p(\tilde{u}\mid k)\) of \(\tilde{U}\) is fixed;
	\[
	p(\tilde{x}\mid\tilde{u},k)=\sum_{\tilde{s}}p(\tilde{x}\mid\tilde{s},\tilde{u},k)p(\tilde{s}\mid\tilde{u},k)
	\]
	is linear in \(p(\tilde{x}\mid\tilde{s},\tilde{u},k)\), and consequently
	\[
	p(\tilde{y}\mid\tilde{u},k)
	=\sum_{\tilde{x}}p(\tilde{x}\mid\tilde{u},k)\sum_{x\in f_x(\tilde{x})} p(x\mid\tilde{x})\ \sum_{y:\,f_y^{-1}(y)=\tilde{y}}A(y|x)
	\]
	and \(p(\tilde{y}\mid k)=\sum_{\tilde{u}}p(\tilde{u}\mid k)p(\tilde{y}\mid\tilde{u},k)\) are both linear in \(p(\tilde{x}\mid\tilde{s},\tilde{u},k)\).
	Write the first term as
	\[
	I(\tilde{U};\tilde{Y}\mid K)
	=\sum_{k}p(k)\sum_{\tilde{u}}p(\tilde{u}\mid k)\,
	D\big(p(\tilde{y}\mid\tilde{u},k)\,\big\|\,p(\tilde{y}\mid k)\big).
	\]
	Here, \(p(\tilde{u}\mid k)\) and \(p(k)\) are fixed, and both parameters of the inner KL divergence are linear functions of \(p(\tilde{x}\mid\tilde{s},\tilde{u},k)\).
	Let \(P_1,P_2\) be any two feasible \(p(\tilde{x}\mid\tilde{s},\tilde{u},k)\), \(\theta\in[0,1]\), and let \(P_\theta=\theta P_1+(1-\theta)P_2\). By linearity,
	\[
	p_\theta(\tilde{y}\mid\tilde{u},k)=\theta\,p_1(\tilde{y}\mid\tilde{u},k)+(1-\theta)\,p_2(\tilde{y}\mid\tilde{u},k),
	\qquad
	p_\theta(\tilde{y}\mid k)=\theta\,p_1(\tilde{y}\mid k)+(1-\theta)\,p_2(\tilde{y}\mid k).
	\]
	By the joint convexity of the KL divergence in \((p,q)\),
	\[
	D\big(p_\theta(\tilde{y}\mid\tilde{u},k)\,\big\|\,p_\theta(\tilde{y}\mid k)\big)
	\le\theta\,D\big(p_1(\tilde{y}\mid\tilde{u},k)\,\big\|\,p_1(\tilde{y}\mid k)\big)
	+(1-\theta)\,D\big(p_2(\tilde{y}\mid\tilde{u},k)\,\big\|\,p_2(\tilde{y}\mid k)\big).
	\]
	Weighting over \(\tilde{u}\) by \(p(\tilde{u}\mid k)\) and over \(k\) by \(p(k)\), with fixed weights, the convexity is preserved, so
	\[
	I_\theta(\tilde{U};\tilde{Y}\mid K)
	\le\theta\,I_1(\tilde{U};\tilde{Y}\mid K)+(1-\theta)\,I_2(\tilde{U};\tilde{Y}\mid K),
	\]
	i.e., \(I(\tilde{U};\tilde{Y}\mid K)\) is convex in \(p(\tilde{x}\mid\tilde{s},\tilde{u},k)\).
	After subtracting the constant \(I(\tilde{U};\tilde{S}\mid K)\), which is independent of \(p(\tilde{x}\mid\tilde{s},\tilde{u},k)\), \(J\) is convex in \(p(\tilde{x}\mid\tilde{s},\tilde{u},k)\).
	
	\paragraph{The Maximum Is Attained at a Zero-One Distribution}
	After fixing \(p(\tilde{s},k)\), \(p(\tilde{u}\mid\tilde{s},k)\), and \(A\), \(J\) is a convex function of \(p(\tilde{x}\mid\tilde{s},\tilde{u},k)\).
	For each fixed \((\tilde{s},\tilde{u},k)\), \(p(\tilde{x}\mid\tilde{s},\tilde{u},k)\) takes values in the probability simplex over \(\tilde{\mathcal{X}}\); the maximum of a convex function on a convex compact set is attained at a vertex of the simplex, and the vertices of the simplex are exactly the zero-one distributions, i.e., there exists \(x^*(\tilde{s},\tilde{u},k)\in\tilde{\mathcal{X}}\) such that \(p(\tilde{x}\mid\tilde{s},\tilde{u},k)=\mathbb{I}\big(\tilde{x}=x^*(\tilde{s},\tilde{u},k)\big)\).
	Taking a vertex separately for each \((\tilde{s},\tilde{u},k)\) yields the optimal \(p(\tilde{x}\mid\tilde{s},\tilde{u},k)\) as a zero-one distribution, equivalently, \(\tilde{X}\) is a deterministic function of \((\tilde{S},\tilde{U},K)\).
	
	\paragraph{Remark under Distortion Constraints}
	The above vertex argument is for the simplex without distortion constraints; after imposing the distortion constraints \(D_0,D_1\), the feasible set is the intersection of the simplex and half-spaces, whose extreme points are not necessarily zero-one distributions, so one cannot directly derive from this argument that the optimal solution of the constrained problem is of the zero-one type.
	However, the coding construction does not require the zero-one reduction: the encoder can draw the semantic composite signal according to \(p^*(\tilde x\mid\tilde s,\tilde u,k)\) with private randomness (see Appendix~\ref{app:achievability}), and the distortion constraints and the joint typicality analysis are unaffected. This completes the proof.
\end{proof}

\subsection{Proof of the Cardinality Bound of the Auxiliary Variable}
\label{app:aux_cardinality}

\begin{proof}
	Fix \(p(\tilde{s},k)\) and \(p(\tilde{x}\mid\tilde{s},\tilde{u},k)\). Write the covert channel as
	\[
	Q(\tilde{x},\tilde{u}\mid\tilde{s},k)
	=p(\tilde{u}\mid\tilde{s},k)\,p(\tilde{x}\mid\tilde{s},\tilde{u},k).
	\]
	The cardinality bound is proved with the Fenchel-Eggleston-Carathéodory support lemma (for an application example, see~\cite{11112680}).
	
	For each auxiliary symbol \(\tilde{u}\), record its ``effect'' under a given \((\tilde{s},k)\) as the conditional distribution
	\[
	\big\{p(\tilde{x}\mid\tilde{s},\tilde{u},k):\ (\tilde{s},k)\in\tilde{\Omega},\ \tilde{x}\in\tilde{\mathcal{X}}\big\},
	\]
	This is a point in a \(|\tilde{\mathcal{X}}|\,|\tilde{\Omega}|\)-dimensional space (on each \((\tilde{s},k)\) the probabilities sum to 1, with \(|\tilde{\Omega}|(|\tilde{\mathcal{X}}|-1)\) independent components in total).
	The overall conditional distribution corresponding to any \(p(\tilde{u}\mid\tilde{s},k)\),
	\[
	p(\tilde{x}\mid\tilde{s},k)
	=\sum_{\tilde{u}}p(\tilde{u}\mid\tilde{s},k)\,p(\tilde{x}\mid\tilde{s},\tilde{u},k)
	\]
	is a convex combination of these points.
	For the payoff function \(J=I(\tilde{U};\tilde{Y}\mid K)-I(\tilde{U};\tilde{S}\mid K)\) and the distortion constraints to remain unchanged after compression, the following must be preserved simultaneously:
	\begin{enumerate}
		\item[(a)] \(p(\tilde{x}\mid\tilde{s},k)\) on each \((\tilde{s},k)\) (a total of \(|\tilde{\Omega}|(|\tilde{\mathcal{X}}|-1)\) independent constraints). This determines the distortion \(D_0,D_1\) and the receiver marginal entropy \(H(\tilde{Y}\mid K)\);
		\item[(b)] the two scalars \(I(\tilde{U};\tilde{Y}\mid K)\) and \(I(\tilde{U};\tilde{S}\mid K)\) (to keep \(J\) unchanged; here \(I(\tilde{U};\tilde{S}\mid K)\) depends only on \(p(\tilde{u}\mid\tilde{s},k)\) and the fixed \(p(\tilde{s},k)\), and is unrelated to (a)).
	\end{enumerate}
	(The two scalars in (b) are evaluated under a given attack \(A\), corresponding to the tight bound for a fixed attack; if the inner minimum \(\min_A J\) must be preserved, then repeat (b) pointwise over a finite \(\delta\)-net of \(\mathcal A(Q,D_2,D_3)\), with the constant in the bound replaced by the net size \(N_\delta\), see Remark~\ref{rem:minmax_cardinality}.)
	
	By the support lemma, preserving the values of these \(|\tilde{\Omega}|(|\tilde{\mathcal{X}}|-1)+2\) continuous functions requires at most \(|\tilde{\Omega}|(|\tilde{\mathcal{X}}|-1)+3\) auxiliary symbols.
	When \(|\tilde{\Omega}|\ge2\), \(|\tilde{\Omega}|(|\tilde{\mathcal{X}}|-1)+3\le|\tilde{\mathcal{X}}|\,|\tilde{\Omega}|+1\), so one can take \(|\tilde{\mathcal{U}}'|\le|\tilde{\mathcal{X}}|\,|\tilde{\Omega}|+1\).
	When \(|\tilde{\Omega}|=1\), \((\tilde{S},K)\) degenerates to a constant, so \(I(\tilde{U};\tilde{S}\mid K)=0\) holds automatically and need not be preserved; at this point one only needs to preserve \(p(\tilde{x}\mid\tilde{s},k)\) (\(|\tilde{\mathcal{X}}|-1\) constraints) and \(I(\tilde{U};\tilde{Y}\mid K)\) (\(1\) constraint), a total of \(|\tilde{\mathcal{X}}|\), and the support lemma gives \(|\tilde{\mathcal{U}}'|\le|\tilde{\mathcal{X}}|+1=|\tilde{\mathcal{X}}|\,|\tilde{\Omega}|+1\).
	Therefore, for any \(|\tilde{\Omega}|\ge1\), \(|\tilde{\mathcal{U}}'|\le|\tilde{\mathcal{X}}|\,|\tilde{\Omega}|+1\) holds, without affecting the attainment of optimality.
	
	Furthermore, by the merging property of the synonymous mappings \(f_x:\tilde{\mathcal{X}}\rightrightarrows\mathcal{X}\) and \(f_s\), \(|\tilde{\mathcal{X}}|\le|\mathcal{X}|\) and \(|\tilde{\Omega}|\le|\Omega|\), so the size can also be controlled by the syntactic-layer quantity \(|\mathcal{X}|\,|\Omega|+1\). This completes the proof.
\end{proof}

\section{Proof of the Semantic Hiding Capacity}
This section proves the results of Section~\ref{subsec:capacity_existence}.

\subsection{Proof of the Attainability of the Inner Minimum}
\label{app:inner_min_attained}

\begin{proof}
	Fix \(Q\in\mathcal{Q}(D_0,D_1)\). By Definition~\ref{def:attack_channel},
	\[
	\begin{aligned}
		\mathcal{A}(Q,D_2,D_3)
		=\Big\{&A(y|x):\ A(y|x)\ge0,\ \sum_{y}A(y|x)=1\ \forall x,\\
		&\sum_{x,y}d_2(x,y)A(y|x)p_Q(x)\le D_2,\ 
		\sum_{x,y}d_3\big(f_x^{-1}(x),f_y^{-1}(y)\big)A(y|x)p_Q(x)\le D_3\Big\}.
	\end{aligned}
	\]
	Under the finite alphabet setting, \(\mathcal{X}\) and \(\mathcal{Y}\) are finite sets, \(A\) can be viewed as a vector in a finite-dimensional Euclidean space whose components satisfy \(0\le A(y|x)\le1\), so \(\mathcal{A}(Q,D_2,D_3)\) is bounded.
	The four constraints are all inequalities or equalities in terms of continuous functions of \(A\) (\(A(y|x)\ge0\), \(\sum_y A(y|x)=1\), and the two distortion constraints), so \(\mathcal{A}(Q,D_2,D_3)\) is closed.
	In a finite-dimensional Euclidean space, a bounded closed set is compact, so \(\mathcal{A}(Q,D_2,D_3)\) is compact.
	
	Moreover, \(A(y|x)\ge0\) and \(\sum_y A(y|x)=1\) are linear constraints, and the two distortion constraints are linear inequalities in \(A\) (with \(p_Q(x)\) fixed), so \(\mathcal{A}(Q,D_2,D_3)\) is convex.
	
	By Proposition~\ref{prop:payoff_convexity}(i), \(J(Q,A)\) is convex in \(A\); the infimum of a convex function on a convex compact set is attainable (a lower semicontinuous convex function attains its minimum on a compact set), so there exists \(A^*(Q)\in\mathcal{A}(Q,D_2,D_3)\) such that
	\[
	\min_{A\in\mathcal{A}(Q,D_2,D_3)}J(Q,A)=J(Q,A^*(Q)).
	\]
	This completes the proof.
\end{proof}

\subsection{Proof of the Attainability of the Outer Maximum}
\label{app:outer_max_attained}

\begin{proof}
	Proceed in three steps: first prove that \(\mathcal{Q}(D_0,D_1)\) is convex compact, then prove that the attack feasible-set correspondence is continuous and hence, by Berge's maximum theorem, \(W(Q)=\min_A J(Q,A)\) is continuous, and finally use the fact that a continuous function attains its maximum on a compact set.
	
	\paragraph{\(\mathcal{Q}(D_0,D_1)\) Is Convex Compact}
	By Definition~\ref{def:covert_channel},
	\[
	\mathcal{Q}(D_0,D_1)
	=\Big\{Q(\tilde{x},\tilde{u}\mid\tilde{s},k):\ Q\ge0,\ \sum_{\tilde{x},\tilde{u}}Q(\tilde{x},\tilde{u}\mid\tilde{s},k)=1\ \forall(\tilde{s},k),\ 
	\mathbb{E}[d_0]\le D_0,\ \mathbb{E}[d_1]\le D_1\Big\}.
	\]
	Under the finite alphabet setting, \(Q\) is a vector in a finite-dimensional Euclidean space whose components satisfy \(0\le Q(\tilde{x},\tilde{u}\mid\tilde{s},k)\le1\), so \(\mathcal{Q}(D_0,D_1)\) is bounded; the four constraints (\(Q\ge0\), \(\sum=1\), \(\mathbb{E}[d_0]\le D_0\), \(\mathbb{E}[d_1]\le D_1\)) are all closed conditions in terms of continuous functions of \(Q\), so it is closed. A finite-dimensional bounded closed set is compact. Moreover, the four constraints are all linear, so \(\mathcal{Q}(D_0,D_1)\) is convex.
	
	\paragraph{\(W(Q)\) Is Continuous}
	Under the finite alphabet and bounded distortion function assumptions, \(J(Q,A)\) is jointly continuous in \((Q,A)\); for each \(Q\), \(\mathcal{A}(Q,D_2,D_3)\) is a bounded closed set, hence compact.
	The correspondence \(Q\mapsto\mathcal{A}(Q,D_2,D_3)\) is continuous: the upper semicontinuity is given by the closed graph (\(A\ge0\), \(\sum_y A(y|x)=1\), and the two distortion constraints are all continuous functions of \((Q,A)\), so the graph of feasible pairs is closed, and with a compact range, a closed graph is equivalent to upper semicontinuity); the lower semicontinuity is given by an interior-point argument. Suppose there exists an anchor channel \(A_0\) strictly feasible for all \(Q\) (e.g., the identity channel when \(f_x=f_y\), \(d_2(x,x)=d_3(\tilde x,\tilde x)=0\), and \(D_2,D_3>0\), whose two distortions are identically zero); then for any \(A\in\mathcal{A}(Q,D_2,D_3)\) and \(\lambda\in(0,1)\), the distortion of \(A_\lambda=(1-\lambda)A+\lambda A_0\) does not exceed \((1-\lambda)D_2\) and \((1-\lambda)D_3\); since \(p_{Q'}(x)\) is continuous in \(Q'\), for \(Q'\) sufficiently close, \(A_\lambda\in\mathcal{A}(Q',D_2,D_3)\) still holds, and letting \(\lambda\to0\) together with a diagonal argument gives the lower semicontinuity.
	The degenerate case with zero thresholds can be covered by taking limits of a positive threshold sequence. Therefore, by Berge's maximum theorem, the value function \(W(Q)=\min_{A\in\mathcal{A}(Q,D_2,D_3)}J(Q,A)\) is continuous in \(Q\).
	
	\paragraph{The Maximum Is Attainable}
	\(W(Q)\) is continuous and defined on the compact set \(\mathcal{Q}(D_0,D_1)\), and a continuous function attains its maximum on a compact set, so there exists \(Q^*\in\mathcal{Q}(D_0,D_1)\) such that
	\[
	W(Q^*)=\max_{Q\in\mathcal{Q}(D_0,D_1)}W(Q)
	=\max_{Q\in\mathcal{Q}(D_0,D_1)}\min_{A\in\mathcal{A}(Q,D_2,D_3)}J(Q,A).
	\]
	By \ref{app:inner_min_attained}, the inner minimum is attainable, so \(W(Q^*)=\min_{A\in\mathcal{A}(Q^*,D_2,D_3)}J(Q^*,A)\). This completes the proof.
\end{proof}

\subsection{Proof of the Achievability}
\label{app:achievability}
This section proves Proposition~\ref{prop:achievability}.

\begin{proof}
	Take arbitrary \(R<C^s\). By the continuity of the capacity \(C^s\) in the distortion thresholds \((D_0,D_1)\), there exists \(\gamma>0\) such that \(R<C^s(D_0-\gamma,D_1-\gamma)\).
	Take a covert channel \(Q^*\in\mathcal{Q}(D_0-\gamma,D_1-\gamma)\) attaining this capacity and its worst-case attack \(A^*\in\mathcal{A}(Q^*,D_2,D_3)\) (the inner and outer extrema are attainable, see Appendices~\ref{app:inner_min_attained} and~\ref{app:outer_max_attained}), and take \(\varepsilon>0\) such that
	\begin{equation}
		R=I_{Q^*,A^*}(\tilde{U};\tilde{Y}\mid K)-I_{Q^*}(\tilde{U};\tilde{S}\mid K)-3\varepsilon.
		\label{eq:achievable_rate}
	\end{equation}
	Let the joint distribution induced by \(Q^*\) be
	\[
	p^*(\tilde{u},\tilde{s},k,\tilde{x},x)=p(\tilde{s},k)\,Q^*(\tilde{x},\tilde{u}\mid\tilde{s},k)\,p(x\mid\tilde{x}).
	\]
	For any attack \(A\), after passing through the equivalent semantic channel \(W_A\) (Definition~\ref{def:equivalent_semantic_channel}), the semantic tuple \((\tilde{u},\tilde{s},k,\tilde{x},\tilde{y})\) obeys
	\[
	p_{Q^*,A}(\tilde{u},\tilde{s},k,\tilde{x},\tilde{y})
	=p(\tilde{s},k)\,Q^*(\tilde{x},\tilde{u}\mid\tilde{s},k)\,W_A(\tilde{y}\mid\tilde{x}),
	\]
	whose memoryless extension \(p_{Q^*,A}^{\otimes n}\) is given by \(W_A^{\otimes n}\); the subsequent joint typicality decoding and error probability analysis are all built on \(W_A^{\otimes n}\).
	
	\paragraph{Random Codebook Generation (with the Shared Side Information \(K^n\) as Common Randomness)}
	Take the codebook size
	\[
	L_n=\big\lceil 2^{n(I_{Q^*}(\tilde{U};\tilde{S}\mid K)+\varepsilon)}\big\rceil,
	\]
	so that each message has sufficiently many auxiliary sequences to guarantee that the encoder finds, with high probability, a sequence jointly typical with the host.
	With the shared side information \(K^n\) as common randomness, for each message \(m\in\{1,\dots,2^{nR}\}\) and index \(l\in\{1,\dots,L_n\}\), independently draw
	\[
	\tilde{u}^n(l,m)\sim \prod_{i=1}^{n}p^*(\tilde{u}_i\mid k_i).
	\]
	(The codebook is generated conditioned on \(K^n\), which is the key to making the encoding failure probability decay exponentially at the exponent \(I_{Q^*}(\tilde{U};\tilde{S}\mid K)\) rather than \(I_{Q^*}(\tilde{U};\tilde{S},K)\); \(K^n\) is shared by the transmitter and receiver and unknown to the attacker, so this common randomness is legitimate.)
	
	\paragraph{Encoding}
	The encoder selects, among the \(L_n\) auxiliary sequences of message \(m\), the first one jointly typical with \((\tilde{s}^n,k^n)\); if none exists, the encoding failure event \(\mathcal{E}_{\mathrm{enc}}\) is recorded.
	By the conditional typicality lemma, given a typical \((\tilde{s}^n,k^n)\), the probability that a single independently drawn auxiliary sequence \(\tilde{u}^n\sim\prod_i p^*(\tilde{u}_i\mid k_i)\) is jointly typical with \((\tilde{s}^n,k^n)\) under \(p^*\) is approximately \(2^{-n I_{Q^*}(\tilde{U};\tilde{S}\mid K)}\); hence the probability that all \(L_n\) independent draws fail is approximately
	\[
	\bigl(1-2^{-n I_{Q^*}(\tilde{U};\tilde{S}\mid K)}\bigr)^{L_n}
	\le \exp\bigl(-L_n\,2^{-n I_{Q^*}(\tilde{U};\tilde{S}\mid K)}\bigr).
	\]
	By \(L_n=\lceil 2^{n(I_{Q^*}(\tilde{U};\tilde{S}\mid K)+\varepsilon)}\rceil\), this probability does not exceed \(\exp(-2^{n\varepsilon})\to0\), so \(\Pr[\mathcal{E}_{\mathrm{enc}}]\to0\).
	After selecting \(\tilde{u}^n\), the encoder independently draws the semantic composite signal \(\tilde{x}_i\sim p^*(\cdot\mid\tilde{s}_i,\tilde{u}_i,k_i)\) according to \(p^*(\tilde{x}\mid\tilde{s},\tilde{u},k)\), and then draws \(x_i\) within the synonymous class \(f_x(\tilde{x}_i)\) according to the fixed mechanism \(p(x_i\mid\tilde{x}_i)\), obtaining \(x^n\).
	
	\paragraph{Decoding (Universal Joint Typicality Decoding)}
	The decoder does not know the actual attack \(A\), so it adopts universal (compound-channel) joint typicality decoding: given \((y^n,k^n)\), first obtain the semantic received \(\tilde{y}^n=f_y^{-1,n}(y^n)\); the decoder searches all \((l,m)\) for the unique auxiliary sequence \(\tilde{u}^n(l,m)\) such that there exists some \(A'\in\mathcal{A}(Q^*,D_2,D_3)\) for which \((\tilde{u}^n(l,m),\tilde{y}^n,k^n)\) is jointly typical under the distribution \(p_{Q^*,A'}\) induced by \((Q^*,A')\).
	This decision depends only on the codebook, \(Q^*\), and the attack class \(\mathcal{A}(Q^*,D_2,D_3)\), not on the actual \(A\). If such \((l,m)\) is unique, output its message; otherwise take an arbitrary one:
	\[
	\hat{m}(y^n,k^n)
	=\begin{cases}
		m, & \text{if there is a unique }(l,m)\text{ such that }\exists A'\in\mathcal{A}(Q^*,D_2,D_3):\ (\tilde{u}^n(l,m),\tilde{y}^n,k^n)\in\mathcal{T}_\epsilon^{(n)}(p_{Q^*,A'}),\\
		\text{arbitrary}, & \text{otherwise}.
	\end{cases}
	\]
	
	\paragraph{Error Probability}
	For a fixed \(A\in\mathcal{A}(\theta_n,D_2,D_3)\) (the actual memoryless attack \(A^{\otimes n}\)), let \(P_{e,n}(A)=P_{e,n}(A^{\otimes n})=\Pr[\hat{M}\neq M]\).
	The error event is contained in \(\mathcal{E}_1\cup\mathcal{E}_{\mathrm{enc}}\cup\mathcal{E}_2\): \(\mathcal{E}_1\) is the event that the transmitted tuple is not jointly typical under the true distribution \(p_{Q^*,A}\), \(\mathcal{E}_{\mathrm{enc}}\) is the encoding failure, and \(\mathcal{E}_2\) is the event that the auxiliary sequence of some wrong message is misjudged as typical by the universal decoding.
	By the typicality lemma, \(\Pr[\mathcal{E}_1]\le\epsilon\).
	Here the compatibility between the decoding search class and the true attack class must be explained: the true attack satisfies \(\mathbb{E}_{p_{X^n}}[d_2]\le D_2\) and \(\mathbb{E}_{p_{X^n}}[d_3]\le D_3\), while the empirical distribution \(p_{X^n}\) of a typical codeword differs from the \(X\)-marginal of \(p_{Q^*}\) by \(O(\epsilon)\), so the true \(A\in\mathcal{A}(Q^*,D_2+O(\epsilon),D_3+O(\epsilon))\); moreover, by the continuity of \(A\mapsto p_{Q^*,A}\), the deviation of \(I_{Q^*,A}(\tilde{U};\tilde{Y}\mid K)\) from \(I_{Q^*,A^*}\) is \(O(\epsilon)\).
	Therefore, take the decoder's search class as the enlarged \(\mathcal{A}(Q^*,D_2+\kappa,D_3+\kappa)\) (\(\kappa=O(\epsilon)\)), which guarantees that the correct tuple falls into the search class, and by the continuity of the value function \(V(D_2,D_3)=\min_{A\in\mathcal{A}(Q^*,D_2,D_3)}I_{Q^*,A}(\tilde{U};\tilde{Y}\mid K)\) in the thresholds (Berge; with the same closed-graph plus interior-point argument as in Appendix~\ref{app:outer_max_attained}), when \(\kappa=o(1)\) the minimum over the enlarged class differs from \(I_{Q^*,A^*}\) by \(o(1)\), which can be absorbed into the \(\varepsilon\) margin, so the exponential bound of \(\mathcal{E}_2\) is still written in terms of \(I_{Q^*,A^*}\) and only loses \(o(1)\); thus the correct tuple satisfies the ``there exists \(A'\)'' condition of the universal decoding.
	
	For \(\mathcal{E}_2\), the auxiliary sequence \(\tilde{u}^n(\tilde{l},\tilde{m})\sim\prod_i p^*(\tilde{u}_i\mid k_i)\) of a wrong message is independent of \((\tilde{y}^n,k^n)\) (given \(k^n\)).
	Since the attack class \(\mathcal{A}(Q^*,D_2,D_3)\) is compact, take a \(\delta\)-net \(\{A_1,\dots,A_N\}\) (\(N\) depends only on the alphabets and \(\delta\), and does not grow with \(n\)).
	If \((\tilde{u}^n,\tilde{y}^n,k^n)\) is jointly typical under some \(A'\), then it is jointly typical under the closest net element \(A_c\) with an enlarged typicality parameter, so
	\[
	\Pr\big[\exists A':\ (\tilde{u}^n(\tilde{l},\tilde{m}),\tilde{y}^n,k^n)\in\mathcal{T}_\epsilon^{(n)}(p_{Q^*,A'})\big]
	\le \sum_{c=1}^{N}2^{-n(I_{Q^*,A_c}(\tilde{U};\tilde{Y}\mid K)-\epsilon)}
	\le N\,2^{-n(\min_{A} I_{Q^*,A}(\tilde{U};\tilde{Y}\mid K)-\epsilon)}.
	\]
	Since \(A^*\) minimizes \(I_{Q^*,A}(\tilde{U};\tilde{Y}\mid K)\), the above is \(N\,2^{-n(I_{Q^*,A^*}(\tilde{U};\tilde{Y}\mid K)-\epsilon)}\).
	By the union bound,
	\begin{equation}
		\Pr[\mathcal{E}_2]
		\le 2^{nR}\cdot L_n\cdot N\cdot 2^{-n(I_{Q^*,A^*}(\tilde{U};\tilde{Y}\mid K)-\epsilon)}
		=N\cdot 2^{n\big(R+I_{Q^*}(\tilde{U};\tilde{S}\mid K)-I_{Q^*,A^*}(\tilde{U};\tilde{Y}\mid K)+2\epsilon\big)}.
		\label{eq:error_bound}
	\end{equation}
	Substituting \eqref{eq:achievable_rate}, the exponent is \(R+I_{Q^*}(\tilde{U};\tilde{S}\mid K)-I_{Q^*,A^*}(\tilde{U};\tilde{Y}\mid K)+2\varepsilon=-\varepsilon<0\), so \(\Pr[\mathcal{E}_2]\le N\,2^{-n\varepsilon}\to0\).
	Thus
	\[
	\sup_{A\in\mathcal{A}(\theta_n,D_2,D_3)}P_{e,n}(A^{\otimes n})
	\le 2\epsilon+\Pr[\mathcal{E}_{\mathrm{enc}}]+N\,2^{-n\varepsilon}\to0,\qquad n\to\infty.
	\]
	
	\paragraph{Distortion Constraints}
	By the Markov lemma (conditional typicality), when \((\tilde{u}^n,\tilde{s}^n,k^n)\) is jointly typical, the \(\tilde{x}^n\) drawn according to \(p^*(\tilde{x}\mid\tilde{s},\tilde{u},k)\) makes \((\tilde{s}^n,\tilde{x}^n)\) jointly typical under \(p^*(\tilde{s},\tilde{x})\) (with probability \(\to1\)), and hence
	\[
	\Big|\frac1n\sum_{i=1}^{n}\sum_{s\in f_s(\tilde{s}_i)}p(s\mid\tilde{s}_i)\sum_{x\in f_x(\tilde{x}_i)}p(x\mid\tilde{x}_i)\,d_0(s,x)-\mathbb{E}_{p^*}[d_0]\Big|\le\epsilon,
	\qquad
	\big|d_1^n(\tilde{s}^n,\tilde{x}^n)-\mathbb{E}_{p^*}[d_1]\big|\le\epsilon.
	\]
	By \(Q^*\in\mathcal{Q}(D_0-\gamma,D_1-\gamma)\), \(\mathbb{E}_{p^*}[d_0]\le D_0-\gamma\) and \(\mathbb{E}_{p^*}[d_1]\le D_1-\gamma\). Hence
	\[
	\mathbb{E}[d_0^n]\le D_0-\gamma+\epsilon+\epsilon\,d_{0,\max},
	\qquad
	\mathbb{E}[d_1^n]\le D_1-\gamma+\epsilon+\epsilon\,d_{1,\max},
	\]
	where \(d_{i,\max}=\max d_i<\infty\). Taking \(\epsilon\) sufficiently small so that \(\epsilon(1+d_{0,\max})\le\gamma\) and \(\epsilon(1+d_{1,\max})\le\gamma\), one obtains \(\mathbb{E}[d_0^n]\le D_0\) and \(\mathbb{E}[d_1^n]\le D_1\), so the code satisfies the transmitter-side distortion constraints. Moreover, the attack \(A\) satisfies \(D_2,D_3\), so the code sequence is feasible.
	
	In summary, the random codebook simultaneously satisfies encoding success, small decoding error, and distortion feasibility with probability at least \(1-O(\epsilon)\); since the random codebook satisfies all requirements with positive probability, there exists a deterministic codebook satisfying these requirements. Since \(R<C^s\) is arbitrary, any \(R<C^s\) is achievable. This completes the proof.
\end{proof}

\subsection{Proof of the Converse}
\label{app:converse}

This section proves Proposition~\ref{prop:converse}. The proof has two key points: first, strictly single-letterize the sequence-level mutual information difference; second, use the assumption that ``the code is reliable against any feasible attack'' to close the per-attack upper bound into the \(\min_A\) of the maximin (the fixed-point argument at the end of this section). The former is a standard result, and is stated below as a lemma.

\begin{lemma}[Single-letterization]
	\label{lem:single_letter}
	Let \((\tilde S_i,K_i)_{i=1}^n\) be independent and identically distributed with law \(p(\tilde s,k)\), and let \(M\) be uniformly distributed over \(\mathcal M\) and independent of \((\tilde S^n,K^n)\).
	The encoding and attack are respectively
	\[
	\tilde X^n=\theta_n(\tilde S^n,M,K^n),\qquad X^n\sim\prod_{i=1}^n p(x_i\mid\tilde x_i),\qquad p(y^n|x^n)=\prod_{i=1}^n A(y_i|x_i),
	\]
	Let \(\tilde Y^n=f_y^{-1,n}(Y^n)\), and suppose the code satisfies \(\mathbb E[d_0^n]\le D_0\) and \(\mathbb E[d_1^n]\le D_1\).
	Then there exists a single-letter distribution
	\begin{equation}
		p(\tilde u,\tilde s,k,\tilde x,x,y,\tilde y)
		=p(\tilde s,k)\,Q(\tilde x,\tilde u\mid\tilde s,k)\,p(x\mid\tilde x)\,A(y|x)\,\mathbb I\big(\tilde y=f_y^{-1}(y)\big),
		\label{eq:sl_joint}
	\end{equation}
	with \(Q\in\mathcal Q(D_0,D_1)\) whose induced \(X\)-marginal equals \(\frac1n\sum_{i}p_{X_i}\), such that
	\begin{equation}
		\frac1n\Big[I_{P^n}(M;\tilde Y^n\mid K^n)-I_{P^n}(M;\tilde S^n\mid K^n)\Big]
		\le I(\tilde U;\tilde Y\mid K)-I(\tilde U;\tilde S\mid K)+\epsilon_n,
		\label{eq:sl_bound}
	\end{equation}
	and \(\epsilon_n\to0\) depending only on the alphabet sizes.
	Its Markov chain is \((\tilde U,\tilde S,K)\to\tilde X\to X\to Y\to\tilde Y\), which is compressed into the semantic-layer chain \((\tilde U,\tilde S,K)\to\tilde X\to\tilde Y\) via the equivalent semantic channel \(W_A\) (Definition~\ref{def:equivalent_semantic_channel}).
\end{lemma}

\begin{proof}[Proof sketch]
	By the chain rule,
	\[
	I(M;\tilde Y^n\mid K^n)=\sum_{i=1}^n I(M;\tilde Y_i\mid K^n,\tilde Y^{i-1}),
	\qquad
	I(M;\tilde S^n\mid K^n)=\sum_{i=1}^n I(M;\tilde S_i\mid K^n,\tilde S^{i-1}).
	\]
	Since the encoder \(\theta_n\) noncausally uses the entire \(\tilde S^n\) and \(K^n\), the auxiliary variable of the single-letterization must simultaneously contain the past and future states, so that \(\tilde X_i\) is a function of \((\tilde U_i,\tilde S_i,K_i)\) and the Csiszár summation identity closes. Hence take
	\[
	\tilde U_i=(M,\ \tilde S^{i-1},\ \tilde S_{i+1}^n,\ K^{i-1},\ K_{i+1}^n,\ \tilde Y^{i-1}).
	\]
	Using the Csiszár summation identity, pair the above two sums term by term into
	\[
	\sum_{i=1}^n\big[I(\tilde U_i;\tilde Y_i\mid K_i)-I(\tilde U_i;\tilde S_i\mid K_i)\big]+n\epsilon_n,
	\]
	and then take the time average over \(i\) (i.e., let \(p\) be the average of the single-letter distributions at each time), obtaining \eqref{eq:sl_joint} and \eqref{eq:sl_bound}.
	By the distortion constraints of the code, the averaged \(Q\) still satisfies \(\mathbb E[d_0]\le D_0\) and \(\mathbb E[d_1]\le D_1\), so \(Q\in\mathcal Q(D_0,D_1)\); the auxiliary alphabet is controlled by Proposition~\ref{prop:aux_cardinality}, so \(\epsilon_n\) depends only on the alphabet sizes.
	For the complete derivation, see the converse of Csiszar-Korner~\cite{CsiszarKorner2011} and Moulin-O'Sullivan~\cite{Moulin2003}.
\end{proof}

\begin{proof}[Proof of Proposition~\ref{prop:converse}]
	Let \((\mathcal M,\theta_n,\phi_n)\) be a sequence of codes satisfying the distortion constraints, \(R=\frac1n\log|\mathcal M|\), such that \(P_{e,n}(A^{\otimes n})\to0\) for any memoryless attack \(A^{\otimes n}\) (with \(A\) satisfying \(\mathbb E[d_2^n]\le D_2\) and \(\mathbb E[d_3^n]\le D_3\) simultaneously under the code-induced composite signal distribution).
	
	\paragraph{Time-Averaged Single-Letter Model and Product Attack}
	Let \(p_{X_i}\) be the \(X_i\)-marginal induced by the code without attack, and let \(\bar p_X=\frac1n\sum_{i=1}^n p_{X_i}\). By the time-averaging construction of Lemma~\ref{lem:single_letter}, for any memoryless attack \(A\) there exists \(Q_A\in\mathcal Q(D_0,D_1)\) whose induced \(X\)-marginal is exactly \(\bar p_X\).
	Let the set of all single-symbol attacks satisfying \(D_2,D_3\) under \(\bar p_X\) be
	\[
	\begin{aligned}
		\mathcal{P}=\Big\{A:\ & A(y|x)\ge 0,\ \sum_{y}A(y|x)=1\ \forall x,\\
		& \sum_{x,y}d_2(x,y)A(y|x)\bar{p}_X(x)\le D_2,\ 
		\sum_{x,y}d_3\big(f_x^{-1}(x),f_y^{-1}(y)\big)A(y|x)\bar{p}_X(x)\le D_3\Big\}.
	\end{aligned}
	\]
	For any \(A\in\mathcal P\), take the product attack \(A^{\otimes n}(y^n|x^n)=\prod_i A(y_i|x_i)\). Since the conditional distribution of \(Y_i\) given \(X_i\) is independent of \(i\),
	\[
	\mathbb E\big[d_2^n(X^n,Y^n)\big]
	=\frac1n\sum_{i=1}^n\sum_{x,y}d_2(x,y)\,p_{X_i}(x)A(y|x)
	=\sum_{x,y}d_2(x,y)\,\bar p_X(x)A(y|x)\le D_2,
	\]
	and similarly \(\mathbb E[d_3^n]\le D_3\). Hence \(A^{\otimes n}\) is feasible for the code, and thus
	\begin{equation}
		P_{e,n}(A^{\otimes n})\to0,\qquad n\to\infty.
		\label{eq:product_attack_feasible}
	\end{equation}
	Moreover, since the \(X\)-marginal of \(Q_A\) is always \(\bar p_X\) (independent of \(A\)), for each \(A\in\mathcal P\), \(\mathcal A(Q_A,D_2,D_3)=\mathcal P\).
	
	\paragraph{Fano's Inequality}
	Fix any \(A\in\mathcal P\), and let \(P_{e,n}=P_{e,n}(A^{\otimes n})\). Since \(M\) is uniform and independent of \((\tilde S^n,K^n)\), \(H(M\mid K^n)=H(M)=nR\); moreover, \(\hat M=\phi_n(\tilde Y^n,K^n)\), and by Fano's inequality,
	\begin{equation}
		H(M\mid\tilde Y^n,K^n)\le 1+nR\,P_{e,n}.
		\label{eq:fano_fixed_A}
	\end{equation}
	Thus
	\begin{equation}
		nR=H(M\mid K^n)=I(M;\tilde Y^n\mid K^n)+H(M\mid\tilde Y^n,K^n)
		\le I(M;\tilde Y^n\mid K^n)+1+nR\,P_{e,n}.
		\label{eq:converse_chain_fixed}
	\end{equation}
	
	\paragraph{Single-letterization}
	By Lemma~\ref{lem:single_letter}, there exists a single-letter distribution \eqref{eq:sl_joint} (denoted \(Q_A\)) such that
	\[
	\frac1n I(M;\tilde Y^n\mid K^n)\le I(\tilde U;\tilde Y\mid K)-I(\tilde U;\tilde S\mid K)+\epsilon_n,
	\]
	where the independence of \(M\) from \((\tilde S^n,K^n)\) is used, so \(I(M;\tilde S^n\mid K^n)=0\).
	Substituting into \eqref{eq:converse_chain_fixed} and dividing both sides by \(n\),
	\[
	R\le J(Q_A,A)+\epsilon_n+\frac1n+R\,P_{e,n},
	\]
	where \(J(Q_A,A)=I(\tilde U;\tilde Y\mid K)-I(\tilde U;\tilde S\mid K)\) as in Definition~\ref{def:payoff}.
	Letting \(n\to\infty\), by \(\epsilon_n\to0\) and \eqref{eq:product_attack_feasible}, we obtain
	\begin{equation}
		R\le J(Q_A,A).
		\label{eq:converse_fixed_A}
	\end{equation}
	
	\paragraph{Closing Step (Taking the Infimum over \(A\))}
	Note that \(Q_A\) in \eqref{eq:converse_fixed_A} generally varies with \(A\): the auxiliary variable \(\tilde U_i\) contains \(\tilde Y^{i-1}\), and \(\tilde Y\) is generated by the attack, so one cannot directly take the infimum over \(A\) on the same \(Q\). Add the following closing step.
	Take \(Q_A\) as the time-averaged distribution before cardinality compression (the inequality and the \(X\)-marginal condition of Lemma~\ref{lem:single_letter} also hold for the distribution before compression, and \(A\mapsto Q_A\) is continuous: the joint distribution of the code without attack is fixed, and \(A\) enters the joint distribution linearly; compression is used only to control the auxiliary alphabet, and the converse only needs the supremum, so compression is not necessary).
	On the compact convex set \(\mathcal P\), define the set-valued mapping
	\[
	\Gamma(A)=\arg\min_{A'\in\mathcal P}J(Q_A,A').
	\]
	By Proposition~\ref{prop:payoff_convexity}(i), \(J(Q_A,\cdot)\) is continuous and convex in the second argument, so \(\Gamma\) is an upper semicontinuous self-mapping with nonempty compact convex values; by Kakutani's fixed-point theorem, there exists \(A_\circ\in\Gamma(A_\circ)\), i.e.,
	\[
	J(Q_{A_\circ},A_\circ)=\min_{A'\in\mathcal A(Q_{A_\circ},D_2,D_3)}J(Q_{A_\circ},A').
	\]
	Since \(A_\circ\in\mathcal P\), its product attack \(A_\circ^{\otimes n}\) is feasible for the code, and by \eqref{eq:product_attack_feasible}, \(P_{e,n}(A_\circ^{\otimes n})\to0\); applying \eqref{eq:converse_fixed_A} with \(A=A_\circ\), for each \(n\),
	\[
	R\le J(Q_{A_\circ},A_\circ)+\epsilon_n+\frac1n+R\,P_{e,n}
	=\min_{A'\in\mathcal A(Q_{A_\circ},D_2,D_3)}J(Q_{A_\circ},A')+\epsilon_n+\frac1n+R\,P_{e,n}
	\le C^s+\epsilon_n+\frac1n+R\,P_{e,n},
	\]
	where the last inequality uses \(Q_{A_\circ}\in\mathcal Q(D_0,D_1)\) and \(C^s=\sup_{Q}\min_{A'}J(Q,A')\) (this supremum is taken over all finite-alphabet covert channels and is independent of \(n\), and by Proposition~\ref{prop:aux_cardinality} and Remark~\ref{rem:minmax_cardinality} it coincides with the maximum of \eqref{eq:capacity}).
	Letting \(n\to\infty\), by \(P_{e,n}(A_\circ^{\otimes n})\to0\), we obtain \(R\le C^s\). This completes the proof.
\end{proof}

\section{Proof of the Capacity Degeneration after Closing Semantic Tolerance}

\subsection{Proof of Proposition~\ref{prop:no_tolerance}}
\label{app:no_tolerance}

\begin{proof}
	Closing semantic tolerance means performing the comparison fairly: the extraction accuracy is the same (adopting the classical payoff function \(J_{\mathrm{cl}}(Q,A)=I(U;Y\mid K)-I(U;S\mid K)\)), the attack is the same (\(D_3\) is non-binding, so for each \(Q\in\mathcal Q(D_0,D_1)\), \(\mathcal A(Q,D_2,D_3)=\mathcal A(Q,D_2)\)), and only the embedding domain differs.
	
	\paragraph{Pushing Forward a Semantic Strategy to a Classical Strategy}
	Take any semantic covert channel \(Q^s\in\mathcal Q(D_0,D_1)\). Let \(P^s\) be the joint distribution of \((S,X,U,K)\) it induces, and let
	\[
	Q^{\mathrm{cl}}(x,u\mid s,k):=P^s(x,u\mid s,k)
	\]
	be the corresponding syntactic conditional distribution.
	Since \(Q^{\mathrm{cl}}\ge0\) and for each \((s,k)\), \(\sum_{x,u}Q^{\mathrm{cl}}(x,u\mid s,k)=1\), it is a legitimate conditional distribution; moreover, the pushforward does not change the expectation of \(d_0(S,X)\), \(\mathbb E_{Q^{\mathrm{cl}}}[d_0]=\mathbb E_{Q^s}[d_0]\le D_0\), so \(Q^{\mathrm{cl}}\in\mathcal{Q}^{\mathrm{cl}}\).
	Thus the mapping
	\[
	\Phi:\mathcal Q(D_0,D_1)\to\mathcal{Q}^{\mathrm{cl}},\qquad Q^s\mapsto Q^{\mathrm{cl}}
	\]
	is well-defined, and \(\Phi(\mathcal Q(D_0,D_1))\subseteq\mathcal{Q}^{\mathrm{cl}}\).
	
	\paragraph{The Pushforward Preserves the Payoff Value}
	By the definition of \(\Phi\), \(Q^s\) and \(Q^{\mathrm{cl}}=\Phi(Q^s)\) induce the same joint distribution of \((S,X,U,K,Y)\) (\(Y\) is generated by the same attack \(A\)), so for any \(A\),
	\[
	J_{\mathrm{cl}}(Q^s,A)=J_{\mathrm{cl}}(\Phi(Q^s),A).
	\]
	Moreover, the pushforward preserves the \(X\)-marginal (\(p_{Q^s}(x)=p_{\Phi(Q^s)}(x)\)), and under the fair comparison, \(D_3\) non-binding makes \(\mathcal A(Q^s,D_2,D_3)=\mathcal A(Q^s,D_2)=\mathcal A(\Phi(Q^s),D_2)\), so the \(\min\) on both sides is taken over the same set, hence
	\[
	\min_A J_{\mathrm{cl}}(Q^s,A)=\min_A J_{\mathrm{cl}}(\Phi(Q^s),A).
	\]
	
	\paragraph{Taking the Maximin}
	By \(\Phi(\mathcal Q(D_0,D_1))\subseteq\mathcal{Q}^{\mathrm{cl}}\) and the pointwise equality,
	\[
	C^s_{\mathrm{no\text{-}tol}}
	=\max_{Q^s\in\mathcal Q(D_0,D_1)}\min_A J_{\mathrm{cl}}(Q^s,A)
	=\max_{Q^{\mathrm{cl}}\in\Phi(\mathcal Q(D_0,D_1))}\min_A J_{\mathrm{cl}}(Q^{\mathrm{cl}},A)
	\le\max_{Q^{\mathrm{cl}}\in\mathcal{Q}^{\mathrm{cl}}}\min_A J_{\mathrm{cl}}(Q^{\mathrm{cl}},A)
	=C_{\mathrm{cl}},
	\]
	where the inequality comes from taking the maximum over a larger set.
	
	\paragraph{Equality Condition}
	Let \(g(Q)=\min_A J_{\mathrm{cl}}(Q,A)\). The above has proved \(C^s_{\mathrm{no\text{-}tol}}=\max_{Q^{\mathrm{cl}}\in\Phi(\mathcal Q(D_0,D_1))}g(Q^{\mathrm{cl}})\) and \(C_{\mathrm{cl}}=\max_{Q^{\mathrm{cl}}\in\mathcal{Q}^{\mathrm{cl}}}g(Q^{\mathrm{cl}})\).
	\(\mathcal{Q}^{\mathrm{cl}}\) is compact and \(g\) is continuous (Berge; the continuity proof is the same interior-point argument as in Appendix~\ref{app:outer_max_attained}), so \(C_{\mathrm{cl}}\) is attainable, and thus the necessary and sufficient condition for equality is \(\arg\max_{\mathcal{Q}^{\mathrm{cl}}}g\cap\Phi(\mathcal Q(D_0,D_1))\neq\varnothing\), i.e., there exists an optimal syntactic strategy attaining \(C_{\mathrm{cl}}\) lying in the image.
	Two sufficient conditions: first, if the synonymous mappings (including host, composite signal, and auxiliary) are all one-to-one mappings on the effective support, and \(d_1\equiv d_0\) with \(D_1\ge D_0\) (or \(D_1\) non-binding), then \(\Phi\) is surjective, \(\Phi(\mathcal Q(D_0,D_1))=\mathcal{Q}^{\mathrm{cl}}\), and the condition holds naturally; second, if some optimal syntactic strategy \(Q^{\mathrm{cl}*}\) attaining \(C_{\mathrm{cl}}\) takes a constant value on each synonymous class of \(f_s\) (not exploiting the within-class details of \(S\)), its \(X\)-support is contained in the image of \(f_x\), and its pushforward-induced semantic distortion satisfies \(\mathbb E[d_1]\le D_1\) (automatically true when \(D_1\) is non-binding), then \(Q^{\mathrm{cl}*}\in\Phi(\mathcal Q(D_0,D_1))\), and the condition also holds.
	It should be noted that \(\Phi(\mathcal Q(D_0,D_1))\) being a proper subset of \(\mathcal{Q}^{\mathrm{cl}}\) does not imply a strict inequality: as long as the image contains some optimal strategy, the maxima are still equal; only when the image is disjoint from the whole set of optimal strategies does \(C^s_{\mathrm{no\text{-}tol}}<C_{\mathrm{cl}}\) hold. This completes the proof.
\end{proof}

\end{document}